\documentclass{emulateapj}
\usepackage{graphics,graphicx}
\usepackage{epstopdf}
\usepackage{amssymb, amsmath, amsthm}

\usepackage{color}
\usepackage{xcolor}

\shortauthors{Lin et al.}

\shorttitle{SED Fittings for OB Cluster-Forming Regions}
\shortauthors{Lin et al.}

\begin{document}

\title{Cloud Structure of Galactic OB Cluster Forming Regions from Combining Ground and Space Based Bolometric Observations}

\author{Yuxin Lin\altaffilmark{1}}
\author{Hauyu Baobab Liu\altaffilmark{2}}
\author{Di Li\altaffilmark{1,3}}
\author{Zhiyu Zhang \altaffilmark{2,4}}
\author{Adam Ginsburg\altaffilmark{2}}
\author{Jaime E. Pineda\altaffilmark{5}}
\author{Lei Qian\altaffilmark{1}}
\author{Roberto Galv\'{a}n-Madrid\altaffilmark{6}}
\author{Anna Faye McLeod\altaffilmark{2}}
\author{Erik Rosolowsky\altaffilmark{7}}
\author{James E. Dale\altaffilmark{8,9}}
\author{Katharina Immer\altaffilmark{2}}
\author{Eric Koch\altaffilmark{7}}
\author{Steve Longmore\altaffilmark{10}}
\author{Daniel Walker\altaffilmark{10}}
\author{Leonardo Testi\altaffilmark{2}}

\affil{$^{1}$National Astronomical Observatories, Chinese Academy of Sciences; \textcolor{blue}{yuxinlin@bao.ac.cn}}
\affil{$^{2}$European Southern Observatory (ESO), Karl-Schwarzschild-Str. 2, D-85748 Garching, Germany}
\affil{$^{3}$Key Laboratory of Radio Astronomy, Chinese Academy of Sciences}
\affil{$^{4}$Institute for Astronomy, University of Edinburgh, Royal Observatory, Blackford Hill, Edinburgh EH9 3HJ, UK}
\affil{$^{5}$Max-Planck-Institut f\"{u}r Extraterrestrische Physik, Gießenbach-str 1, D-85748, Garching bei M\"{u}nchen, Germany}
\affil{$^{6}$Instituto de Radioastronom\'ia y Astrof\'isica, UNAM, Apdo. Postal 3-72 (Xangari), 58089 Morelia, Michoac\'an, M\'exico.}
\affil{$^{7}$Department of Physics, University of Alberta, 4-181 CCIS, Edmonton, AB T6G 2E1, Canada}
\affil{$^{8}$Universit\"{a}ts--Sternwarte M\"{u}nchen, Scheinerstr. 1, 81679 M\"{u}nchen, Germany}
\affil{$^{9}$Excellence Cluster `Universe', Boltzmannstr. 2, 85748 Garching, Germany}
\affil{$^{10}$Astrophysics Research Institute, Liverpool John Moores University, 146 Brownlow Hill, Liverpool L3 5RF, UK}

\begin{abstract}
We have developed an iterative procedure to systematically combine the millimeter and submillimeter images of OB cluster-forming molecular clouds, which were taken by ground based (CSO, JCMT, APEX, IRAM-30m) and space telescopes (Herschel, Planck).
For the seven luminous ($L$$>$10$^{6}$ $L_{\odot}$) Galactic OB cluster-forming molecular clouds selected for our analyses, namely W49A, W43-Main, W43-South, W33, G10.6-0.4, G10.2-0.3, G10.3-0.1, we have performed single-component, modified black-body fits to each pixel of the combined (sub)millimeter images, and the Herschel PACS and SPIRE images at shorter wavelengths.
The $\sim$10$''$ resolution dust column density and temperature maps of these sources revealed dramatically different morphologies, indicating very different modes of OB cluster-formation, or parent molecular cloud structures in different evolutionary stages.
The molecular clouds W49A, W33, and G10.6-0.4 show centrally concentrated massive molecular clumps that are connected with approximately radially orientated molecular gas filaments.
The W43-Main and W43-South molecular cloud complexes, which are located at the intersection of the Galactic near 3-kpc (or Scutum) arm and the Galactic bar, show a widely scattered distribution of dense molecular clumps/cores over the observed  $\sim$10 pc spatial scale.
The relatively evolved sources G10.2-0.3 and G10.3-0.1 appear to be affected by stellar feedback, and show a complicated cloud morphology embedded with abundant dense molecular clumps/cores. 
We find that with the high angular resolution we achieved, our visual classification of cloud morphology can be linked to the systematically derived statistical quantities (i.e., the enclosed mass profile, the column density probability distribution function, the two-point correlation function of column density, and the probability distribution function of clump/core separations).
In particular, the massive molecular gas clumps located at the centre of G10.6-0.4 and W49A, which contribute to a considerable fraction of their overall cloud masses, may be special OB cluster-forming environments as a direct consequence of global cloud collapse.
These centralized massive molecular gas clumps also uniquely occupy much higher column densities than what is determined by the overall fit of power-law column density probability distribution function. 
We have made efforts to archive the derived statistical quantities of individual target sources, to permit comparisons with theoretical frameworks, numerical simulations, and other observations in the future.
\end{abstract}


\section{Introduction}
\footnote{This paper and all related analysis code are available on the web at https://github.com/yxlinaqua/paper\_cloud\_structures\_OB.git} The {\it Herschel} far infrared and submillimeter imaging observations on the Galactic plane have significantly advanced our knowledge about the morphology as well as the temperature and density distributions of molecular clouds \citep{Andre10,Molinari10}, including those of OB cluster forming regions \citep{Motte10}.
Recent {\it Planck} images at (sub)millimeter and millimeter wavelengths further revealed an extremely cold population of dense molecular cores, namely the Planck Cold Clumps \citep[PCCs,][]{PlanckClumps11,PlanckClumps15}. 
These observations have provided a large sample of star-forming molecular clumps/cores in a broad range of evolutionary stages. 
In addition, the better angular resolution of the {\it Spitzer} near-infrared ($\sim$2$''$ at 3.6-8 $\mu$m) and {\it Herschel} 70 and 160 $\mu$m ($\sim$6$''$ and 12$''$) bands can help identify heating sources (normally young stars) in the molecular clouds \citep{Churchwell09,Stutz13}, which help our understanding of their star-formation history.
Studies of the parent molecular cloud properties require observations at (sub)millimeter wavelengths, which are typically more sensitive to the thermal emission of the 10-30 K interstellar medium from the optically thin part of the spectrum.
The quantitative analyses including spectral energy distribution (SED) fitting, need to be adjusted to the angular resolution of the longest wavelength observations (e.g. 37$''$ for {\it Herschel} observations at 500 $\mu$m wavelength).
For cold molecular clumps which typically have spatial scales of $\sim$0.5 pc, the population will be resolved for molecular clouds within d$\sim$3 kpc, such as the Rosette and Carina molecular clouds  (e.g. Schneider et al. 2012; Rebolledo et al. 2016), while the majority of the distant population of e.g. $\sim$17$''$ at $d$$\sim$6 kpc \citep{BerginTafalla07} and represents the initial state of star-formation in giant molecular clouds (GMCs), and are therefore easily missed in the derived smeared and poor angular resolution temperature and column density maps.
The contrast of the localized heating sources can also be significantly suppressed due to this spatial smearing.

Ground-based millimeter and submillimeter bolometric observations can probe long wavelength emission of cold clumps at high angular resolution 
compared to space telescopes \citep{Schuller09,Aguirre11,Ginsburg13,Merello15}. 
However, ground-based observations are often subject to extended and strong atmospheric emission at these wavelengths, which is not always distinguishable from the extended emission of the molecular clouds.
The atmospheric foreground subtraction procedures may lead to significant missing flux from the molecular cloud, which will bias the observed fluxes of bright sources, and may leave the adjacent fainter sources partially or fully immersed in regions of negative brightness, known as negative `bowls'.
The details of how the extended structures will be missed in the ground-based bolometric imaging will also depend on the sky condition during the observations, which can not be easily predicted with accuracy.  
Therefore, it is not trivial to correctly fit the SED solely with the ground-based bolometric images for the longer wavelengths, and may prevent the identification and quantitative studies of some faint structures.

Sadavoy et al. (2013) have proposed that artificially filtering out the extended structures from both the ground-based (sub)millimeter bolometric images and from the shorter wavelength images taken by the space telescopes can alleviate the bias when fitting the SED of compact sources. 
However, in this way, the properties of extended molecular cloud structures cannot be derived, and the impact of negative brightness bowls still exists.
Alternatively, Liu et al. (2015) proposed that merging ground-based and space telescope observations at the same wavelengths can avoid the drawbacks of each technique.  
This is because the space telescope observations, which normally have poorer angular resolution than the ground-based ones, are not subject to the atmospheric foreground emission.
The extended structure recovered from space-telescope observations can therefore be used to complement the ground based images to achieve high angular resolution images without missing flux.
Csengeri et al. (2016) also applied this technique to combine the Atacama Pathfinder Experiment (APEX) telescope Large Area Survey of the Galaxy (ATLASGAL) 870 \micron\ images with the Planck/HFI 353 GHz images.

\begin{figure}
\vspace{-0.5cm}
\hspace{2.0cm}
\begin{tabular}{ p{0.45\linewidth} }
\includegraphics[scale=0.5]{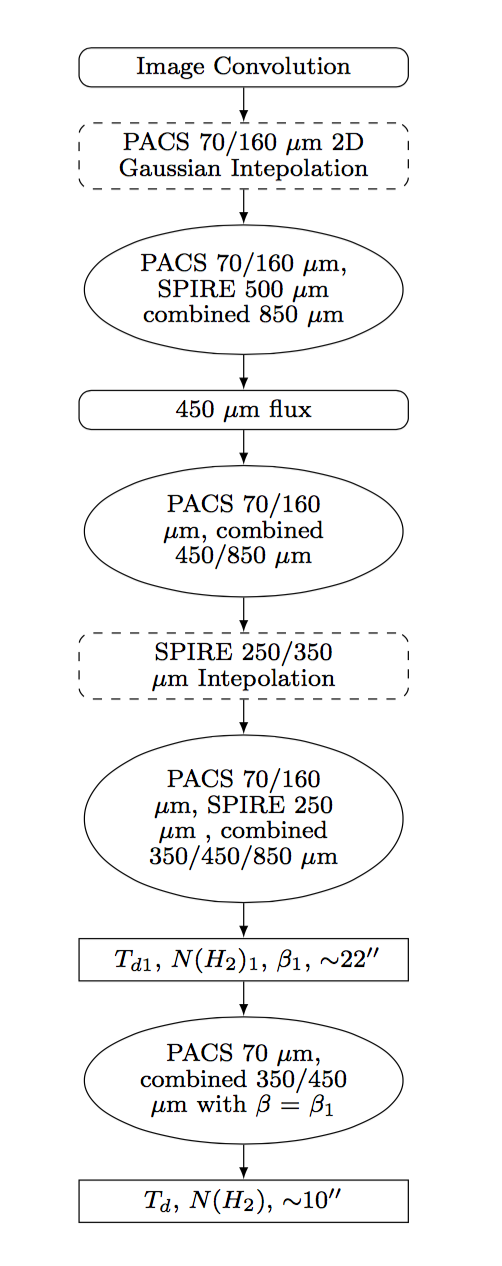} \\
\end{tabular}
\caption{The image combination and SED fitting procedure, including the iterative processes to derive the saturated pixels or interpolate the corresponding space telescope image for several wavelengths observations. Each ellipse in the flow chart represents a SED fitting process. For more description see Section \ref{subsection:procedure}.
}
\label{fig:floatchart}
\vspace{0.5cm}
\end{figure}

In this paper, we use the technique proposed by Liu et al. (2015) to analyze the dust emission SED for a sample of very luminous OB cluster-forming regions, including the mini-starburst W49A \citep{Nagy12,GM13,Wu16} and W43 \citep{Bally10,NguyenLuong11,Louvet14}, the W33 complex \citep{Immer14,Messineo15}, the regions G10.2--0.3 and G10.3--0.1 in the W31 complex \citep{Blum01,KimKoo02,Beuther11}, and G10.6--0.4 \citep{Liu10a,Liu10b,Liu11,Liu12}.

We demonstrate that the proposed technique can provide high quality maps of the distribution of dust temperature and column density. 
The maps can be used to probe the origin of large-scale heating due to, e.g., illumination from the embedded OB-stars or shocks caused by supernovae or cloud-cloud collisions.
We also investigate the column density distribution function, density profiles, and the spatial correlation of dense structures.
These quantities are fundamental to our understanding of massive star cluster formation, since the structures of molecular clouds can reflect their initial condition, and their potential subsequent evolution due to gravitational contraction (Lada \& Lada 2003; V{\'a}zquez-Semadeni et al. 1995, 2007, 2009; Li et al. 2016).
Therefore, they may be linked to the origin of different modes of star-formation, and may additionally reveal signatures of interaction between newly formed stars with their natal clouds.
We update the identification of dense cores/clumps in these regions, and derive the mass, averaged temperature, and bolometric luminosity of cores and clumps.  
Finally, we investigate the spatial distribution/segregation of the cold and hot cores/clumps, which may give clues on the fragmentation processes of these molecular clouds.

Our observations and data analysis procedures are outlined in Section \ref{section:obs}.
Our results are provided in Section \ref{section:results}.
Comparison among the observed star-forming regions is given in Section \ref{section:discussion}.
Our conclusion and ending remarks are in Section \ref{section:conclusion}.

\begin{table*}\footnotesize{ 
\caption{Source information}
\label{tab:source_info}
\hspace{-0.7cm}
\begin{tabular}{lcccccccc}\hline\hline

Target Source&RA&DEC & Distance & Mass$^{\mbox{\tiny{a}}}$ & Luminosity$^{\mbox{\tiny{a}}}$& Reference$^{\mbox{\tiny{b}}}$ \\
 &(J2000)  &(J2000)  &(kpc)& $(\mathrm{M_{\odot}})$ & $(\mathrm{L_{\odot}})$\\

\hline

W49A &19$^{\mbox{\scriptsize{h}}}$10$^{\mbox{\scriptsize{m}}}$13$^{\mbox{\scriptsize{s}}}$.000&09$^{\circ}$06$'$00$''$.00&$11.4^{+1.2}_{-1.2}$&2.30$\times10^{5}$&3.67$\times10^{7}$& Zhang et al. (2013)\\     

W43-main &18$^{\mbox{\scriptsize{h}}}$47$^{\mbox{\scriptsize{m}}}$36$^{\mbox{\scriptsize{s}}}$.427&-01$^{\circ}$59$'$02$''$.48&$5.49^{+0.39}_{-0.34}$&1.32$\times10^{5}$
&1.23$\times10^{7}$& Zhang et al. (2014)\\

W43-south &18$^{\mbox{\scriptsize{h}}}$46$^{\mbox{\scriptsize{m}}}$02$^{\mbox{\scriptsize{s}}}$.084&-02$^{\circ}$43$'$00$''$.83&$5.49^{+0.39}_{-0.34}$&6.43$\times10^{4}$
&6.29$\times10^{6}$& Zhang et al. (2014)\\

G10.2-0.3&18$^{\mbox{\scriptsize{h}}}$09$^{\mbox{\scriptsize{m}}}$23$^{\mbox{\scriptsize{s}}}$.000&-20$^{\circ}$16$'$17$''$.00&$4.95^{+0.51}_{-0.43}$&1.03$\times10^{5}$&6.43$\times10^{6}$&Corbel \& Eikenberry (2004)$^{\mbox{\tiny{d}}}$\\

G10.6-0.4 &18$^{\mbox{\scriptsize{h}}}$10$^{\mbox{\scriptsize{m}}}$29$^{\mbox{\scriptsize{s}}}$.26&-19$^{\circ}$55$'$59$''$.5&$4.95^{+0.51}_{-0.43}$&2.33$\times10^{4}$&3.20$\times10^{6}$& Sanna et al. (2014)\\

W33  &18$^{\mbox{\scriptsize{h}}}$14$^{\mbox{\scriptsize{m}}}$13$^{\mbox{\scriptsize{s}}}$.65& -17$^{\circ}$55$'$38$''$.9&$2.40^{+0.17}_{-0.15}$&3.63$\times10^{4}$&2.43$\times 10^{6}$& Immer et al. (2013)\\

G10.3-0.1&18$^{\mbox{\scriptsize{h}}}$08$^{\mbox{\scriptsize{m}}}$58$^{\mbox{\scriptsize{s}}}$.000&-20$^{\circ}$05$'$15$''$.00&$3.22^{+0.12}_{-0.12}$/$2.56^{+0.28}_{-0.28}$&2.70$\times10^{4}$/1.70$\times 10^{4}$&1.54$\times10^{6}$/9.71$\times10^{5}$&Kazi Rygl, Katharina Immer$^{\mbox{\tiny{c}}}$\\

\hline                     
\end{tabular}
\vspace{0.1cm}
\footnotesize{Note. 
}
\par
\scriptsize{
\begin{itemize}
\item[$^{\mbox{\scriptsize{a}}}$] Total masses were summed from our derived column density maps (Figure \ref{fig:TN_W43M}-\ref{fig:TN_G10p2}) above a common threshold of $7\times 10^{21}$cm$^{-2}$. Considering the threshold chosen for adding up total masses, we note that these mass values should be considered as lower limits of these sources.
Total bolometric luminosity is calculated by integrating from 0.1 \micron\ to 1 cm of the obtained SED for each pixel, and adding all the values in each field.
\item[$^{\mbox{\scriptsize{b}}}$]Distances were quoted from the Bar and Spiral Structure Legacy Survey (BeSSeL; e.g. Brunthaler et al. 2011) water maser trigonometric parallaxes if without further notes.
\item[$^{\mbox{\scriptsize{c}}}$] Preliminary results based on 6.7 GHz masers, which requires further confirmation (Kazi Rygl, Katharina Immer, private communication). In this work, we used the larger distance value for analysis on source G10.3-0.1. \vspace{-0.15cm}
\item[$^{\mbox{\scriptsize{d}}}$] Since the distance of G10.2-0.3 is still uncertain, we now adopted the same distance as with G10.6-0.4. Corbel \& Eikenberry (2004) point out that the two sources are likely to be located at approximately the same distance in the -30 kms$^{-1}$ spiral arm (sometimes called the 4 kpc arm, Menon \& Ciotti 1970; Greaves \& Williams 1994). \vspace{-0.15cm}

\end{itemize}
}
}
\vspace{0.4cm}
\end{table*}

\section{Observations}\label{section:obs}
We provide a brief description of our target sources in Section \ref{subsection:target}.
Details of our Caltech Submillimeter Observatory (CSO) Submillimetre High Angular Resolution Camera II (SHARC2) 350 $\mu$m observations and data reductions are given in Section \ref{subsection:sharc2}.
Section \ref{subsection:archive} outlines the archival data we included for the SED analysis.
Our procedures for producing the final images and SED fitting are given in Section \ref{subsection:procedure}.

\subsection{Target Sources}\label{subsection:target}
Table \ref{tab:source_info} summarizes the basic properties of the selected target sources.
More details on the individual sources are given in the following subsections.

\subsubsection{W49A mini-starburst}
W49A is located in the Galactic disk at coordinates 
$(l,b)$=$(43.1^\circ,0.0^\circ)$. 
The previous measurement of its bolometric luminosity 
\citep{Sievers91} gives $L_\mathrm{bol}\approx10^{7.2}~L_\odot$ at the refined 
parallax distance of $d=11.4\pm1.2$ kpc \citep{Zhang13}.
The associated GMC has an extent of $ l\sim 100$ pc, 
but most star formation resides in the central $\sim 20$ pc 
\citep{GM13}. This inner region contains the well known massive star formation 
regions W49 north (W49N), W49 south 
(W49S, $\sim 2\arcmin$ southeast of W49N), and W49 southwest 
(W49SW, $\sim 1.5\arcmin$ southwest of W49N). W49N is by far the densest, 
most massive `hub' within W49A \citep{GM13}. 
The deeply embedded population of young massive 
stars in W49N is only revealed by dozens of radio-continuum hypercompact  and ultracompact 
H\textsc{ii} regions (Welch et al. 1987; De Pree et al. 1997; McLeod et al., in prep.) clustered within a radius of a few pc. At somewhat lower extinction, part of the stellar population can be seen at infrared wavelengths \citep{HomeierAlves05, Saral16}. Recently, several stars more massive than $100~M_\odot$ have been confirmed by infrared spectrophotometry in W49N \citep{Wu14,Wu16}.

\subsubsection{W43 mini-starburst}
The W43 molecular cloud complex is located at the connection point of the Galactic near 3-kpc arm (Scutum arm) and the near end of the Galactic bar, at a distance of 5.5 kpc from 
the Sun \citep{Zhang14}. It represents the nearest example of extreme star formation possibly caused by the interaction of the Galactic bar with the spiral arms \citep{NguyenLuong11}. This complex consists of two main, connected 
clumps: W43-main $(l,b)$=$(30.8^\circ,0.02^\circ)$ and W43-south  $(l,b)$=$(29.96^\circ,-0.02^\circ)$. A series of recent studies mapping the molecular gas in CO and denser gas tracers have shown that W43 possesses starburst conditions \citep{Motte03, Carlhoff13,Louvet14}.
The giant H\textsc{ii} region near W43-main is estimated to have a far-infrared continuum luminosity of $\sim 3.5\times\ 10^{6}\ L_{\odot}$ (Smith et al. 1978; Blum et al. 1999; Bik et al. 2005). Near-infrared spectroscopy studies have revealed that there are hot, massive stars at the core of this H\textsc{ii} region with the brightest one in the cluster identified as a Wolf-Rayet star (Blum et al. 1999).

\subsubsection{G10.2-0.3 and G10.3-0.1}
G10.2-0.3 is a giant H\textsc{ii} region that produces $> 10^{50}$ Lyman continuum photons per second.
A dense stellar cluster is revealed by near-infrared images, in which four of the brightest members are identified as early O-type stars with other embedded young stellar object (YSO) candidates located in the heart of the cluster (Blum et al. 2001). 
The distance of this region remains ambiguous, with a spectrophotometry distance determined from O-stars of 3.4$\pm$0.3 kpc, and a kinematic distance based on high resolution CO spectroscopy combining radio recombination lines of 4.5 kpc (Corbel \& Eikenberry). 

G10.3-0.1 is a bipolar H\textsc{ii} region exhibiting an ionized central region with lobes extending perpendicular to the dense elongated filament (Kim \& Koo 2001; Deharveng et al. 2015).
Multiple class 0/\textsc{I}/\textsc{II} YSO candidates are 
associated with the H\textsc{ii} region, and triggered star formation from the interaction of the H\textsc{ii} region with the surrounding dense material is suggested based on a multi-wavelength analysis (Deharveng et al. 2015).
 
Large amounts of dense gas (e.g. traced by line emission of CS molecule) are detected towards G10.2-0.3 and G10.3-0.1, which indicate their fertility of forming massive stars (Kim \& Koo 2001).  Different evolutionary stages are suggested for the two regions: G10.2-0.3 is more evolved with more widely distributed YSO candidates, as it has had more time to stir up its natal molecular clouds resulting in a larger line width (Beuther et al. 2011). 

\subsubsection{G10.6-0.4}
The molecular cloud G10.6-0.4 is a $\sim10^{6}$$L_{\odot}$ \color{black} OB cluster-forming region, located at a distance of 4.95 \color{black} kpc (Sanna et al. 2014).
Several H\textsc{ii} and ultra-compact HII (UC H\textsc{ii}) regions are already present across the central $\sim$10 pc area, which suggests simultaneous massive star-formation over the entire dense molecular cloud (Ho et al. 1986; Sollins \& Ho 2005).
The high angular resolution observations of molecular lines and dust continuum emission towards this region have resolved an overall hierarchically collapsing, hub-filament morphology (Myers 2009, 2011) connecting from the $\sim$10 pc scale down to the central $\sim$1 pc scale flattened rotating accretion flow (Keto et al. 1987; Liu et al. 2010a, 2011, 2012a).

A condensed cluster of bright infrared sources (Liu et al. 2012a) and several high velocity molecular outflows were found around the central $\sim$1 pc scale where the large-scale filaments converge.
Detailed interferometric studies of the HCN 3-2 absorption line features further detected signatures of molecular cloud/core collapsing from several localized regions over the central $\sim$1 pc scale area (Liu et al. 2013a).

\subsubsection{W33 molecular cloud complex}
Westerhout (1958) detected the W33 complex as a thermal radio source in their 1.4 GHz survey.
Submillimeter observations of W33 with the Atacama Pathfinder Experiment (APEX) telescope (Schuller et al. 2009) resolved the complex into three larger (W33\,B, W33\,A, W33\,Main) and three smaller molecular clouds (W33\,B1, W33\,A1, W33\,Main1).
Water and methanol masers were detected towards W33\,B, W33\,A, and W33\,Main (e.g. Genzel \& Downes 1977; Jaffe et al. 1981; Haschick et al. 1990).
Parallax observations of these water masers yield a distance of 2.4 kpc to the W33 complex, locating it in the Scutum spiral arm (Immer et al. 2013). 
A cluster of zero age main sequence (ZAMS) stars with spectral types between O7.5 and B1.5 (revised for a distance
of 2.4 kpc) was detected in the most massive molecular cloud W33\,Main (Dyck \& Simon 1977; Haschick \& Ho 1983). 
The total bolometric luminosity of the complex is $\sim$10$^{6}$ L$_{\sun}$ (Stier et al. 1984; revised for a distance of 2.4 kpc).
Spectral line observations by Immer et al. (2014) of the six clouds in W33 with the APEX telescope and the Submillimeter Array (SMA) showed that the sources follow an evolutionary sequence from quiescent pre/protostellar clouds (W33\,Main1, W33\,A1, W33\,B1) over hot cores (W33\,B, W33\,A) to HII regions (W33\,Main).

\subsection{CSO-SHARC2 Observations}\label{subsection:sharc2}
High angular resolution, ground based continuum observations at 350 \micron\ towards W43-Main, W43-South, W49A, W33, G33.92+0.11, G10.2-0.3, G10.3-0.1, and G10.6-0.4 were carried out using the SHARC2 bolometer array, installed on the CSO Telescope.
The array consists of 12$\times$32 pixels \footnote{Approximately 85\% of these pixels work well according to the online documentation http://www.astro.caltech.edu/~sharc/}.
The simultaneous field of view (FOV) provided by this array is 2$'$.59$\times$0$'$.97, and the diffraction limited beam size is $\sim$8$''$.8.

The data of W43-Main, W43-South, W49A, and G33.92+0.11 were acquired on March 24th 2014 ($\tau_{\mbox{\scriptsize{225 GHz}}}$$\sim$0.06), with an on-source exposure time of 20, 20, 42, and 30 minutes respectively. 
W33 was observed on March 27th 2014 ($\tau_{\mbox{\scriptsize{225 GHz}}}$$\sim$0.05), with 50 minutes of on-source exposure time.
The data of G10.3-0.1 was acquired on June 9th 2015 ($\tau_{\mbox{\scriptsize{225 GHz}}}$$\sim$0.04), and the data of G10.6-0.4 and G10.2-0.3 were acquired on April 5th 2014 ($\tau_{\mbox{\scriptsize{225 GHz}}}$$\sim$0.03); the integration time was 30 minutes for each of the three sources.

The telescope pointing and focusing were checked every 1.5-2.5 hours.
Mars was observed for absolute flux calibration.
We used the standard 10$'$$\times$10$'$ on-the-fly (OTF) box scanning pattern, and the scanning center for each source is listed in Table \ref{tab:source_info}.

Basic data calibration was carried out using the CRUSH software package (Kov\'{a}cs 2008).
We used the {\tt -faint} option of the CRUSH software package during data reduction, which optimized the reconstruction of the faint and compact sources with the cost of the more aggressive filtering of extended emission.
Nevertheless, the extended emission components will ultimately be complemented by the observations of space telescopes (more in Section \ref{subsub:combinesharc}).
The final calibrated map was smoothed by a 2/3 beam FWHM ({\tt -faint} option) yielding an angular resolution of 9$''$.6 for optimized sensitivity and source reconstruction. The rms noise levels we measured from the approximately emission-free areas of W43-main, W43-south, W49A , G10.2-0.3, G10.3-0.1, and W33 were $\sim$76, 42, 67, 42, 102, and 68 mJy\,beam$^{-1}$, respectively. 
The observations of G33.92+0.11 were published separately by Liu et al. (2015), however, were left out from the present paper since there are no high quality ground based 450 $\mu$m and 850 $\mu$m observations for this target sources.

\begin{figure*}
\vspace{-1.5cm}
\hspace{-0.1cm}\includegraphics[width=20cm]{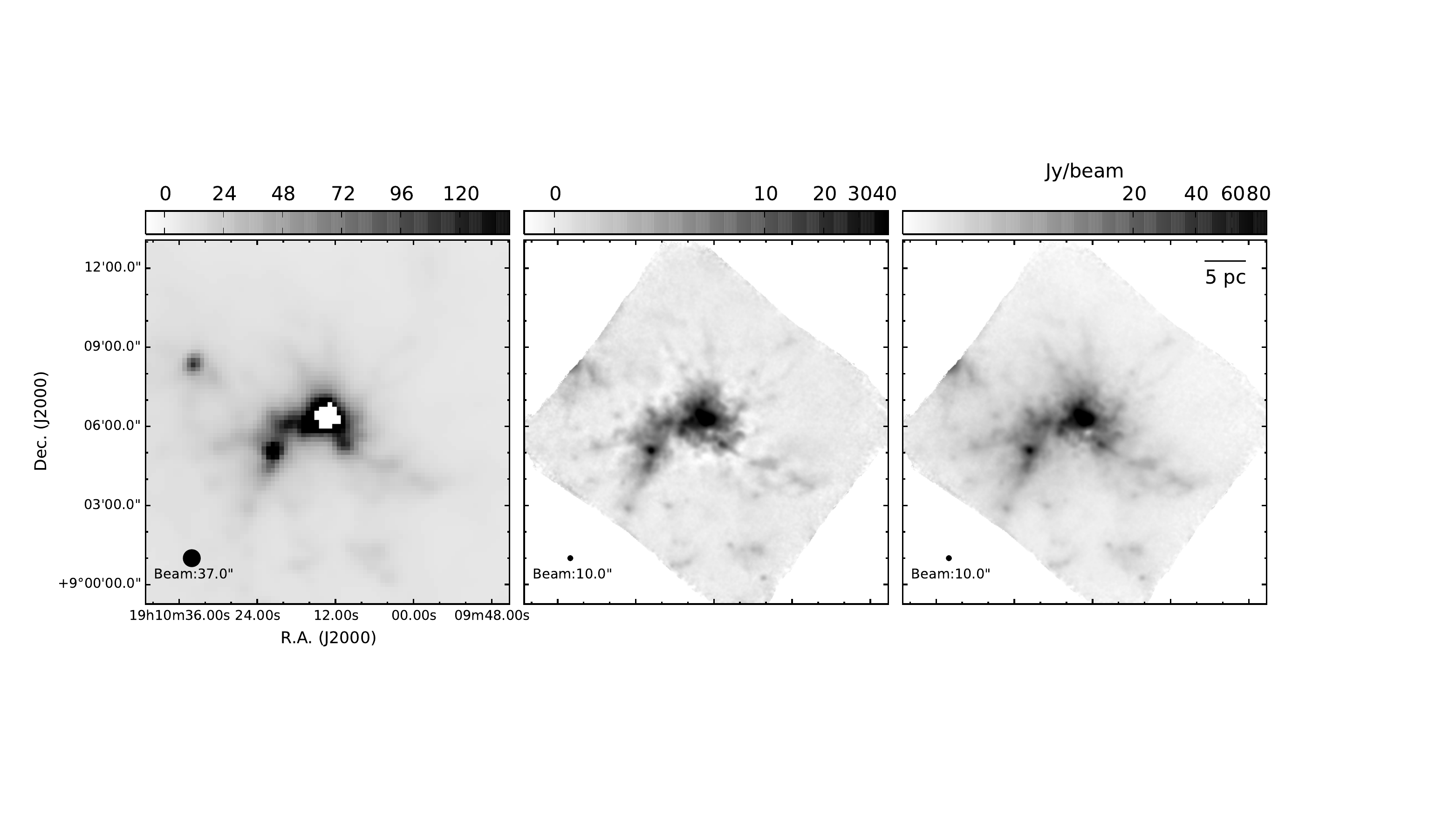}
\vspace{-2.5cm}
\caption{The 350 \micron\ image of the W49A mini-starburst region, from left to right, SPIRE 350 \micron\ image, SHARC2 350 \micron\ image and image generated by combining the CSO-SHARC2 image with the Herschel-SPIRE image. Color scale for the middle and right images are stretched to illustrate the difference between the original SHARC2 map and the combined image.
}
\vspace{0.3cm}
\label{fig:w49n_350comb}
\end{figure*}

\subsection{Herschel, Planck, JCMT and APEX data}\label{subsection:archive}
We retrieved the available nightly observations of James Clerk Maxwell Telescope (JCMT)\footnote{The James Clerk Maxwell Telescope is operated by the East Asian Observatory on behalf of The National Astronomical Observatory of Japan, Academia Sinica Institute of Astronomy and Astrophysics, the Korea Astronomy and Space Science Institute, the National Astronomical Observatories of China and the Chinese Academy of Sciences (Grant No. XDB09000000), with additional funding support from the Science and Technology Facilities Council of the United Kingdom and participating universities in the United Kingdom and Canada. The James Clerk Maxwell Telescope has historically been operated by the Joint Astronomy Centre on behalf of the Science and Technology Facilities Council of the United Kingdom, the National Research Council of Canada and the Netherlands Organization for Scientific Research. Additional funds for the construction of SCUBA-2 were provided by the Canada Foundation for Innovation.}  Submillimetre Common-User Bolometer Array 2 (SCUBA2)
 (Dempsey et al. 2013;
\ Chapin et al. 2013;
\ Holland et al. 2013
) at 450 \micron\ and 850 \micron\ from the online data archive. (Program ID:S13AU02 (W43; W31), MJLSJ02 (W33; G10.2-0.3; G10.3-0.1; G10.6-0.4), M13BU27 (W49A) ).

We retrieved the level 2.5/3 processed, archival {\it Herschel}\footnote{Herschel is an ESA space observatory with science instruments provided by European-led Principal Investigator consortia and with important participation from NASA.} images that were taken by the Herschel Infrared Galactic Plane (Hi-GAL) survey (Molinari et al 2010) at 70/160 \micron\ using the PACS instrument (Poglitsch et al. 2010) and at 250/350/500 \micron\ using the SPIRE instrument (Griffin et al. 2010). (obsID: W43: 1342186275, 1342186276; G10.2-0.3, G10.3-0.1, G10.6-0.4: 1342218966; W49A: 1342207052, 1342207053; W33: 1342218999, 1342219000.)

\begin{table}\footnotesize{ 
\begin{center}
\hspace{-10.0 cm}
\vspace{-0.5 cm}

\caption{Observational parameters of multi-band data}
\begin{tabular}{lcccccccc}\hline\hline
$\lambda$ (\micron) & Beam \emph{FWHM} & Pixel size  & Flux unit\\
 Camera            & (arcsec) &(arcsec)                \\
 
 \hline
 
70/PACS &  $5.8\times12.1$&3.2      & Jy/pixel\\
160/PACS &  $11.4\times13.4$ &3.2 & Jy/pixel\\
250/SPIRE&  18.1&6.0&     MJy/sr\\
350/SPIRE &  25.2&9.72&  MJy/sr\\
500/SPIRE &  36.9&14.0&  MJy/sr\\
450/SCUBA2 &  8.0&3.0&   mJy/$arcsec^{2}$\\
850/SCUBA2 &  14.0&3.0&  mJy/$arcsec^{2}$\\
350/SHARC2 & 8.0&1.5&   Jy/beam\\
870/LABOCA &    19.2 & 6.0  & Jy/beam\\
1200/MAMBO-2&11.0&3.50&     mJy/beam\\
217GHz/PLANCK&292.2&  60.0&   K$_{cmb}$\\
353GHz/PLANCK&279.0& 60.0 &K$_{cmb}$\\
\hline
      
\end{tabular}
\end{center}
}
\vspace{0.1cm}

\end{table}

Since we are interested in the extended structures, we adopt the extended emission products, which have been absolute zero-point corrected based on the images taken by the {\it Planck} space telescope.

We retrieved the Planck/High Frequency Instrument (HFI) 353 GHz images (and also 217 GHz images when a MAMBO2 image is available for a particular source) which are in units of $K_{CMB}$.
We convert the Planck images to Jy\,beam$^{-1}$ units based on the conversion factors provided by 
Zacchei et al. (2011) and the Planck HFI Core Team (2011a and 2011b). 

For the combination of 850 \micron\ images, we also used the APEX-LABOCA ATLASGAL survey (Siringo et al. 2009) observations for our sources. 
The ATLASGAL survey (Schuller et al. 2009) data products are reduced in a way of optimizing the compact sources recovery with flux calibration uncertainty of $\sim 15 \%$.

\begin{figure*}
\includegraphics[scale=0.5]{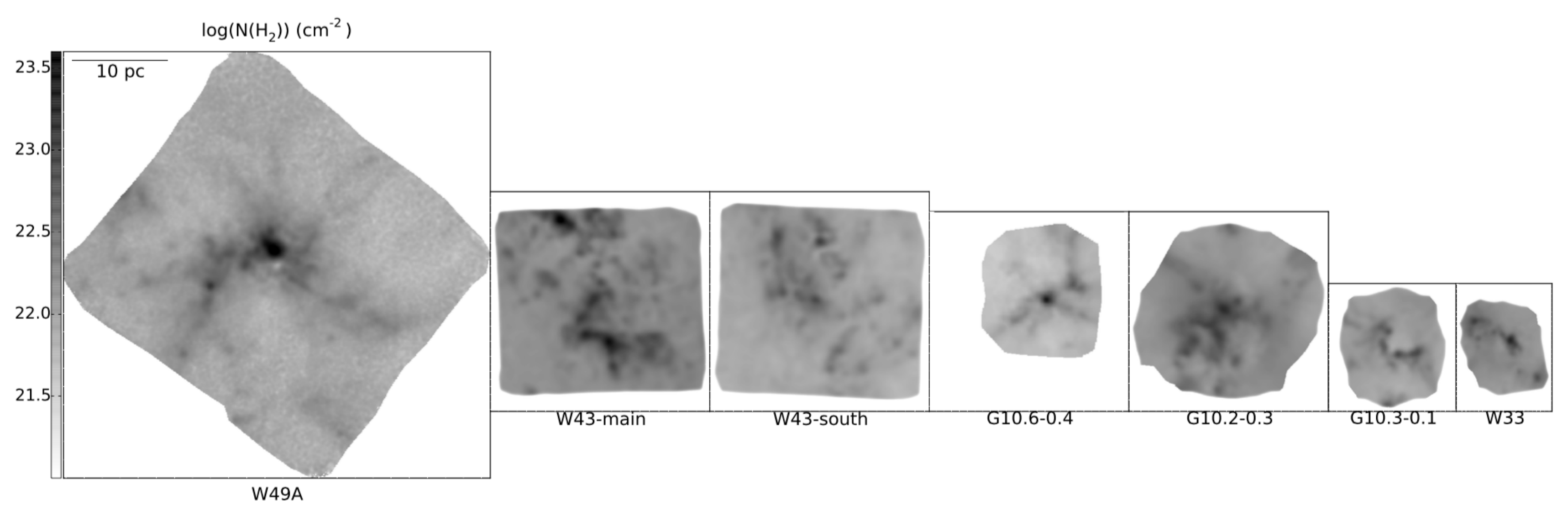}
\caption{Column density maps of all target sources smoothed and reprojected to the resolution of W49A, which is the most distant source within our sample.}
\vspace{0.4cm}
\label{fig:rep_all_w49}
\end{figure*}

\subsection{An Overall Flow of Our Image Analysis}\label{subsection:procedure}
In our procedure we first replace the saturated pixels in the archival Herschel images by interpolating (see Section \ref{subsection:saturate}).
The interpolated Herschel images are then convolved with the kernels provided in Aniano et al. (2011), to suppress the defects caused by the non-Gaussian beam shapes of the Herschel space telescope.
Afterwards, images were linearly combined with the ground based observations of CSO-SHARC2 at 350 \micron\, and the observations of JCMT-SCUBA2 at 450 \micron\ (Section \ref{subsub:combinesharc}, \ref{subsub:combinejcmt}), yielding high resolution millimeter and submillimeter images that have little or no loss of extended structures.
We combined the JCMT-SCUBA2 850 \micron\ images and the IRAM-30m-MAMBO2 1200 \micron\ images with those taken by the {\it Planck} space telescope, in the cases that the SCUBA2 and MAMBO2 images are available. 
Finally, we performed the pixel-by-pixel SED fitting incorporating the {\it Herschel}-PACS 70 and 160 \micron\ images, the {\it Herschel}-SPIRE 250 \micron\ images, and our combined images at 350, 450, 850, and 1200 \micron\ , assuming a single black-body component in each line of sight (if available; Section \ref{subsection:SED}). 
A flow chart of the overall procedure is given in Figure \ref{fig:floatchart}.

There are fundamental limits to the accuracy of our SED fitting, which is  typically  dominated by the noise level of the 450 \micron\ and 850 \micron\ images, and the degeneracy when optimizing the dust opacity index ($\beta$) and the dust temperature $T_{d}$. 
For the case that the brightness sensitivity of the ground based image is a lot worse than that of the space telescope image, flux scales of the bright structures can still be corrected after combining with the space telescope image.
However, in such cases, the extended diffuse structures in the combined image may still be immersed in thermal noise.  
We are subject to this issue because we used the archival JCMT-SCUBA2 450 \micron\ and 850 \micron\ images, which were not all planned to achieve a matched brightness sensitivity with the Herschel Hi-GAL images.
For our selected target sources, these problems are not very serious for dense structures in the molecular clouds, which are our major interests. 
We will only analyze structures above certain column density thresholds to avoid confusion by the Galactic foreground/background emission, which further alleviate the effects of the aforementioned defects.

\begin{figure*}
\hspace{-0.3cm}
\vspace{-0.1cm}
\includegraphics[scale=0.67]{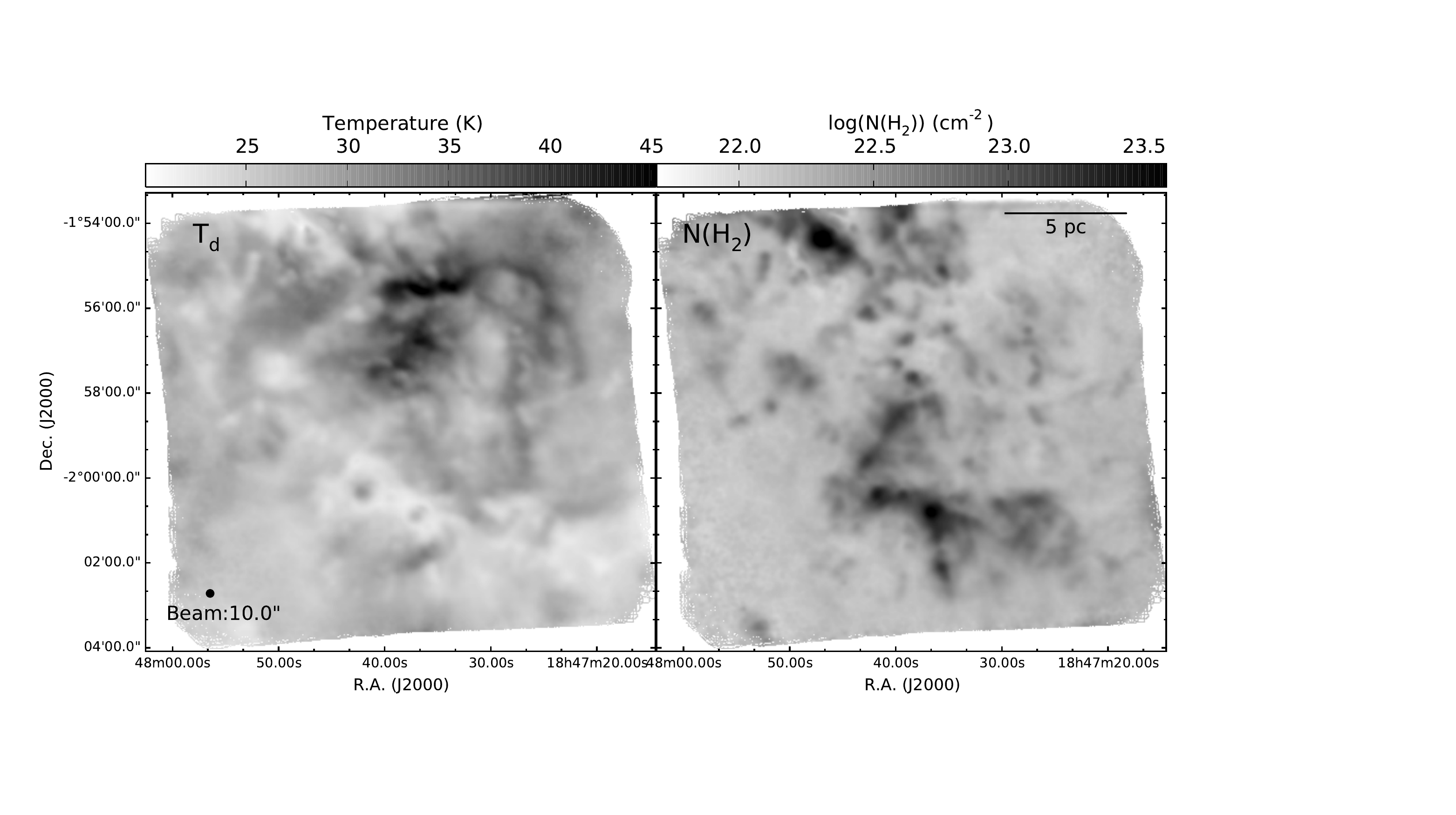}
\vspace{-2cm}
\caption{The dust temperature and column density distributions of W43-Main, derived based on fitting modified black-body spectra iteratively to the Herschel-PACS 70/160 \micron\ , Herschel-SPIRE 250 \micron\ , and the combined 350/450/850 \micron\ images. The detailed procedures can be found in Section \ref{subsection:procedure}.
}
\label{fig:TN_W43M}
\end{figure*}

\begin{figure*}
\hspace{-0.3cm}
\vspace{-0.1cm}
\includegraphics[scale=0.67]{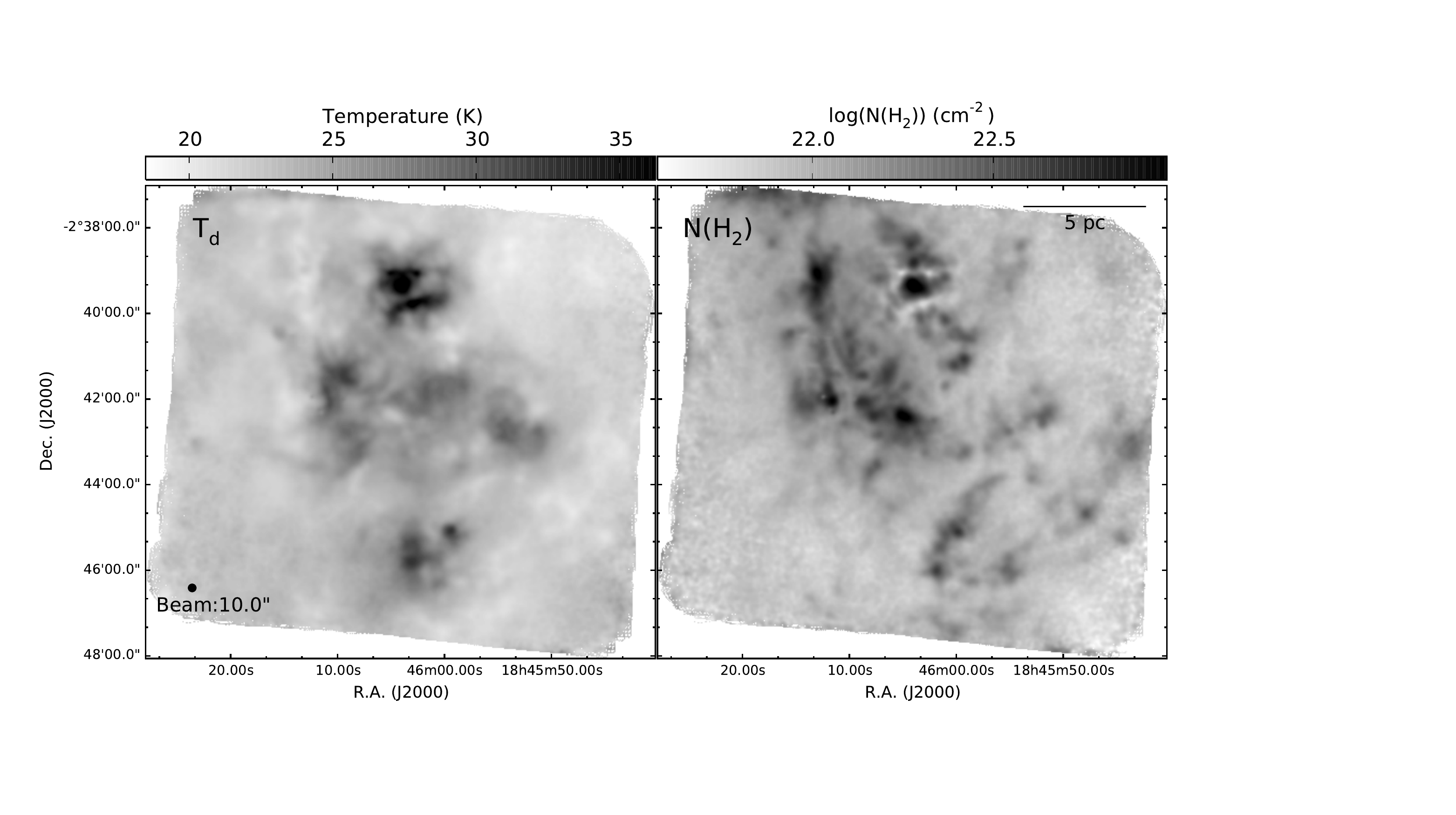}
\vspace{-2cm}
\caption{Similar to Figure \ref{fig:TN_W43M}, but for the target source W43-South.}
\label{fig:TN_W43S}
\end{figure*}

\begin{figure*}
\hspace{-0.3cm}
\vspace{-0.1cm}
\includegraphics[scale=0.67]{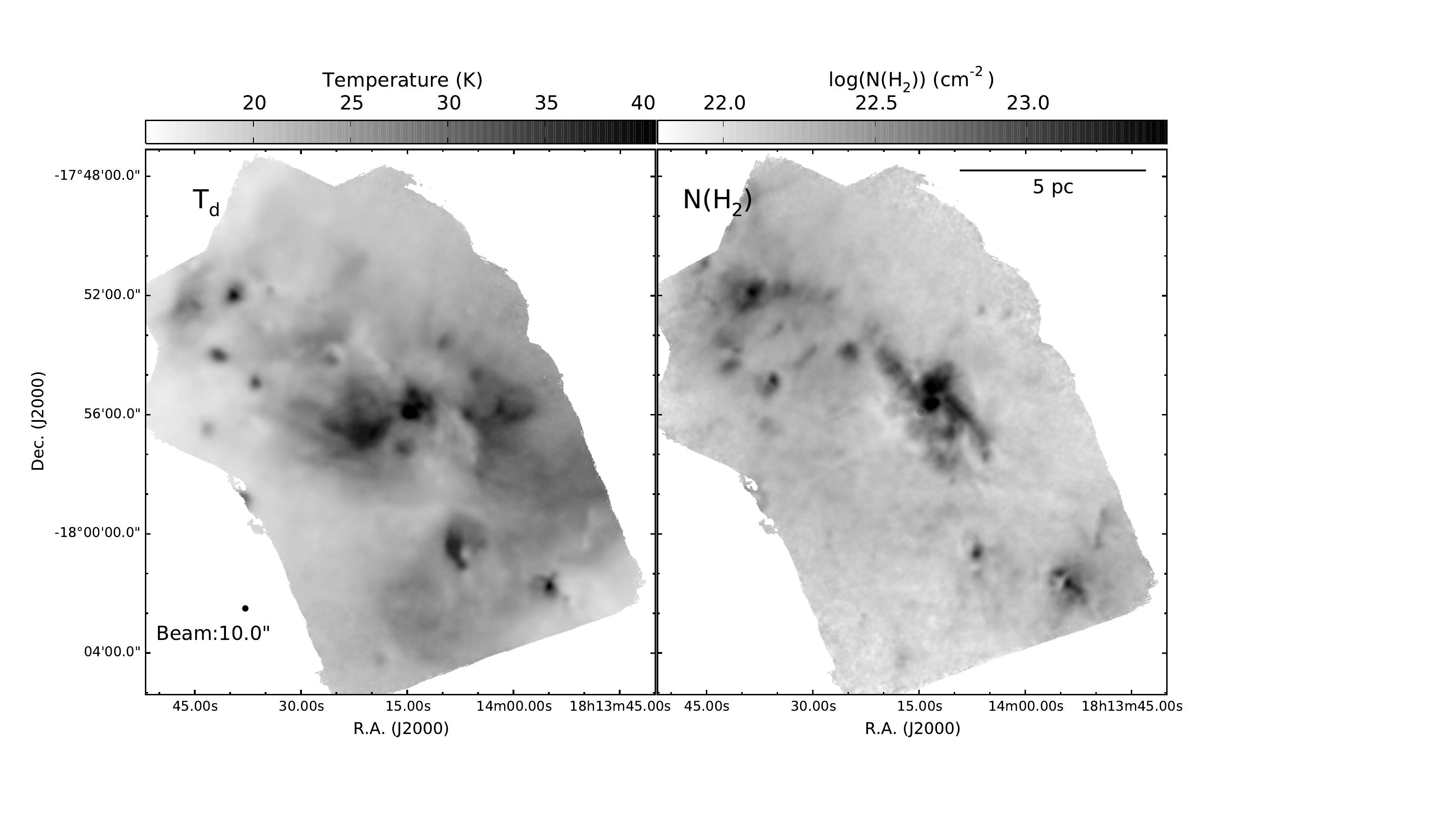}
\vspace{-1.5cm}
\caption{Similar to Figure \ref{fig:TN_W43M}, but for the target source W33.
}
\label{fig:TN_W33}
\end{figure*}

\begin{figure*}
\hspace{-0.3cm}
\vspace{-0.1cm}
\includegraphics[scale=0.67]{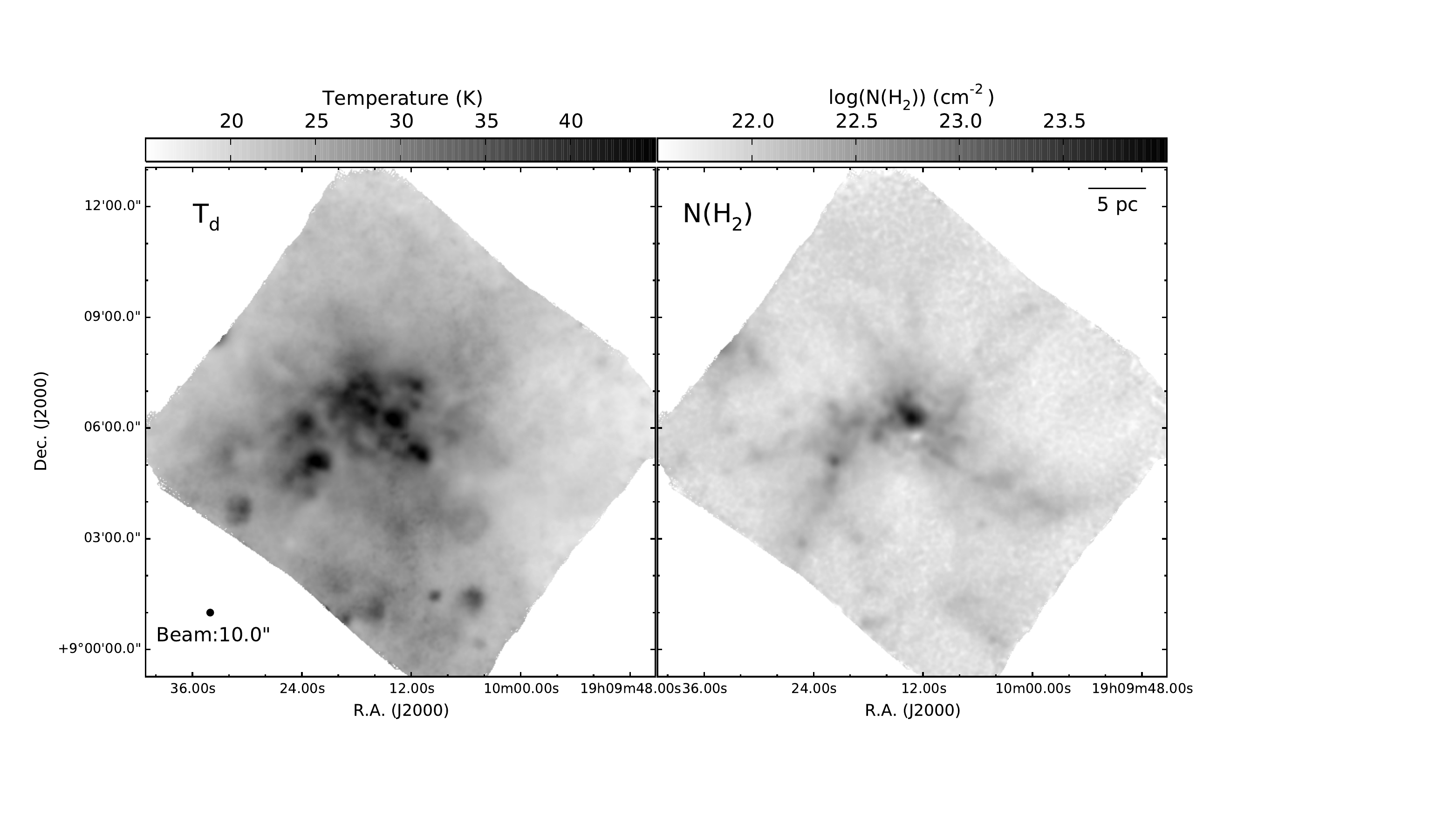}
\vspace{-1.5cm}
\caption{Similar to Figure \ref{fig:TN_W43M}, but for the target source W49A.}
\label{fig:TN_W49N}
\end{figure*}

\begin{figure*}
\hspace{-0.3cm}
\vspace{-0.1cm}
\includegraphics[scale=0.67]{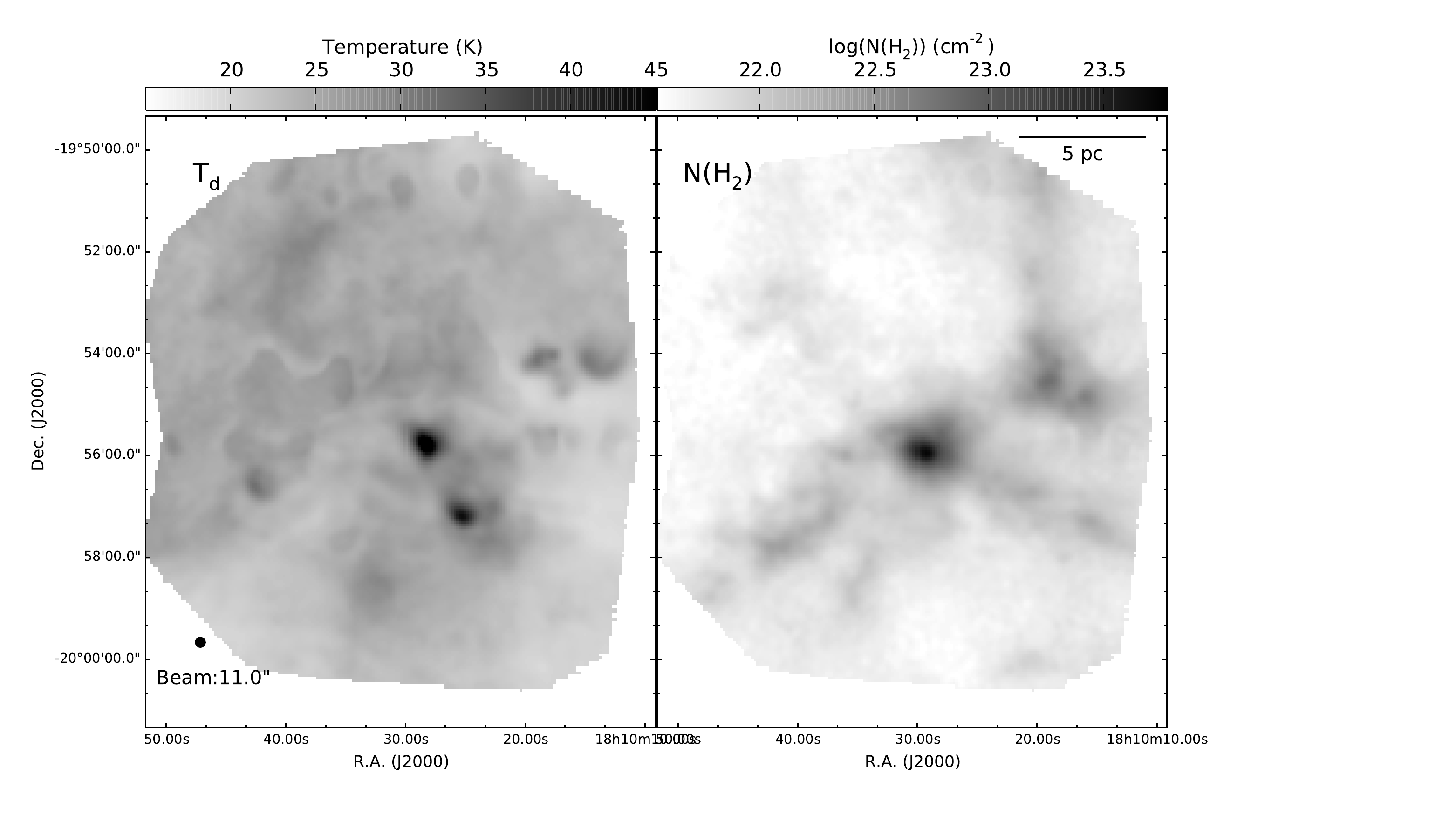}
\vspace{-1cm}
\caption{Similar to Figure \ref{fig:TN_W43M}, but for the target source G10.6-0.4.}
\label{fig:TN_G10p6}
\end{figure*}

\begin{figure*}
\hspace{-0.3cm}
\vspace{-0.1cm}
\includegraphics[scale=0.67]{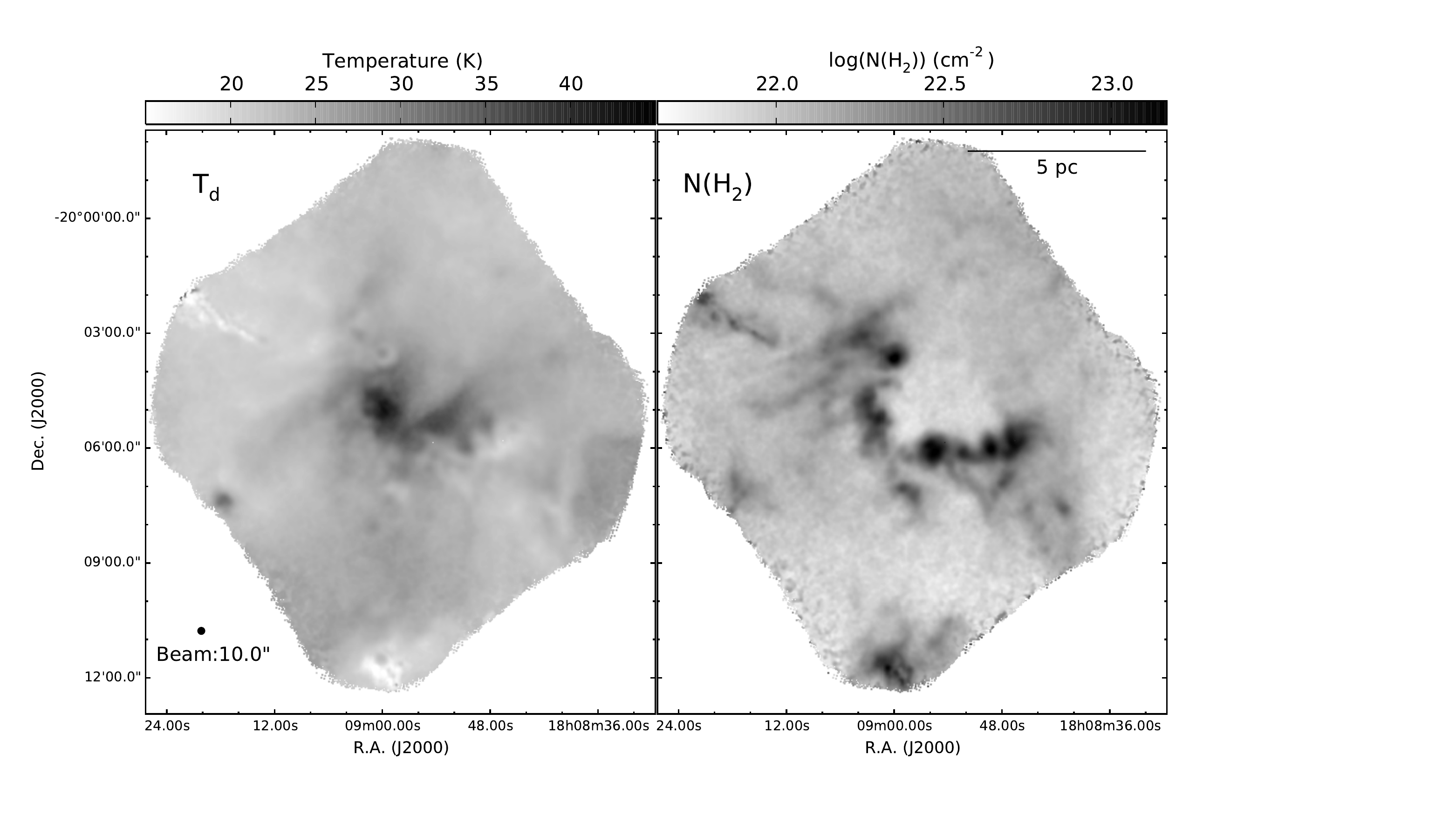}
\vspace{-1cm}
\caption{Similar to Figure \ref{fig:TN_W43M}, but for the target source G10.3-0.1.}
\label{fig:TN_G10p3}
\end{figure*}

\begin{figure*}
\hspace{-0.3cm}
\vspace{-0.1cm}
\includegraphics[scale=0.67]{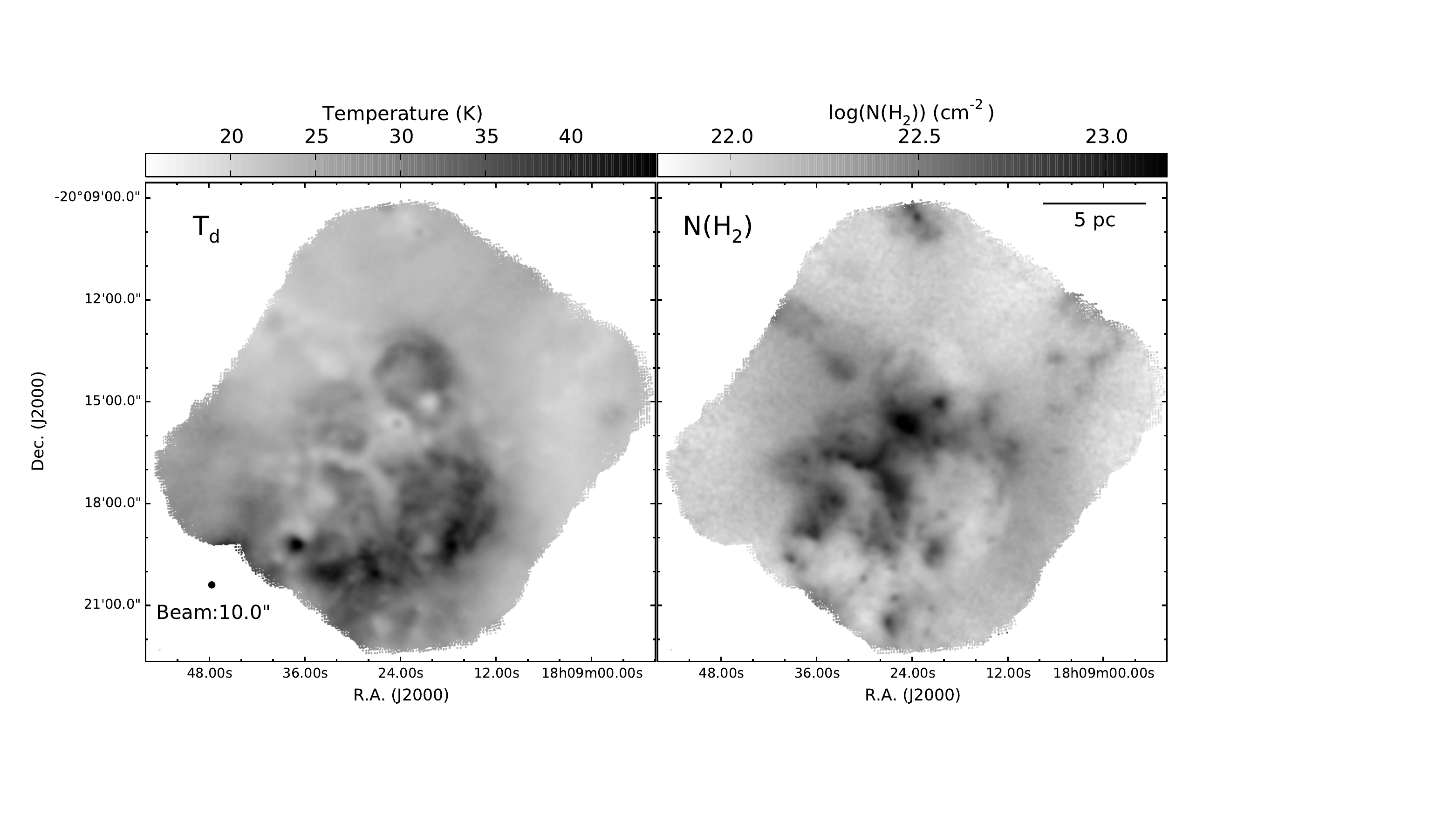}
\vspace{-1.5cm}
\caption{Similar to Figure \ref{fig:TN_W43M}, but for the target source G10.2-0.3.}
\label{fig:TN_G10p2}
\end{figure*}

\begin{figure*}
\hspace{-0.3cm}
\vspace{-0.1cm}
\includegraphics[scale=0.67]{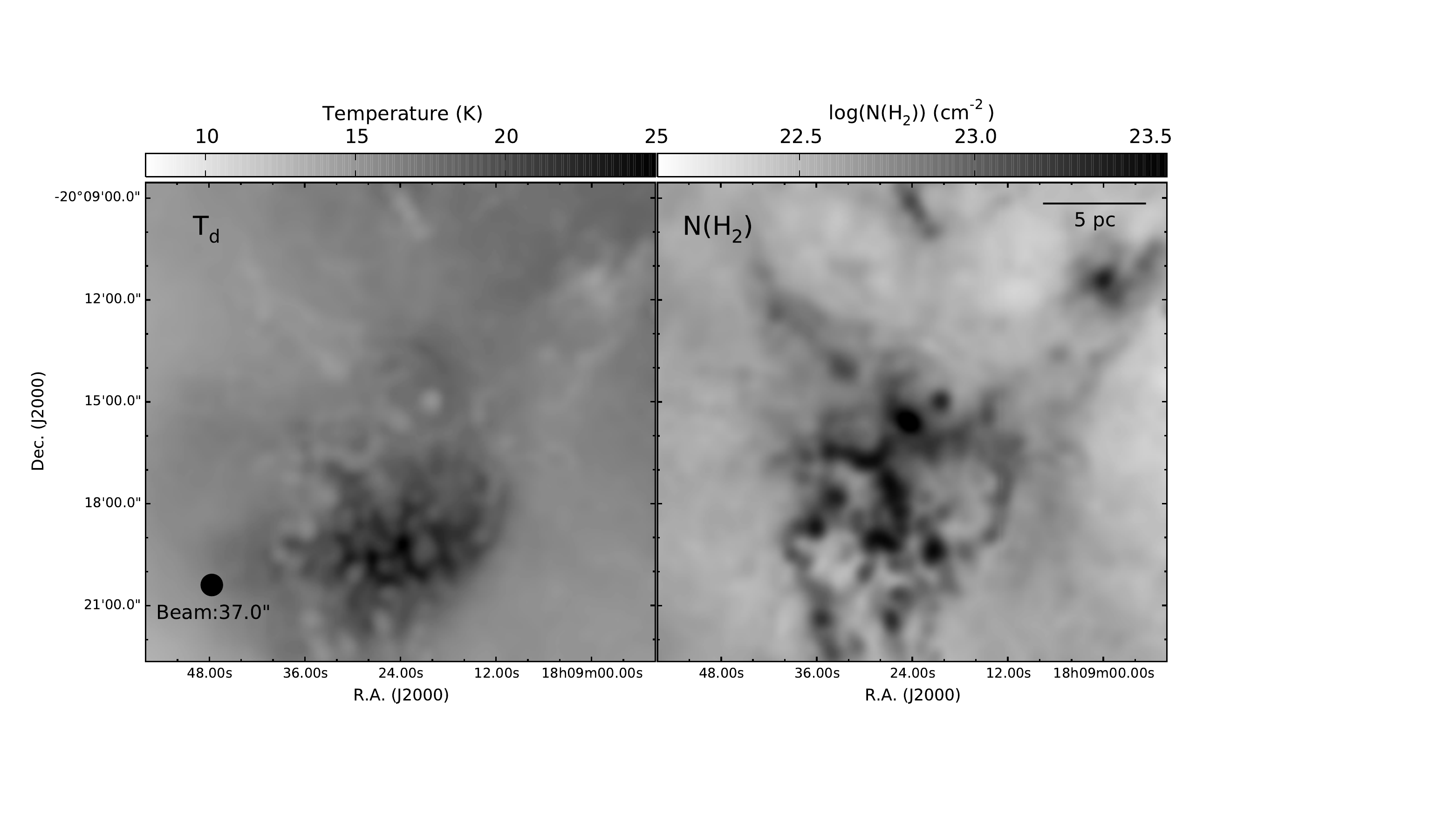}
\vspace{-1.5cm}
\caption{The dust temperature and column density distributions of G10.2-0.3, derived based on fitting modified black-body spectra to the Herschel-PACS 160 \micron\ , Herschel-SPIRE 250 and 350 \micron\ images with fixed dust opacity index $\beta$ = 2.0.}
\label{fig:TN_herschel_G102}
\end{figure*}

\subsubsection{Interpolating saturated pixels}\label{subsection:saturate}
Some Herschel images of our target sources were saturated around the bright sources (see Appendix A).
Before combining these Herschel images with the ground based observations, or before using these Herschel images in the SED fitting, we replaced the saturated pixels using interpolated values, which in some cases required iterative processes. 
Before the SED fitting (see also Figure \ref{fig:floatchart}), the {\it Herschel} images were convolved with the kernels provided by  Aniano et al. (2011) to suppress the effect of the non-Gaussian beam shape, except where otherwise noted.

For the target sources W43-main, W49A, and G10.6-0.4, one, or both of the PACS 70 and 160 \micron\ images show saturated pixels.
The saturated pixels in these PACS images were replaced by the interpolated values from the two-dimensional Gaussian fits to the adjacent pixels. 

For the sources G10.2-0.3, G10.3-0.1 and W43-south, only the SPIRE 250 \micron\ images were saturated. 
Their saturated pixels were replaced following the procedure described below. 
First, we derived the approximated 450 \micron\ intensity image at the angular resolution of the SPIRE 500 \micron\ images (36$\farcs$9), by performing the pixel-by-pixel modified black-body SED fitting (more in Section \ref{subsection:SED}) to the PACS 70/160 \micron, the SPIRE 350/500 \micron, and the combined 850 \micron\ images.
Then, the approximated, 36$\farcs$9 resolution 450 \micron\ images were linearly combined with the SCUBA2 450 \micron\ images, following a procedure that will be introduced in Section \ref{subsub:combinejcmt}.
Afterwards, we performed the pixel-by-pixel modified black-body SED fitting to the combined, higher angular resolution 450 \micron\ image, the PACS 70/160 \micron\ images, and the combined 350/450/850 \micron\ images, to derive approximated 250 \micron\ images that have an 18$''$ angular resolution. 
In the end, we use the values in the approximated 250 \micron\ image to replace those of the saturated pixels in the original Herschel 250 \micron\ images. 

For the cases that both the SPIRE 350 \micron\ and SPIRE 250 \micron\  images are saturated (e.g., Figure \ref{fig:w49n_350comb}, see also Appendix A), 450 \micron\ intensity images were derived based on SED fits to PACS 160 \micron , SPIRE 500 \micron\ and combined 850 \micron\ images. Adding the combined 450 \micron\ (Section \ref{subsub:combinejcmt}) and PACS 70 \micron\ images to the SED fitting with previous bands, we derived the fluxes for the saturated pixels in the SPIRE 350 \micron\ and SPIRE 250 \micron\ images.
The detailed procedure is similar to that of replacing the saturated 250 \micron\ pixels, so we do not repeat it here. 
However, we additionally took advantage of our available SHARC2 350 \micron\ images to double check whether the replaced pixel values in the saturated SPIRE images are reasonable or not by comparing the intensity distributions.

\subsubsection{Deriving dust temperature and column density based on SED fitting}\label{subsection:SED}
Before performing any SED fitting, we smoothed all images to a common angular resolution, which is slightly bigger than the FHWM of the largest telescope beam for each iteration.
In addition, all images were re-gridded to have the same pixel size, and were converted to the unit of Jy/pixel. 
We weighted the data points by the measured noise level in the least-squares fits.
We adopted a dust opacity law similar to what was introduced by Hildebrand (1983). 

In this case, flux density $S_{\nu}$ at a certain observing frequency $\nu$ is given by
\begin{equation}
S_{\nu} = \Omega_{m} B_{\nu}(T_{d})(1-e^{-\tau_{\nu}}),
\end{equation}
and 
\begin{equation}
N_{\rm H_{2}} = \frac{\tau_{\nu}}{\kappa_{\nu}\mu m_{\rm H}},
\end{equation}
where
\begin{equation}
B_{\nu}(T_{d})=\frac{2h\nu^{3}}{c^{2}}\frac{1}{\rm exp(h\nu/kT_{d})-1},
\end{equation}
is the Planck function for a given temperature $T_{d}$,
\begin{equation}
\kappa_{\nu}=\kappa_{230}(\frac{\nu}{230\ GHz})^{\beta}
\end{equation}
\noindent is the dust opacity, $\beta$ is the dust opacity index, $\Omega_{m}$ 
is the considered solid angle, $\mu = 2.8$ is the mean molecular weight, $m_{\mu}$ is the mass of a hydrogen atom, and $\kappa_{230}$ = $0.09\ \rm cm^{2}g^{-1}$ is the dust opacity per unit mass at 230\ GHz. 
The opacity $\kappa_{\nu}$ is interpolated from Ossenkopf \& Henning (1994). 
We assumed a gas-to-dust mass ratio of 100.

Our iterative fitting procedure first used the available images at all wavelength bands, to simultaneously fit $\beta$, $T_{\rm d}$ and $N_{\rm H_{2}}$.
In this step, the PACS 70/160 \micron, the SPIRE 250 \micron, and the combined 350/450/850 \micron\ images were smoothed to 22$''$ resolution, which is slightly larger than the measured beam size of the SPIRE 250 \micron\ image.
We used the obtained 22$''$ resolution images of  $\beta$, $T_{\rm d}$ and $N_{\rm H_{2}}$, to initialize the second fitting iteration.
However, in the second fitting iteration, we use the values of $\beta$  from the last iteration, and only fit to the PACS 70 \micron\ images, the combined 350 and 450 \micron\ images \footnote{For W43-South which lacks of 450 \micron\ SCUBA2 data, we used 70 \micron\ image with combined 350 \micron\ image only for the final fitting.}, to yield the $T_{d}$ and the $N_{\rm H_{2}}$ images with our best achievable angular resolution of $\sim$10$''$.
The initialization of $\beta$, $T_{\rm d}$ and $N_{\rm H_{2}}$ using results from the previous fitting iteration helped improve the convergence of the fitting.

The quality of the SHARC2 350 \micron\ image of G10.6-0.4 was poor, which may be due to a temporarily poorer weather condition, or imperfect focus. 
Therefore, it was excluded from our analysis.
The SED fits of this particular source were complemented by the combined MAMBO2 1.2 mm and Planck/HFI 217 GHz images, when deriving the final, $\sim$11$''$ resolution column density and temperature maps.

Our present scientific discussion focuses on the column density distribution N$_{\rm H_{2}}$.
We noticed that the possibly biased $\beta$ vs. $T_{d}$ relation that comes from the least-squares fits due to the noise of measurements is suggested in previous works (e.g. Shetty et al. 2009a; Juvela et al. 2013). However, the anti-correlated $\beta$ vs. $T_{d}$ is also observed in some molecular clouds (e.g. Hill et al. 2006; D{\'e}sert et al. 2008). Besides, longer wavelengths in the Rayleigh-Jeans limit may recover $\beta$ more accurately (e.g. Shetty et al. 2009b) and with more wavelengths in the SED fits the variation of $\beta$ is alleviated, as seen in simulations (Malinen et al. 2011).
Considering the large dynamical range of column density we are measuring towards these regions (the possible change of $\beta$) and the advantage of utilizing $\geq 5$ wavelengths, we fit $\beta$ with $T_{d}$ and column density simultaneously.
In future works, we will adopt a multi-level Bayesian technique in SED fits (Kelly et al. 2012) to further improve the accuracy of fitting $\beta$.

We caution that in the cases that there are multiple temperature components in the line of sight, our fits will be approximate, with ambiguous determinations for temperature and dust opacity index. We expect that this only results in errors of gas column density by a fraction of the actual gas column density that we are probing. Such errors are very small as compared with the ranges of gas column density that we are probing and therefore will not likely impact our statistical analyses significantly. We refer to Marsh et al. (2015) for an approach to fit multiple temperature components.
We will update the dust model in the future works, when we come to the more detailed analysis about the temperature distribution and the variation of dust properties (e.g. Galametz et al. 2016).

\subsubsection{Combining \emph{SHARC2} and \emph{SPIRE} image}\label{subsub:combinesharc}
Ground based millimeter and submillimeter continuum imaging observations are often confused with extended atmospheric thermal emission.
The procedures to remove the atmospheric emission components often limit the maximum recoverable angular scales of the ground based observations to narrower than the simultaneous field of view of the bolometric receiver array or camera being used. 
This situation is analogous to the {\it missing short spacing} issue of interferometric observations.
For the case of our CSO-SHARC2 350 \micron\ that were reduced with a {\tt -faint} option (Section \ref{subsection:sharc2}), the maximum recoverable angular scale is limited to one half of the simultaneous field of view ($\sim$1$'$).

We used the task {\tt immerge} of the Miriad software package (Sault et al. 1995) to linearly merge the SHARC2 and the SPIRE 350 \micron\ images in the Fourier domain.
The {\tt immerge} procedure assumes that the lower resolution image better represents the emission profile at the extended angular scales.
This method, which we briefly outline below, is routinely applied by the community of radio astronomy for merging images taken by single dish telescopes and interferometers.

The original SPIRE 350 \micron\ images were first re-gridded to the same pixel size and field of view of the SHARC2 images.
Before combining the images, we aligned the SHARC2 image with the {\it Herschel} images by making cross-correlations.
We manually re-scaled the observed flux density of SHARC2 images, such that the observed fluxes of compact sources match the {\it Herschel} SPIRE 350 \micron\ observations.
In addition, we smoothed the SPIRE 350 \micron\ images to a resolution of 36.3$''$ to enhance the signal to noise ratio and to suppress the effects of the side lobe responses of the {\it Herschel} telescope.
Finally, we used {\tt immerge} to yield the combined images, which have the same angular resolution as the original SHARC2 images but recover the extended emission features. Figure \ref{fig:w49n_350comb} presents an example of the combined 350 \micron\ image of the target source W49A. 
Combined images of other sources are provided in Appendix A.

\subsubsection{Combining the JCMT-SCUBA2 image with the Herschel, APEX/LABOCA and Planck/HFI observations}\label{subsub:combinejcmt}
\paragraph{450 \micron\ } We made interpolations to the 450 \micron\ images, based on the modified black-body fits to the PACS 160 \micron\, and the SPIRE 250/350/500 \micron\ with combined SCUBA2 and Planck 850 \micron\ images\footnote{Based on observations obtained with Planck (http://www.esa.int/Planck), an ESA science mission with instruments and contributions directly funded by ESA Member States, NASA, and Canada.}.
We pre-smoothed all images for the modified black-body fits to the same angular resolution of the SPIRE 500 \micron\ image ($\sim$37$''$).
We then smoothed the interpolated 450 \micron\ images to a resolution of 51.9$''$ (approximate to smoothing by one beam size), to suppress the defects caused by the side lobe responses of the {\it Herschel} telescope.
Finally, the interpolated and smoothed low angular resolution 450 \micron\ image is combined with the JCMT-SCUBA2 450 \micron\ image, following a similar process as outlined in Section \ref{subsub:combinesharc}.

We note that the JCMT-SCUBA2 observations covered a simultaneous field of view of $\sim$480$''$, which in principle can preserve extended emission of the observed molecular clouds on a comparable angular scale from being removed together with the atmospheric emission.
Our final combined 450 \micron\ images therefore do not suffer from loss of structures on any angular scales above the angular scale of the JCMT beam size.

\paragraph{850 \micron\ } The ranges of angular scales probed by the JCMT-SCUBA2 850 \micron\ observations, and by the Planck 353 GHz observations are very different.
This is partially related to the algorithm used to reduce the JCMT-SCUBA2 850 \micron\ observational data, and is being investigated.
Empirically, we found that combining the APEX-LABOCA 870 \micron\ images can complement our sampling of angular scales, and thereby improve the image quality.
As a tentative approach before we develop an optimized routine to re-process the JCMT-SCUBA2 850 \micron\ data to serve our purpose of image combination, our present procedure first combine {\it Planck} 353 GHz images with the LABOCA 870 \micron\  images.
The combined {\it Planck}+LABOCA images were then combined with the SCUBA2 images to achieve the best possible angular resolution. 
We adopted the color correction factors for the {\it Planck} and the LABOCA images, provided by Csengeri et al. (2016).

\begin{figure*}
\begin{tabular}{ p{0.45\linewidth} p{0.45\linewidth} }
\hspace{-1cm}\includegraphics[scale=0.5]{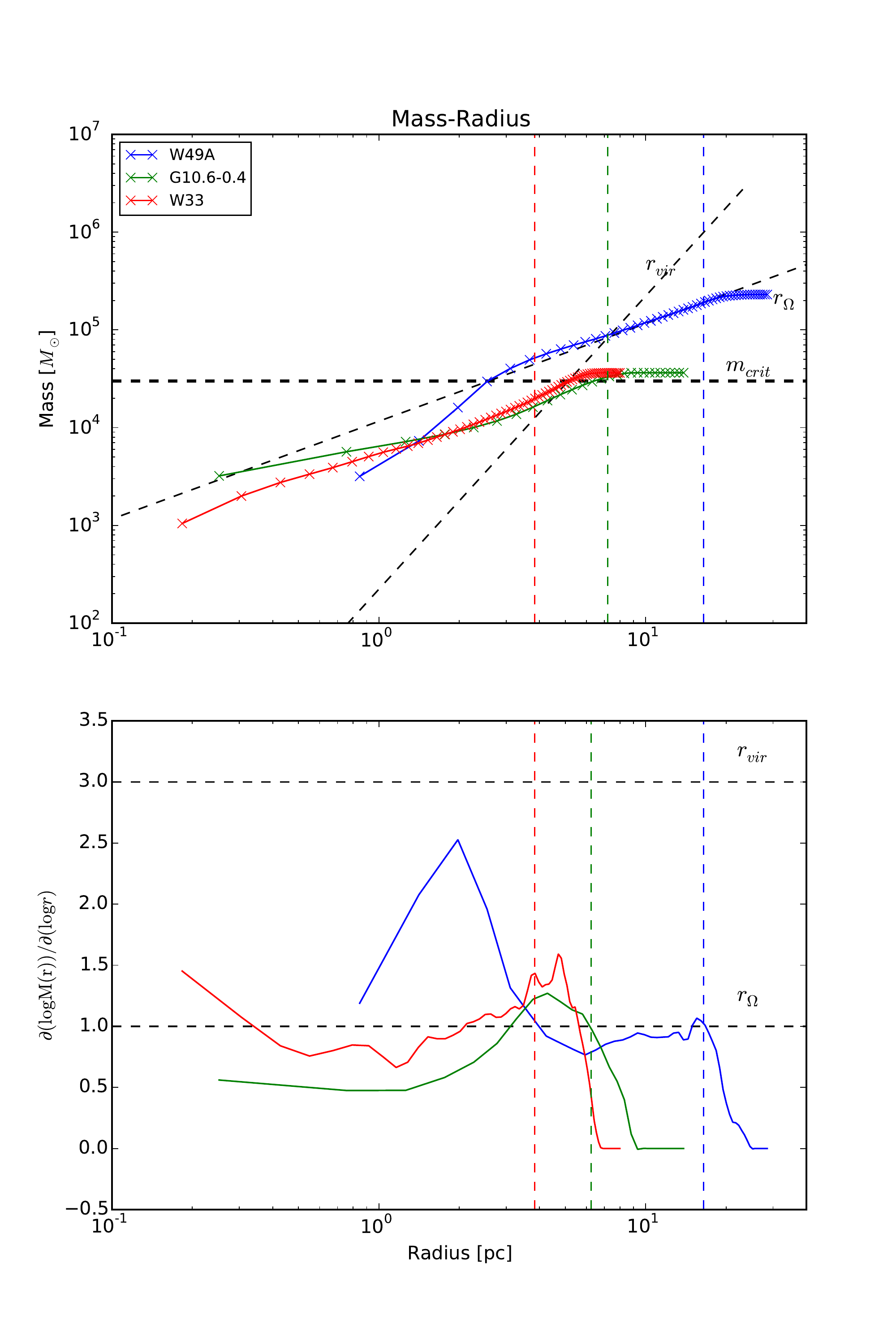}&\includegraphics[scale=0.5]{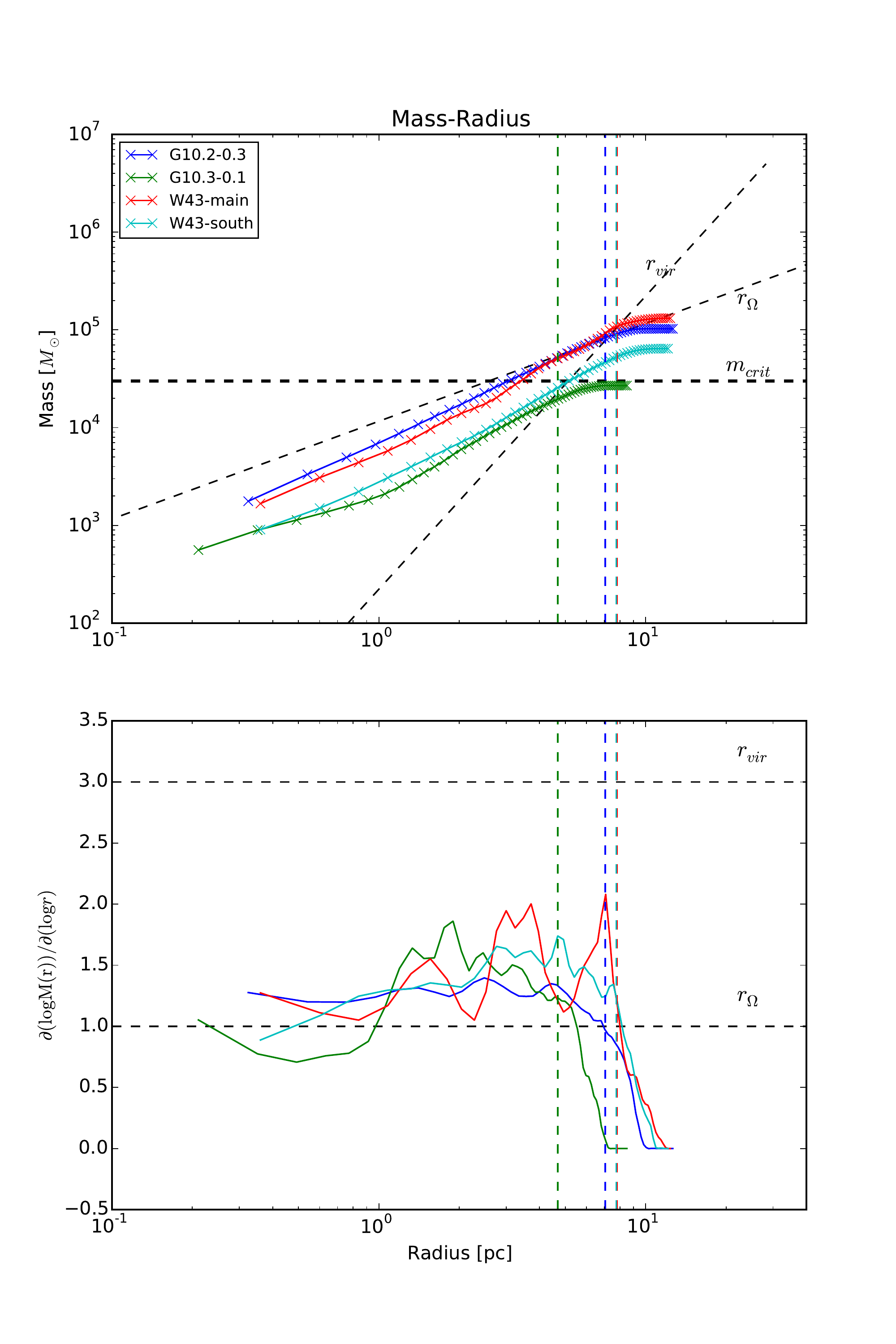}\\
\end{tabular}
\vspace{-0.8cm}
\caption{Enclosed mass as a function of radius. {\it Upper panel:} Horizontal dashed line shows the minimum mass for a potential young massive cluster progenitor $m_{crit}\sim 3\times10^{4} M_{\odot}$, $r_{vir}$ and $r_{\Omega}$ are the virial radius derived with fixed crossing time of 1 Myr and radius derived by setting the potential energy equal to the kinetic energy assuming a bound HII spherical clump model with a sound speed of photo-ionized gas of 10 kms$^{-1}$, respectively. (Bressert et al. 2012) {\it Bottom panel:} The rate of change of log(M) with respect to log(r), with dashed lines showing the constant values in the cases of $r_{vir}$ and $r_{\Omega}$ separately. In all the plots, r $>$ 5 $''$ (angular resolution of 10 $''$) values are plotted with vertical dashed lines indicating the box field (r $\sim$ 5$'$) for each target source. 
}
\vspace{0.3cm}
\label{fig:enclosed}
\end{figure*}

\section{Results}\label{section:results}
This section presents the derived observational quantities.
The derived spatial distributions of dust and gas column density ($N_{\rm H_{2}}$), dust temperature ($T_{\rm d}$), and dust opacity index ($\beta$), are presented in Section \ref{subsection:maps}.
The enclosed masses of the observed sources as a function of radius are shown in Section \ref{subsection:enclosedmass}.
The gas column density probability distribution functions ($N$-PDF) are provided in Section \ref{subsection:pdf}.
The two-point correlation (also known as auto-correlation) functions of the dust/gas column density distribution functions are given in Section \ref{subsection:2pt}.
We applied the dendrogram algorithm (Rosolowsky et al. 2008) to identify dense gas structures, which are presented in Section \ref{subsection:dendrogram}.
The cross-comparison of the results presented in these sections, and the discussion of the physical implications, will be given in Section \ref{section:discussion}.

We follow the existing nomenclature in the literature (e.g., Motte et al. 2007, Zhang et al. 2009; Liu et al. 2012a, 2012b). 
In this way, \textit{massive molecular clumps} refer to structures with sizes of $\sim$0.5-1 pc, \textit{massive molecular cores} refer to the $<$0.1 pc size structures embedded within a clump, and \textit{condensations} refer to the distinct molecular substructures within a core. 
\textit{Fragmentation} refers to the dynamical process that produces or enhances multiplicity. 
\textit{Molecular filaments} refer to the geometrically elongated molecular structures, and \textit{molecular arms} refer to segments of molecular filaments that are located within the $\lesssim$1 pc radii of molecular clumps and may not be fully embedded within molecular clumps.

\subsection{Dust column density, temperature, and dust opacity index distribution}\label{subsection:maps}
Figure \ref{fig:rep_all_w49} shows the derived gas column density maps ($N_{H_{2}}$) of all selected sources, which were smoothed and reprojected to the distance of W49A.
Figures 4 to 10 show the derived high angular resolution ($\sim$10$''$) dust temperature ($T_{d}$) and $N_{H_{2}}$ maps, obtained according to the procedures introduced in Section \ref{subsection:procedure}.
The distributions of the fitted column density, dust temperature and spectral index of each pixel for each region are provided in Appendix B. 

As a consistency check of our method with the existing studies based on {\it Herschel} PACS and SPIRE images (e.g. Nguyen-Luong et al. 2013; Schneider et al. 2015), we present gas column density ($N_{\rm H_{2}}$) and dust temperature fits ($T_{\rm d}$) of G10.2-0.3, obtained by using {\it Herschel} 160 \micron\ , 250 \micron\ and 350 \micron\ images only ($\beta$ fixed to a constant 2.0) in Figure \ref{fig:TN_herschel_G102}.
We use G10.2-0.3 as an example, since the {\it Herschel} images of this source are subject to the least amount of saturation.
Figures \ref{fig:TN_G10p2} and \ref{fig:TN_herschel_G102} show consistent $T_{d}$ and $N_{\rm H_{2}}$ distributions. 
However, the less smeared, higher (10$''$) angular resolution $T_{\rm d}$ map presents a better contrast, which we also believe is representative of the more accurate fitting of both $T_{\rm d}$ and $N_{\rm H_{2}}$.
In particular, the 10$''$ resolution $T_{\rm d}$ map reveals an embedded heated source in the highest column density molecular clump that cannot be seen merely from the SED fits of the {\it Herschel} images (Figure \ref{fig:TN_herschel_G102}).
The high resolution $N_{\rm H_{2}}$ image also separates the localized dense clumps/cores from the fluffy or filamentary cloud structures in a more clear way that is crucial for our structure identification (Section \ref{subsection:dendrogram}) and diagnoses the hierarchical cloud morphology.
The improved sensitivity of the low resolution $N_{\rm H_{2}}$ image detects the more diffuse cloud structures on larger angular scales. 
However, without velocity information, it is very difficult to distinguish the extended and low column density cloud structures from the foreground and background contamination.
The $T_{\rm d}$ maps can help visually identify some (if not all) heated diffuse dust and gas by the OB clusters around our target sources. 
However, in some heated but low $N_{\rm H_{2}}$ regions, our assumption of a single dominant modified black-body component in the SED fitting break down (Section \ref{subsection:SED}).
The hotter but lower $N_{\rm H_{2}}$ component in the line of sight can dominate the luminosity, bias the fitting to a higher averaged $T_{\rm d}$, and therefore can lead to a locally underestimated $N_{\rm H_{2}}$.
This effect may appear as jumps of column density in the $N_{\rm H_{2}}$ maps.
Due to the confusion with the foreground/background, and the aforementioned issues of the single-component modified black-body fitting, the low $N_{\rm H_{2}}$ structures were excluded from our quantitative analysis by setting the column density thresholds (e.g. Section \ref{subsection:pdf}).

The high resolution $N_{\rm H_{2}}$ images of our target sources show distinct morphologies, which qualitatively can be separated into molecular clouds with widely varying spatial distributions of sub-structures (e.g. W43-Main, and W43-South, and G10.2-0.3), molecular clouds showing a high concentration of mass at the central region (e.g. W49A, G10.6-0.4), and clouds with morphology in between these two cases. 
We resolved extended heated sources and localized ones that are embedded in local overdense regions.
A more detailed description of the individual sources is provided below.

The dominant dense component of W43-main exhibits a $\sim$15 pc scale \color{black} Z-shaped filamentary structure, with several embedded internal heating sources.
In addition, we resolved a large number of dense clumps/cores that are widely spread over the field of view of our SHARC2 350 \micron\ image.
The $T_{d}$ in this Z-shaped dense structure is in general lower than $\sim$25 K.
The $>$30 K localized heated sources may be associated with newly formed high-mass stars, or star-clusters. 
On large scales, the heated dust (and gas) appears spatially not well correlated with the dense structures and presents a filamentary or irregular structure. 
W43-South was also resolved with numerous, widely spread dense clumps/cores.
In W43-South, only some high column density structures show high $T_{d}$.
There are some heated, extended diffuse structures. 

For the sources W49A, G10.6-0.4, and W33, the column density peaks reside at the center of the molecular clouds, which are connected with exterior filamentary dense gas structures.
In particular, the massive molecular gas clumps located at the center of W33 are connected with parsec scale molecular gas arms, which are similar to those resolved by the Submillimeter Array (SMA) and the Atacama Large Millimeter Array (ALMA) observations of the OB cluster-forming region G33.92+0.11 (Liu et al. 2012b, 2015).
The highest temperature regions of these sources are spatially associated with the column density peaks.

The sources G10.2-0.3 and G10.3-0.1 may be relatively evolved, as they contain extended H\textsc{ii} regions in the Very Large Array (VLA) centimeter band observations (e.g. Kim \& Koo 2002).
The source G10.3-0.1 shows a clumpy, incomplete ring-like geometry over a $\sim$6 pc scale region.
The dominant heating sources are likely encircled by this incomplete ring.
From the $T_{d}$ map, we see some heated dust inside the ring, which is connected to a bi-conical heated shell on large scales. 
The column density distribution of G10.2-0.3 appears very irregular, and presents a large number of marginally resolved, or non-spatially resolved dense cores. 
For this region, the highest column density structures are seen preferentially with low $T_{d}$.
Some high column density structures, including the peak column density, are embedded with localized heating sources. 
There is extended, diffuse heated gas/dust that shows a shell-like or irregular morphology.
On the $\sim$5 pc scale, the marginally resolved curvature of the cool and dense molecular gas structures seem to follow the shell-like geometry of the heated gas, indicating strong effects of stellar feedback on the parent molecular cloud structures.

We attempt to quantify our visual impression of the target sources using  statistical methods, which are introduced in Section \ref{subsection:enclosedmass}, \ref{subsection:pdf}, \ref{subsection:2pt}, and \ref{subsection:dendrogram}.
The links of the derived quantities to the underlying physics, unfortunately, are not yet certain, and deserve future comparisons with numerical hydrodynamics simulations.

\begin{table*}\footnotesize{ 
\begin{center}
\hspace{-10.5 cm}
\vspace{-0.5 cm}

\caption{Fitting results of column density probability distribution functions (N-PDFs)}
\label{tab:PDFs_fit}
\begin{tabular}{lccccccc}\hline\hline
Target Source &Measured Threshold & Mean Column Density&Power-law Slopes with Starting $\eta$&\\
& log$_{10}$(N(H$_{2}))$&log$_{10}$(N(H$_{2}))$&$s_{0}$&$s_{1}$&$\eta_{1}$&$s_{2}$&$\eta_{2}$\\
\hline

W49A &21.87&22.08&-3.10(0.01)&-2.96(0.01)&-0.42 &-2.36(0.03)&1.31\\
G10.6-0.4&21.73&22.00&-2.54(0.01)& -2.55(0.04)&0.51& -1.57(0.24)&2.80\\
W43-south&21.88&22.06&-2.96(0.01)& -3.91(0.03)&0.66& -3.76(0.07)&1.26\\
W43-main&22.18&22.47&-2.86(0.01)& -2.88(0.01)&0.32& -2.89(0.02)&0.52\\
W33&22.10&22.41&-3.18(0.01)& -2.87(0.01)&0.02& -2.89(0.01)&0.34\\
G10.3-0.1&21.95&22.15&-2.98(0.01)& -3.01(0.01)&-0.15& -2.91(0.01)&0.29\\
G10.2-0.3&22.09&22.35&-2.69(0.01)&-2.84(0.01)&-0.52 &-4.68(0.05)&0.93\\
 \hline            
\end{tabular}
\end{center}
}\end{table*}

\subsection{Enclosed masses as functions of radius}\label{subsection:enclosedmass}
The overall masses of the observed molecular clouds range from a fraction to a few times of 10$^{5}$ $M_{\odot}$, and are summarized in Table \ref{tab:source_info}.
The degree of matter concentration in the molecular clouds can be quantified by the enclosed gas and dust mass as a function of radius, $M(r)$. 
Generally, OB cluster-forming clouds are more massive at a given radius than low-mass star-forming regions. 
The power-law form of the mass versus radius relation can be a consequence of the power-law density profile, since a radial density profile of $n(r)\propto r^{-p}$ gives a mass vs. radius relation of $M(r)\propto r^{3-p}$. 
Kauffmann et al. (2010) found that the cluster-bearing clouds roughly follow $M(r)\propto r^{1.27}$ , indicating that the slope of the mass versus radius relation, though largely uncertain, should be lower than 1.5.
The radial density profile for 51 massive star-forming regions based on dust continuum maps is found to be of $n(r)\propto r^{1.8\pm 0.4}$ at $r$$\sim$0.2 pc scales (Mueller et al. 2002), which corresponds to $M(r)\propto r^{1.2\pm 0.4}$.   

Figure \ref{fig:enclosed} shows the enclosed gas and dust mass profiles of our seven resolved target sources.
In addition, we show the first (radial) derivative of the enclosed mass profiles in the bottom row of Figure \ref{fig:enclosed}.
When calculating the enclosed mass profiles, we define the centers of W49A, G10.6-0.4, and W33, approximately at the centers of their centralized massive molecular gas clumps.

For W43-main, W43-south, G10.3-0.1 and G10.2-0.3, the centers were defined at the most massive molecular clumps located nearest to the center of the 350 \micron\ observational field of views.
The definitions of the centers of W43-main, W43-south, G10.3-0.1, and G10.2-0.3 are ambiguous to some extent due to their irregular cloud morphology.
Nevertheless, for these four sources, the enclosed mass profiles are not sensitive to the definitions of the centers, for the same reason. 

In Figure \ref{fig:enclosed}, the left column shows the results for the sources that have the most significant matter concentration, while the right column shows the results for those with relatively extended or randomized spatial distribution of dense structures (see Section \ref{subsection:maps} for more descriptions). 
This presentation strategy is mainly for the sake of avoiding crowded data points, but not for discriminating sources subjectively and artificially.

The $\partial (\log M(r))/\partial (\log r)$ of all presented sources resides in the range of 0.0-2.5 over all spatial scales.
As expected, the sources showing a significant centralized concentration of mass (i.e. the sources presented in the left panel of Figure \ref{fig:enclosed}; see discussion Section \ref{subsection:maps}) systematically show lower values of $\partial (\log M(r))/\partial (\log r)$.
Qualitatively, the low value of $\partial (\log M(r))/\partial (\log r)$ occurs when a marginally spatially resolved or unresolved massive molecular gas clump located at the center of the molecular cloud contributes to a very significant fraction of the enclosed mass. 
Outside of the centralized molecular gas clumps, the surrounding molecular gas structures are relatively diffuse and therefore the enclosed mass does not increase rapidly with radius.
On the $\lesssim$3 pc scale, the mini-starburst region W49A is the only case in which the centralized massive molecular gas clump is spatially resolved. 
A peak of $\partial (\log M(r))/\partial (\log r)$ is expected on the approximate spatial scales of the massive molecular gas clump.
This massive molecular gas clump appears to be dominating the mass in the central region of W49A, such that the $\partial (\log M(r))/\partial (\log r)$ drops quickly to below 1.0 on $\sim$3 pc scales.
The massive molecular gas clump in the inner $\sim$3 pc of W49A is embedded with a spatially well resolved, $\sim$2 pc ring-like distribution of UC H\textsc{ii} regions that are orbiting about the center of the ring (Peng et al. 2010; Galv\'an-Madrid et al. 2013).
This massive molecular gas clump also appears at the junction of the North and South filaments in the inner few pc (Galv\'an-Madrid et al. 2013), which are also spatially resolved in our column density image.
The azimuthal asymmetry due to the presence of these dense gas filaments can lead to a poorly defined ``center" when evaluating $M(r)$ that further leads to jumps in $M(r)$ and the apparent hump in $\partial (\log M(r))/\partial (\log r)$.

For all sources, the decrease in $\partial (\log M(r))/\partial (\log r)$ on the large spatial scales is due to the finite cloud size.
Our observational field of views of W43-Main, W43-South, and W33, are comparable with the angular scale of these molecular clouds. 
We double-checked the size of these molecular clouds from the {\it Herschel} SPIRE images. 
When the integrated mass of the extended cloud structures are dominate over the masses of the centralized massive gas clumps/cores (e.g. on $\sim$1-10 pc scales), the observed sources show a weak increasing trend of $\partial (\log M(r))/\partial (\log r)$.

Intriguingly, for most of the observed OB cluster-forming regions, the measured $\partial (\log M(r))/\partial (\log r)$ on parsec scales is smaller than that of a virialized molecular cloud, and appears close to that of a gravitationally bound H\textsc{ii} gas cloud that has a $\sim$10 km\,s$^{-1}$ sound speed.
We refer to Bressert et al. (2012), where $r_{vir}$ is defined as the virial radius based on a crossing time of 1 Myr, corresponding to the age found from recent high resolution studies of several young massive clusters (YMCs).
The enclosed mass profile of the aforementioned bound H\textsc{ii} cloud ($r_{\Omega}$) is calculated assuming an overdensity of gravitationally bound gas, with a sound speed of $\sim 10\ \rm kms^{-1}$, and equal gravitational potential and kinetic energies. The critical initial gas mass to form a YMC, assuming a star formation efficiency (SFE) of 30$\%$, is then $m_{crit}\sim 3\times 10^{4} M_{\odot}$.

\subsection{Column density distribution}\label{subsection:pdf}
The column density probability distribution function (N-PDF) of molecular clouds is a widely applied statistical measurement. 

\begin{figure*}
\hspace{2.5cm}
\begin{tabular}{ p{6.5cm} p{6.5cm} }
\includegraphics[width=7.5cm]{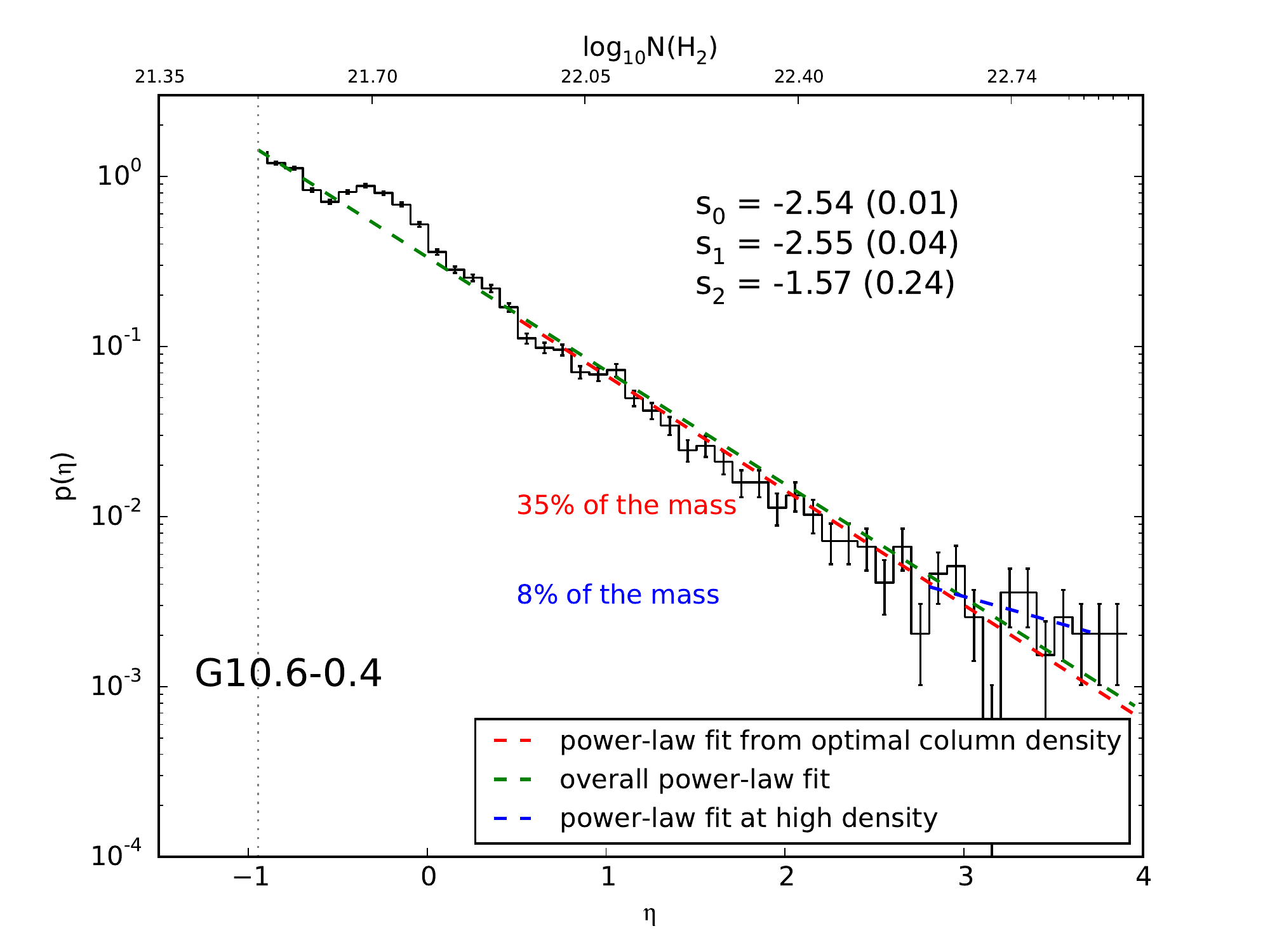} & \includegraphics[width=7.5cm]{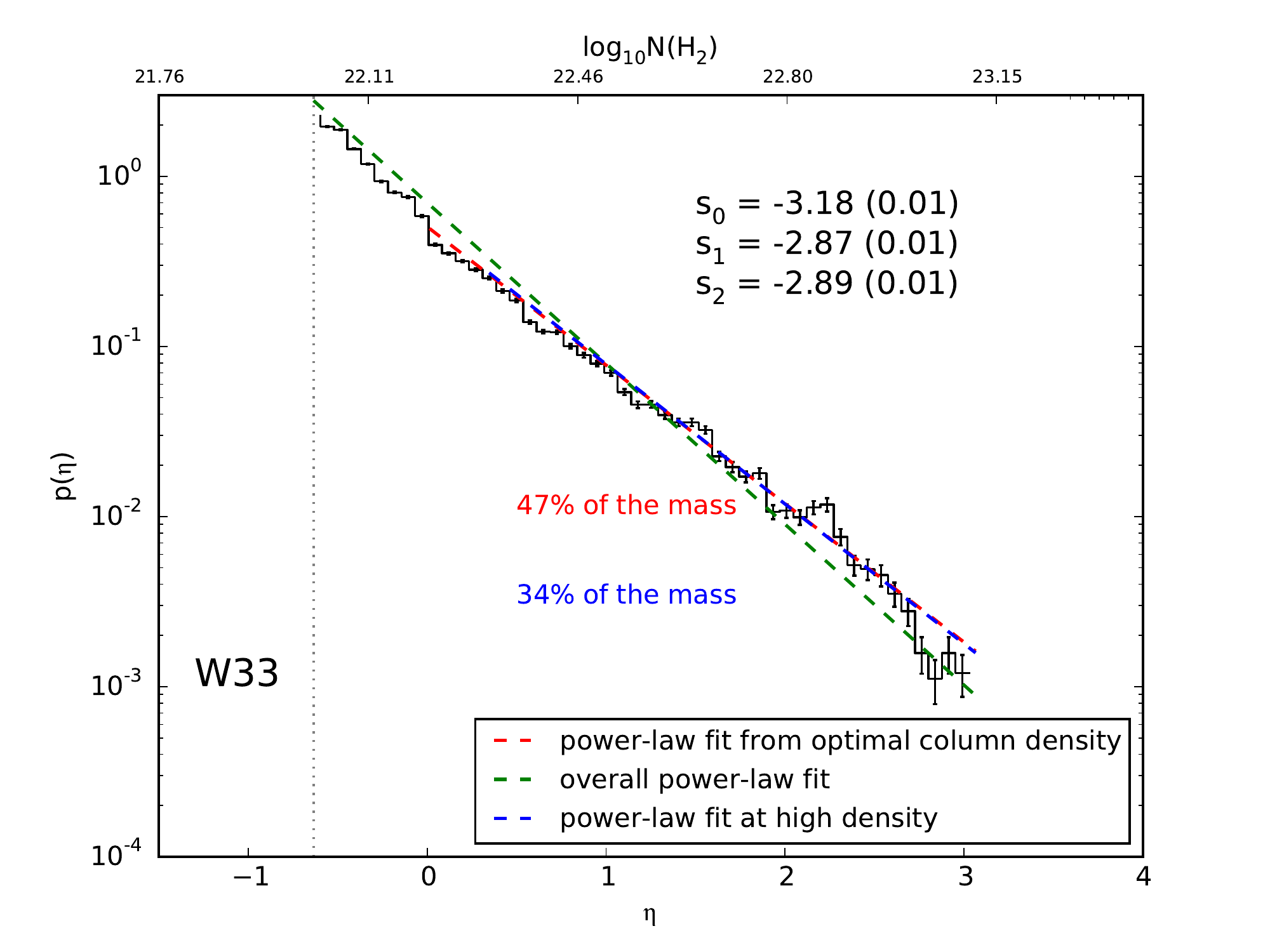} \\
\includegraphics[width=7.5cm]{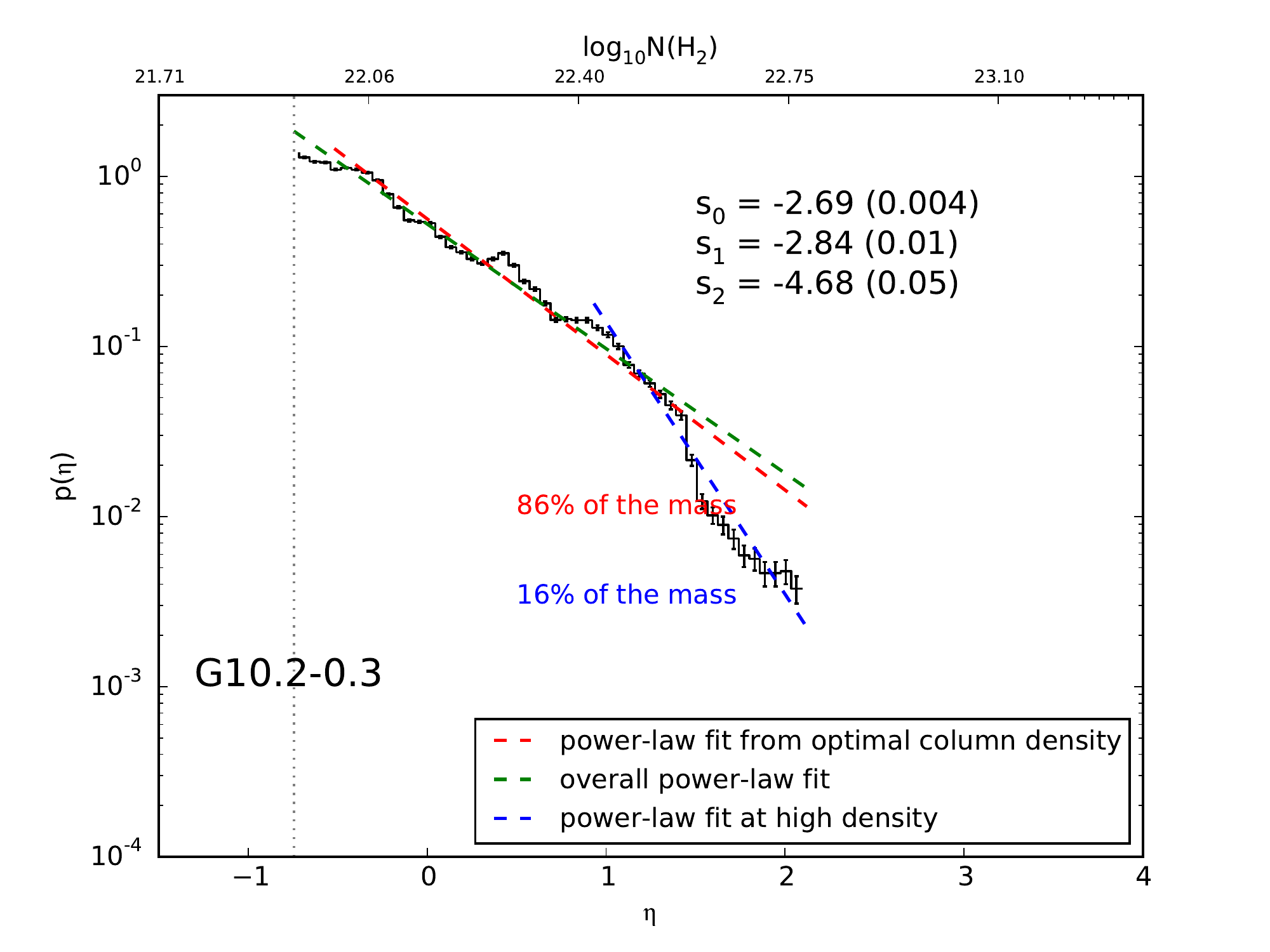} & \includegraphics[width=7.5cm]{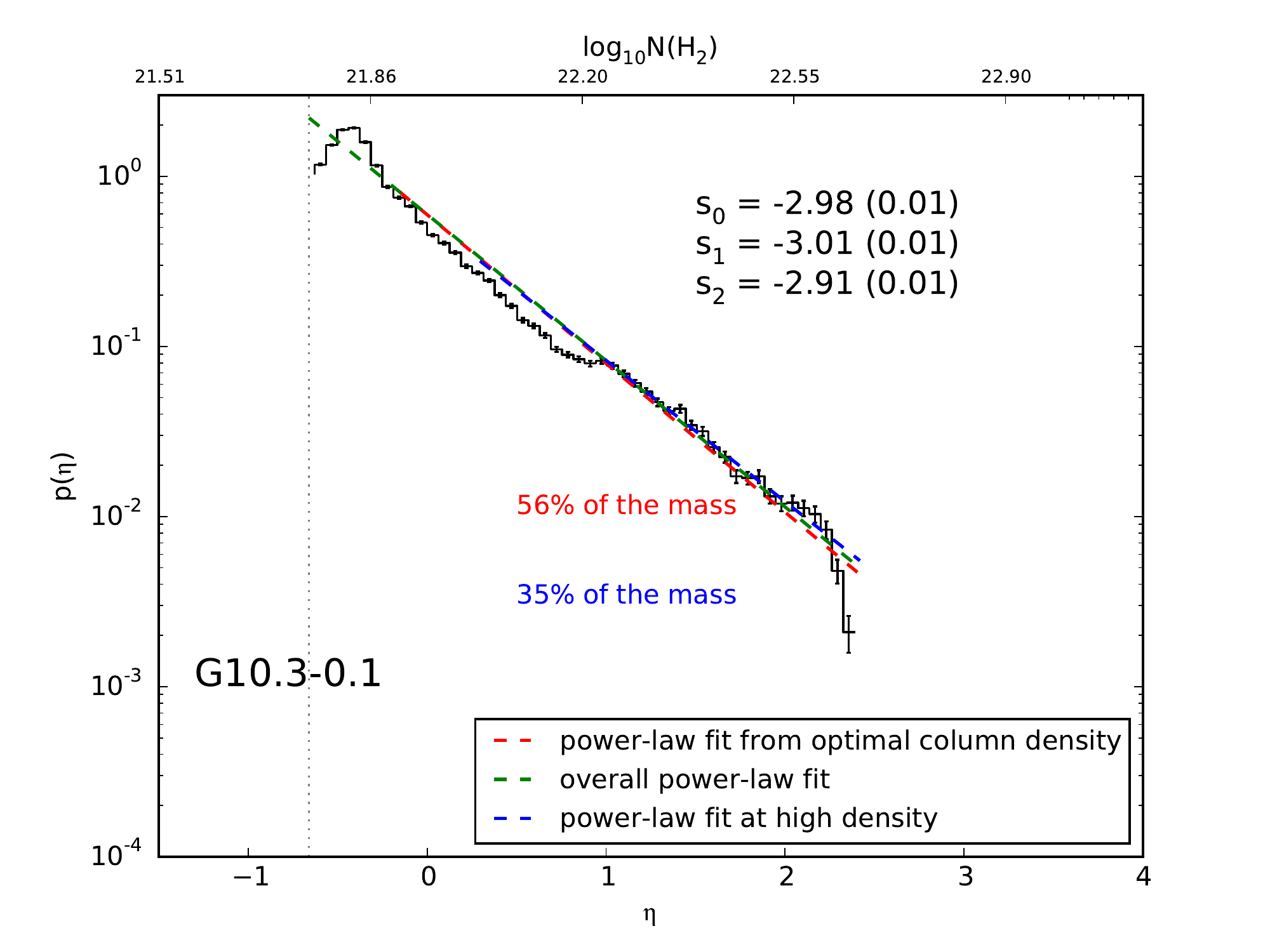} \\ 
\includegraphics[width=7.5cm]{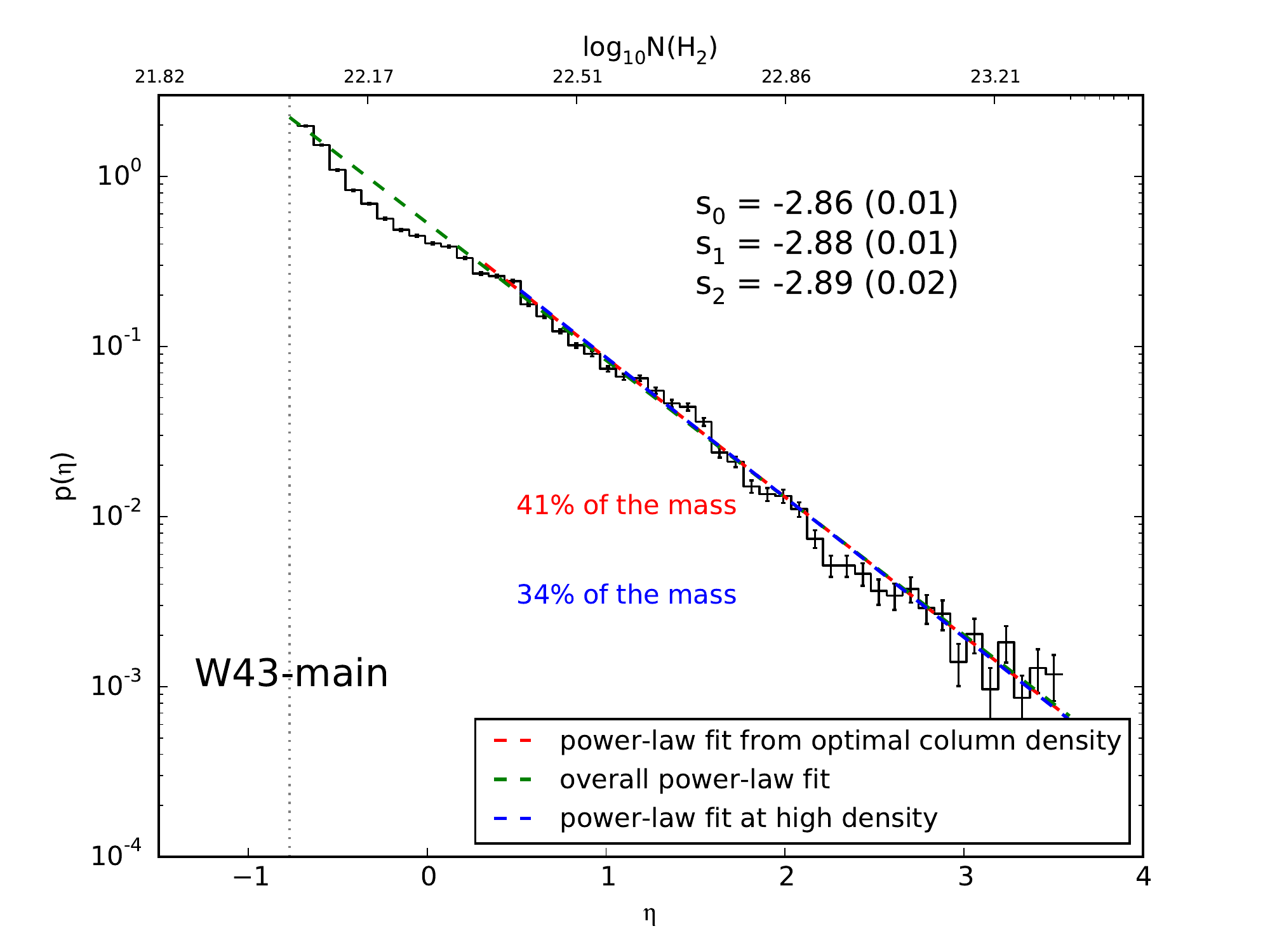} & \includegraphics[width=7.5cm]{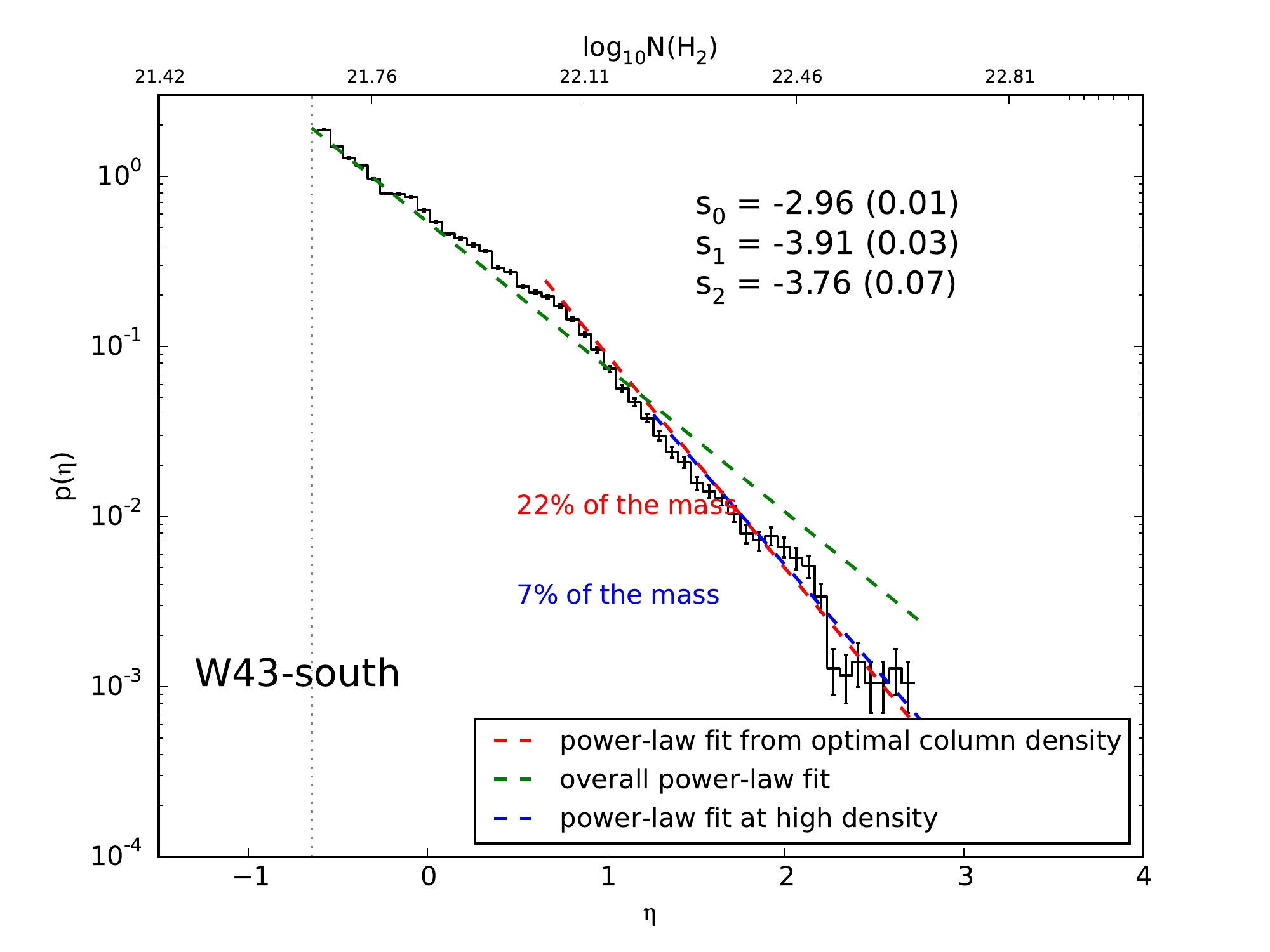} \\
\includegraphics[width=7.5cm]{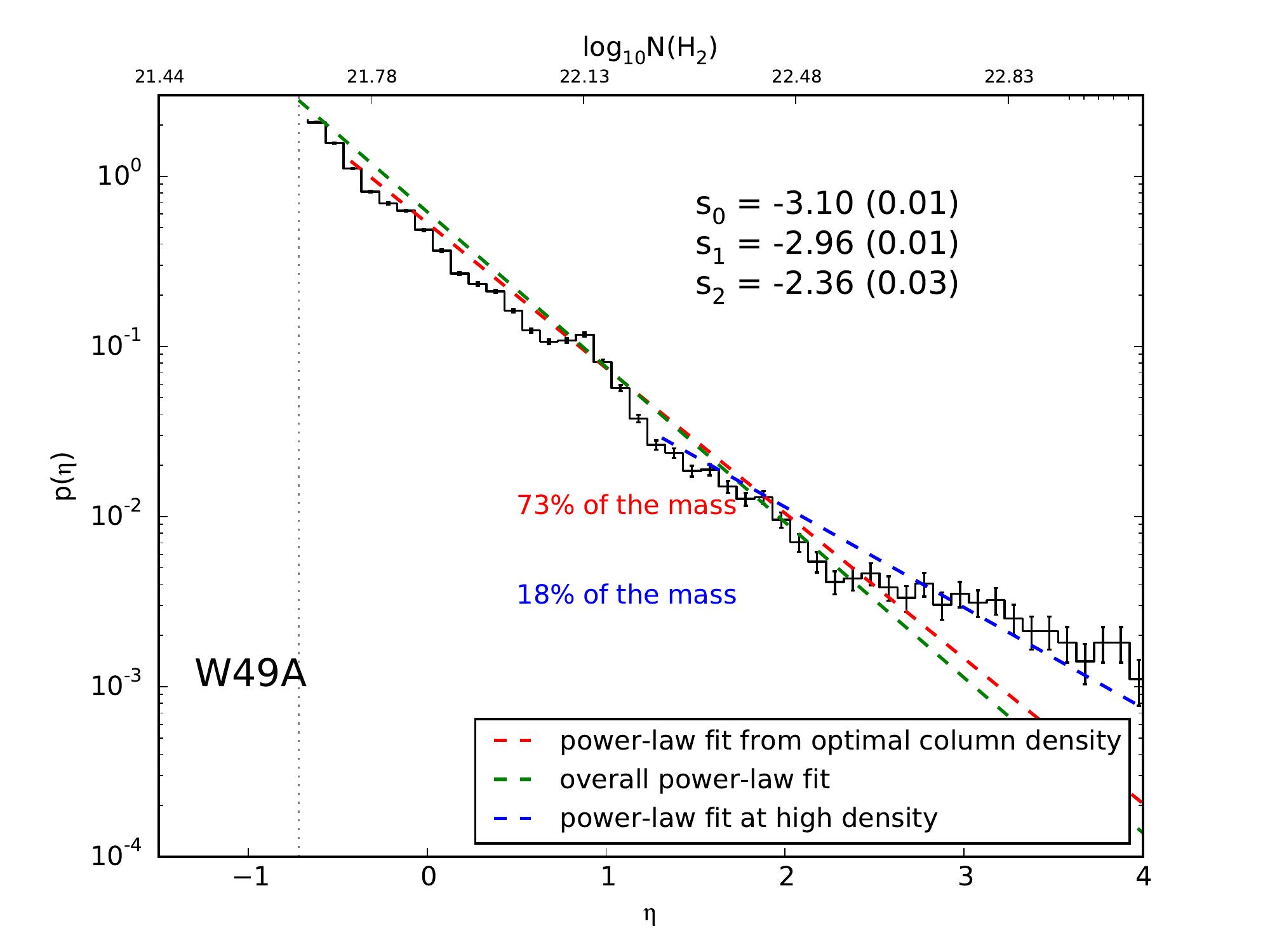} \\
\end{tabular}
\caption{Column density probability distribution functions (N-PDF).The y-axis gives the normalized probability p($\eta$), the lower x-axis is the logarithm normalized column density and the upper x-axis is the logarithm of column density. The dashed vertical line shows the threshold for each source, the 1-$\sigma$ level. Error bars are calculated from Poisson statistics. The green, red and blue dashed lines are the overall power-law fit of slope $s_{0}$ starting from the threshold, the power-law fit of slope $s_{1}$ from the optimal minimum column density value and the power-law fit of $s_{2}$ at the high column density end that describes an excess or deficit that deviates from a single power-law. Mass percentages of the considered column density ranges for the latter two power-law fits are shown with the respective colors in the plots.
}
\label{fig:pdf}
\end{figure*}

We follow the frequently used formalism of N-PDFs from previous numerical works, where the natural logarithm of the ratio of column density and mean column density is 
$\eta = ln(N_{\rm H_{2}}/\langle N_{\rm H_{2}}\rangle)$, and the normalization of the probability function is given by 
$\int_{-\infty}^{+\infty}p(\eta)d\eta = \int_{0}^{+\infty}p(N_{\rm H_{2}})d(N_{\rm H_{2}})=1 $; for more details see Schneider et al. (2015a).
Figure \ref{fig:pdf} shows the N-PDF above the selected column density thresholds that were determined by the 1$\sigma$ noise level in the $N_{\rm H_{2}}$ images.
We found that the N-PDFs of these sources can be approximated by power-laws.
However, the high column density ends of the N-PDF may deviate from the overall power-laws.
For examples, the sources G10.6-0.4 and W49A show excess at the high column density ends. 
On the other hand, the sources G10.2-0.3 and W43-south show a deficit at high column densities.
There are other minor variations in the observed N-PDFs, which are beyond the focus of the present paper.
Part of these minor variations in the intermediate range of column density may be related to the limited field of view (more below).
We note that although the excess of the very high $N_{\rm H_{2}}$ pixels in G10.6-0.4 looks marginal from Figure \ref{fig:pdf}, it is difficult to ignore such excess given that they comprise the most robustly detected component in the image domain (Figure \ref{fig:TN_G10p6}). 
Based on the previous observations of Liu et al. (2010a), (2012a), we expect this excess to become more obvious with improved angular resolution. 
Therefore, this excess of high $N_{\rm H_{2}}$ pixels needs to be included in any quantification of the N-PDF. 

Interestingly, the similarities in the N-PDF features seem to be linked to the similarities of their overall cloud morphology.
The sources that show significant excess of high column density pixels appear to be those which are also showing significant spatial concentration of high column density structures at the center.
On the other hand, for the sources which are showing widely spread or irregular dense structures, there is a deficit of high column density pixels.
These two types of sources can be separated by their enclosed mass profiles, and the slopes of the profiles in Section 3.2.

We used the {\tt powerlaw} package (Alstott et al. 2014) to fit the observed column density distributions with power-laws and an additional high column density power-law tail (Figure \ref{fig:pdf}) using Maximum Likelihood Estimation (MLE).
In this way, it is not required to pre-bin the observed column density distributions, which has been proven to provide better accuracy compared to linear regression on histograms of the observed column density (Clauset et al. 2009).

We estimated the background and foreground contamination as of an offset value determined from the outskirts of a larger column density map derived from {\it Herschel} data, and subtracted this value from our column density data (Schneider et al. 2015b; Ossenkopf et al. 2016). 
Since the lower column density part can be influenced dramatically by the noise level, boundary bias, variation along the line-of-sight or enclosed contours (Lombardi et al. 2015), we only considered the values larger than the threshold measured for each source, as listed in Table 3. 

We note that parts of the diffuse cloud structures of some sources may not be covered by the relatively small field of view of our CSO-SHARC2 350 \micron\ images.

We have examined this from the JCMT-SCUBA2 and {\it Herschel} images that have a wider field of view.
We found that those structures outside of our CSO-SHARC2 fields, above the column density threshold of our N-PDF analysis, mostly have small masses and sizes. 
Therefore, we do not expect this effect to dramatically bias our quantification of the N-PDFs.
It is also not trivial to distinguish these features from foreground/background clouds.

For the quantitative description of our approximated results, we obtained an overall power-law fit (i.e. $p(\eta)\propto\eta^{s}$) to all the data larger than the threshold values, and a power-law fit that finds the optimal minimum column density between the empirical distribution and power-law model distributions. 
Due to deviations from a single power-law, we also fit a second power-law component at high column densities for G10.6-0.4, G10.2-0.3, W43-South, and W49A.
The power-law slopes for the three fits are denoted as $s_{0}$, $s_{1}$ and $s_{2}$, and are summarized in Table \ref{tab:PDFs_fit}.
We caution that the power-law index for G10.6-0.4 at the high density end was only tentatively determined, and needs to be improved with better (e.g., higher angular resolution, better sensitivity, and more frequency bands) observational data to improve the statistical reliability.
A more detailed discussion of the N-PDFs is deferred to Section \ref{discussion:npdf}.

\subsection{Deriving the two-point correlation function of column density maps}\label{subsection:2pt}
The two-point correlation (2PT) function\footnote{This is equivalent to the azimuthally averaged autocorrelation function (ACF) (see e.g., Kleiner \& Dickman 1984).} is a powerful tool to systematically diagnose the characteristic spatial scales in the molecular cloud morphology, and helps to understand the spatial distribution and clustering properties of high $N_{H_{2}}$ structures.
We followed the procedure outlined in Szapudi et al. (2005) to perform fast estimates of two-point correlation function via Fast Fourier Transforms (FFTs) by zero-padding the original maps. 
The normalization of the geometry is calculated by filling 0's to the masked pixels and 1's to the valid ones. 
This form of the two-point correlation function is the same as the unbiased measure described in Kleiner \& Dickman (1984), though we used the column density of each pixel without subtracting the mean column density.
In this way, we are more sensitive to the correlation of the density field, rather than density fluctuations. 
We adopted a flat weighting, where all pixel pairs were assigned with an identical weight.
This otherwise unbiased estimator may be biased when the scales of the lags (i.e., the spatial separation of two pixels in a pair) are comparable with the observational field of view (Kleiner \& Dickman 1984).

\begin{figure}
\hspace{-0.5cm}
\includegraphics[width=9.5cm]{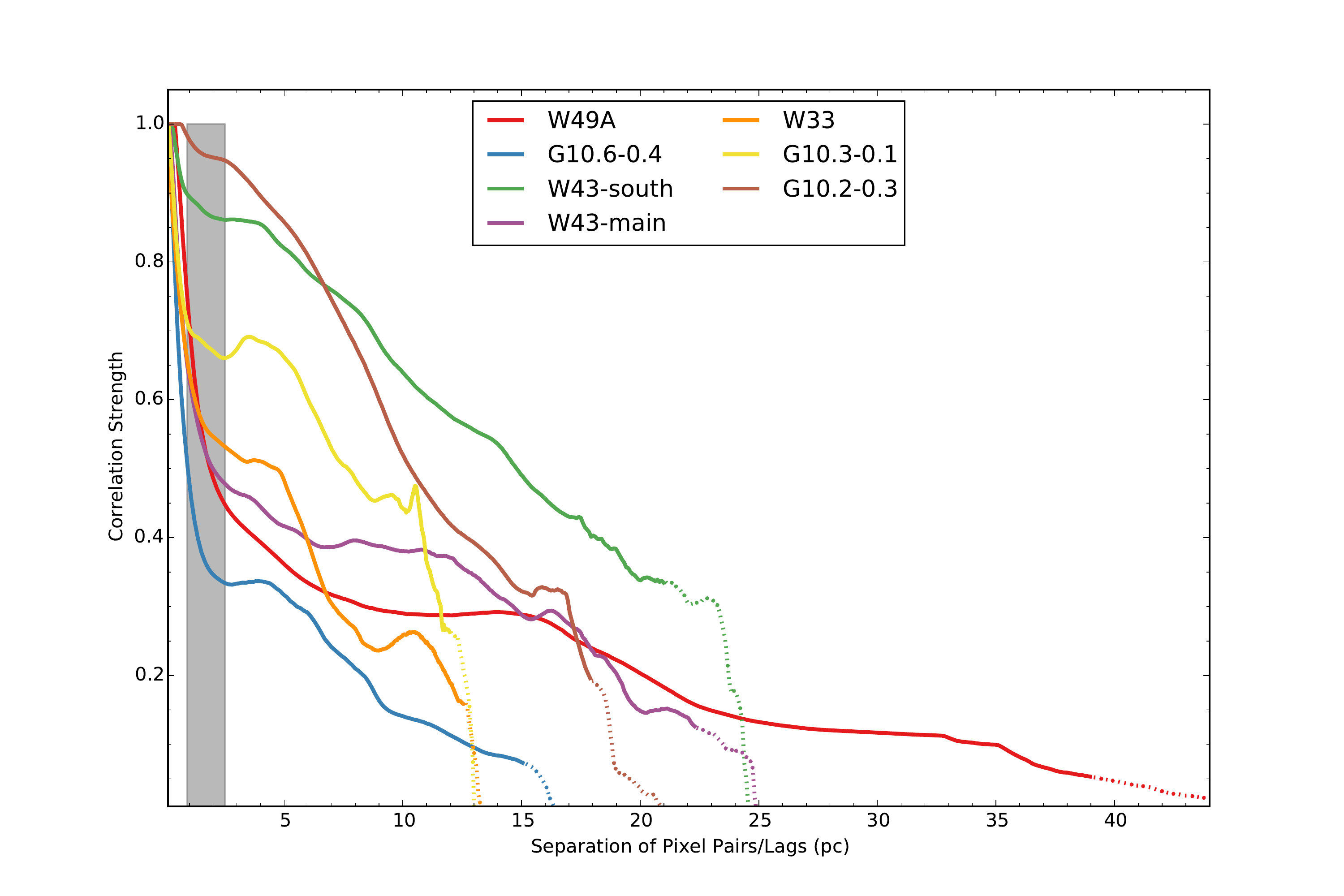}
\caption{Normalized 2PT functions for all the sources, showing the correlation strengths as a function of spatial separation. For separations with pixel pairs of less than $\sim90\%$ total pixel number, the correlation strengths are plotted with dashed lines. The grey filled region indicates approximately where the rapidly decreasing (RD) components end.}
\label{fig:2pt_all}
\end{figure}

Figure \ref{fig:2pt_all} shows the obtained 2PT correlation functions with the directional information averaged out.

The averaged 2PT correlation functions of W49A, G10.6-0.4, W33, W43-Main, and G10.2-0.3, have a common feature: a rapidly decreasing correlation strength at short lags, and a sudden transition to a shallower decrease at larger lags.

The rapidly decreasing (RD) component corresponds to localized clumps/cores, while the shallower decreasing (SD) component comes from the broader molecular cloud.
We note that all the derived 2PT correlation functions show a steep decrease in correlation strength at large lags because of the limited field of view.
To avoid this bias in our analysis, we exclude the range of lags that have pixel pairs less than $\sim 90\%$ of the number of total pixels.

For sources like W43-south and G10.2-0.3, where the structures of column density are composed of widely scattered overdensities, the correlation strengths at shorter lags do not decrease as significantly as the correlation strengths of the more centrally condensed sources, like W49A, W33 and G10.6-0.4. 

They also do not show an obvious rapidly decreasing component at short lags.

There are some bumps seen in the 2PT functions for several sources at larger lags, which may present the characteristic scales of some localized overdensities. 
We do not quantify these features in this paper.
The average azimuthal profiles of our obtained 2PT functions are intended to capture the major similarities and differences between the target sources.

\begin{table*}\footnotesize{ 
\begin{center}
\hspace{-10.0 cm}
\vspace{-0.5 cm}

\caption{Fitting results of two-point correlation functions: Three segments}
\label{tab:2pt_fit_pl_seg3}
\begin{tabular}{lcccccccc}\hline\hline

Target Source&$\alpha_{1}$&$\alpha_{2}$ & $\alpha_{3}$ & $x_{break_{1}}$  & $x_{break_{2}}$& $y_{break_{1}}$  & $y_{break_{2}}$\\
               &                   &                     &                   &                   (pc)                       &                         (pc)  &          &  &\\
\hline 
W49A &$-0.41(0.0016)$&$-0.09(0.013)$&$-1.85(0.032)$&$7.41(2.00(0.026))$&$16.48(2.84(0.018))$&$0.30$&$0.28$\\

G10.6-0.4 &$-0.45(0.010)$&$-0.12(0.037)$&$-1.52(0.029)$&$2.03(0.71(0.10))$&$6.05(1.80(0.019))$&$0.35$&$0.31$\\

W43-south &$-0.06(0.001)$&$-0.51(0.01)$& $-1.19(0.015)$&$6.30(1.84(0.01))$&$14.03(2.64(0.01))$&$0.81$&$0.54$\\

W43-main &$-0.30(0.009)$&$-0.22(0.01)$& $-2.71(0.07)$&$1.77(0.57(0.35))$&$15.74(2.76(0.01))$&$0.53$&$0.32$\\

W33  &$-0.20(0.01)$&$-0.06(0.09)$& $-0.93(0.02)$&$2.80(1.03(0.18))$&$4.41(1.49(0.02))$&$0.51$&$0.48$\\

G10.3-0.1 &$-0.18(0.01)$&$-0.011(0.019)$& $-0.83(0.03)$&$0.81(0.21(0.16))$&$3.53(1.26(0.02))$&$0.68$&$0.67$\\

G10.2-0.3 &$-0.017(0.003)$&$-0.32(0.0033)$& $-1.18(0.016)$&$3.61(1.28(0.059))$&$7.45(2.01(0.018))$&$0.93$&$0.74$\\

\hline            
\end{tabular}
\end{center}
}
\vspace{0.1cm}
\footnotesize{Note.--- For $x_{break_{1}}$  and $x_{break_{2}}$, we listed the fitted values with the format of corresponding separation value(fitted value(error)), since we conducted the fits in log-log space.
}
\par
\end{table*}

\begin{table}\footnotesize{ 
\begin{center}
\hspace{-10.0 cm}
\vspace{0.2 cm}

\caption{Fitting results of two-point correlation functions: Two segments}
\label{tab:2pt_fit_seg_2}
\begin{tabular}{lccccc}\hline\hline
Target Source & $k_{1}$ & $k_{2}$ & $x_{break}$ & $y_{break}$ \\
\hline
W49A & -0.2981 & -0.0095 & 2.2310 & 0.3349 \\
G10.6-0.4 & -0.6089 & -0.0252 & 1.0530 & 0.3588 \\
W43-south & -0.1284 & -0.0304 & 0.6598 & 0.9153 \\
W43-main & -0.3741 & -0.0017 & 1.3757 & 0.4854 \\
W33 & -0.3950 & -0.0391 & 1.0277 & 0.5941 \\
G10.3-0.1 & -0.2866 & -0.0236 & 0.9489 & 0.7280 \\
G10.2-0.3 & -0.0529 & -0.0529 & 0.0 & 1.0 \\

 \hline            
\end{tabular}
\end{center}
}
\end{table}

We hypothesize that the steeply decreasing component and the shallower one represent structures that were created or supported with different physical mechanisms, and thereby have different characteristic spatial scales and spatial distributions.
More discussion about the physical implications is deferred to Section \ref{discussion:2pt}.

For the sake of quantifying and archiving the observed 2PT functions for comparison with other observations, theories, and numerical simulations, we perform segmented linear fits of these 2PT functions in log (i.e., power-law fitting) and linear space.
We note that fits in both the log and the linear space involve large simplifications of the real data. 
Empirically, the fits in the linear space better capture the transition point from the RD to the SD component, without being sensitive to small variations of the data or the fit parameters.
However, fitting in log space presently has a more straightforward link to physical interpretations (more in Section \ref{discussion:2pt}).
The nature of the 2PT functions, including the azimuthal asymmetry, demand more careful future observational and theoretical studies.

We used two-component and three-component segmented linear models to quantify the derived 2PT correlation functions in log-log space, adopting a Levenberg-Marquardt nonlinear least-squares minimisation implemented in {\tt lmfit} (Newville et al. 2014).  
The free parameters are the slopes (power-law indices in linear space) of two/three segments and the break points of lags. We tentatively used these two models since we noticed that, for several sources like W33, G10.3-0.1 and G10.6-0.4, there are apparent plateaus of correlation strength at lags between $\sim$2-8 pc that may bias the slopes at small lags to smaller values. We further compared AIC and BIC statistics, which provide a measure of the quality of fits between different models.
Since that three-component models provide a better fit for all the sources, we have included the results of the three-component model fits listed in Table \ref{tab:2pt_fit_pl_seg3}.
The functional form of the three-component model is,
\begin{equation*}
\begin{split}
\log(S_{tr}(r)) = \\
\begin{cases}
\alpha_{1}\log(r) &r < x_{break_{1}},\\
\alpha_{2}\log(r) + (\alpha_{1}-\alpha_{2})\log(x_{break_{1}})& x_{break_{1}}< r < x_{break_{2}},\\
\alpha_{3}\log(r) + (\alpha_{1}-\alpha_{2})\log(x_{break_{1}}) \\
\       + (\alpha_{2}-\alpha_{3})\log(x_{break_{2}}) &r > x_{break_{2}}.
\end{cases}
\end{split}
\end{equation*}
where $S_{tr}(r)$ represents the correlation strength at separation of \emph{r}.
The first break point is the transition from the RD component to the SD component. 
The power-law indices $\alpha_{1}$, $\alpha_{2}$, $\alpha_{3}$ indicate the decreasing rate of correlation strengths at different separation scales. 
We excluded the fit parameters of the second break point and third segmented component in the analysis since they are subject to the decreasing statistical instability at larger separations.

\begin{figure}
\hspace{-0.5cm}
\includegraphics[width=9.5cm]{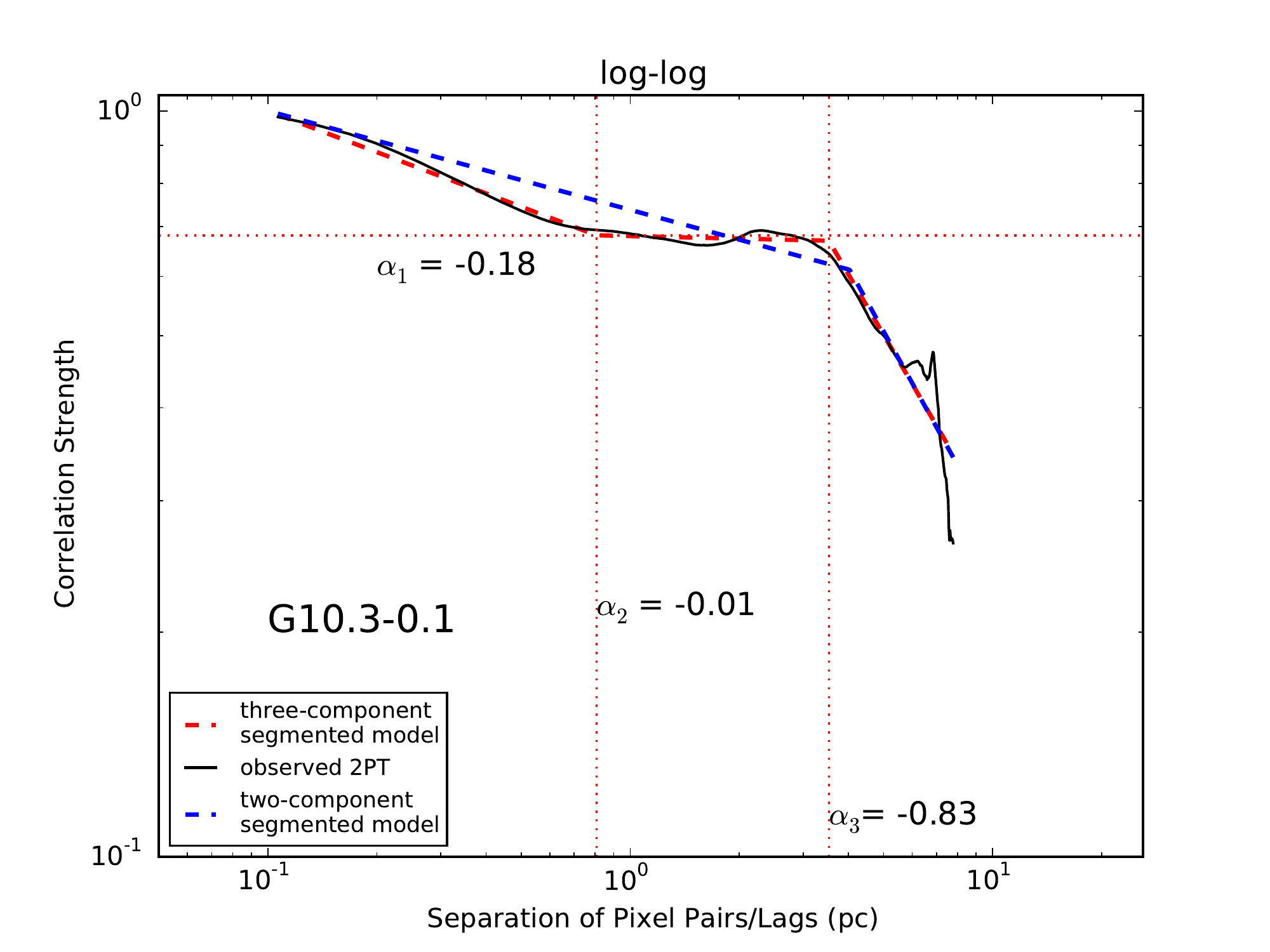}
\caption{2PT correlation function of G10.3-0.1, fitted with a three-component segmented linear model in log-log space. Slopes are marked for different component with red dashed vertical lines showing the break points. Blue dashed line is the two-component segmented linear model fits and red dashed line is the three-component segmented linear model fits.}
\label{fig:2pt_G103}
\vspace{0.2 cm}
\end{figure}

The slopes of the power-law at small scales range from -0.2 to -0.4, while the values of slopes for larger scales are more scattered, with a range of -0.8 to -2.7. 
One example of our fitted three-component model and two-component model of 2PT is shown in Figure \ref{fig:2pt_G103}.

For the segmented linear model, we fit the functional form of
\begin{equation*}
\hspace{-0.5cm}
S_{tr}(r) = \begin{cases}
k_{1}r + 1 &r < x_{break},\\
k_{2}r + (k_{1}-k_{2})x_{break} + 1&  r > x_{break}.\\
\end{cases}
 \end{equation*}
The three fit parameters are the two slopes $k_{1}$ and $k_{2}$ of the two components and the break point $x_{break}$, together with the break correlation strength, listed in Table \ref{tab:2pt_fit_seg_2}.

\subsection{Structure Identification: dendrograms}\label{subsection:dendrogram}
Many structure identification techniques have been developed to describe the  hierarchical nature of the molecular cloud morphology. 
In this paper, we focus on analyzing the structures identified using the tree algorithm implemented in the {\tt dendrograms} realization (Rosolowsky et al. 2008). 
Dendrograms are multi-level structures defined by set of iso-surfaces, where the bottom structure represents the low-density gas that fills the vast volume of a cloud. 
The defined boundary in our case is the threshold we measured with the column density map. 
Emerging from the trunk are different levels of branches that are distinguishable from their parent structures (either a trunk or lower-level branch) as they are denser above a defined increment. 
The leaves are the peak levels of a dendrogram, which are essentially the densest small regions containing local maxima.
This method is able to track the multi-scale density structures without relying on any assumption of the emission profile or morphology, thus provides an unbiased view of the underlying hierarchy in the clouds. 
Since our derived column density maps are able to recover extended emission, the impact of spatial filtering, as pointed out by  Kauffmann et al. (2010), is expected to be alleviated when we are quantifying the properties of the dendrogram output. 

Other commonly adopted structure identification techniques include the {\tt clumpfind} (Williams et al. 1994), and the {\tt gaussclumps} (Stutzki 
\& Guesten 1990) algorithms. 
For our high angular resolution $N_{\rm H_{2}}$ images that present rich and hierarchical structures, analyzing using the {\tt clumpfind} algorithm leads to artificial fragmentation (Pineda et al. 2009).
On the other hand, many localized structures we resolved deviate from the assumed 2-dimensional Gaussian geometry, and lead to poor results with the {\tt gaussclumps} algorithm.
A comparison of the structures identified by these algorithms, and the defects in the analysis, will be elaborated in the thesis of the first author, and is omitted from the present manuscript. 

When deriving the dendrograms of each region, we trimmed $\sim20$$''$ near the edges of all the column density maps to suppress the non-uniform noise mainly propagated from the SHARC2 observations. We set the lowest contour level as $7\times 10^{21}$ $\rm cm^{-2}$, a minimum increment of 5-$\sigma$ for a branch to be identified from its parent structure, and a minimum size of 7 pixels (comparable to the beam size) for a leaf to be considered as an independent entity. 
For leaves that stem from a parent structure, we further calculated a `corrected' mass value obtained by subtracting the merge level, which is defined as the mean column density of the parent structure (see more details from Ragan et al. 2013).
Note that the choice of setting the `increment' value will mainly impact the amount of identified leaves and is also the major factor that influences the original and the corrected mass. 
We caution that this merge level subtraction method can bias the masses of the relatively insignificant leaves, and may over-subtract in some cases depending on the localized column density.
In particular, for the leaves that are only about 5-$\sigma$ more significant than their parent structure, the subtraction process may remove the majority of their mass.
We have to compromise between identifying the most reliable leaves and taking advantage of our high angular resolution maps when doing a dendrogram analysis.

The effective radius is defined as $r_{eff} = (A/\pi)^{1/2}$ where $A$ is the area of a certain structure. 
We calculated the bolometric luminosity ($L_{bol}$)
and bolometric temperature ($T_{bol}$) for each leaf. $L_{bol}$ is calculated by integrating our obtained SEDs,
\begin{equation}\label{eq_lbol}
 L_{bol} = 4\pi d^2 \int_0^{\infty} S_{\nu}d\nu \qquad,
\end{equation}
where $d$ is the distance to our target source,
and the bolometric temperature follows Myers \&\ Ladd (1993):

\begin{equation}
T_\mathrm{bol} = 1.25 \times 10^{-11}~\frac{\int~\nu S_{\nu}d\nu}{\int~S_{\nu}d\nu}~~\mathrm{K}.
\end{equation}
The integration was from 0.1 \micron\ to 1 cm.

Our dendrogram analysis recovered 420 clumps/cores in these 7 regions. 
The measured $T_{bol}$ is less than 40 K for all the leaves. 
Most of the clumps/cores we measured have masses between $10^{2}-8\times10^{3}\ M_{\odot}$with temperatures of $20-30$ K. 
The properties of all leaves for our target sources are summarized in Tables \ref{tab:tab_W43_main_leaves}, \ref{tab:tab_W43_south_leaves}, \ref{tab:tab_W33_leaves}, \ref{tab:tab_G102_leaves}, \ref{tab:tab_G103_leaves}, \ref{tab:tab_G106_leaves}, \ref{tab:tab_W49A_leaves}. 
For each leaf, we list the center (position of the local maximum), effective radius, the un- and corrected mass values, the mean column density, the merge level (if present) and the mean dust opacity index, together with the bolometric luminosity and bolometric temperature.

Without the merging level correction, 149 of the identified leafs meet the threshold for massive star formation proposed by Kauffmann et al. (2010), which accounts for $\sim35\%$ of all the leaves.  

We examined the corrected mass and radius relationship and found that the slopes are between $1.4-2.0$.  
These relations depend highly on the core/clump boundary we measured. 
To investigate the global trend of the mass-radius relation in each region, we plot the diagrams that connect each leaf up to the parent structures it emerges from in Appendix D. 
We also plot several reference lines, including the empirical threshold proposed by Kauffmann et al. (2010) for non-massive and massive star-forming regions.  
More details of the properties of the identified leaves (i.e. dense clumps/cores), and the spatial separations of these structures, are discussed in Section \ref{sub:dendro}.
Some description of the dendrogram results that are not directly related to the major science results of this manuscript, are provided in Appendix D.

\begin{figure}
\hspace{-0.3cm}
\includegraphics[width=11cm]{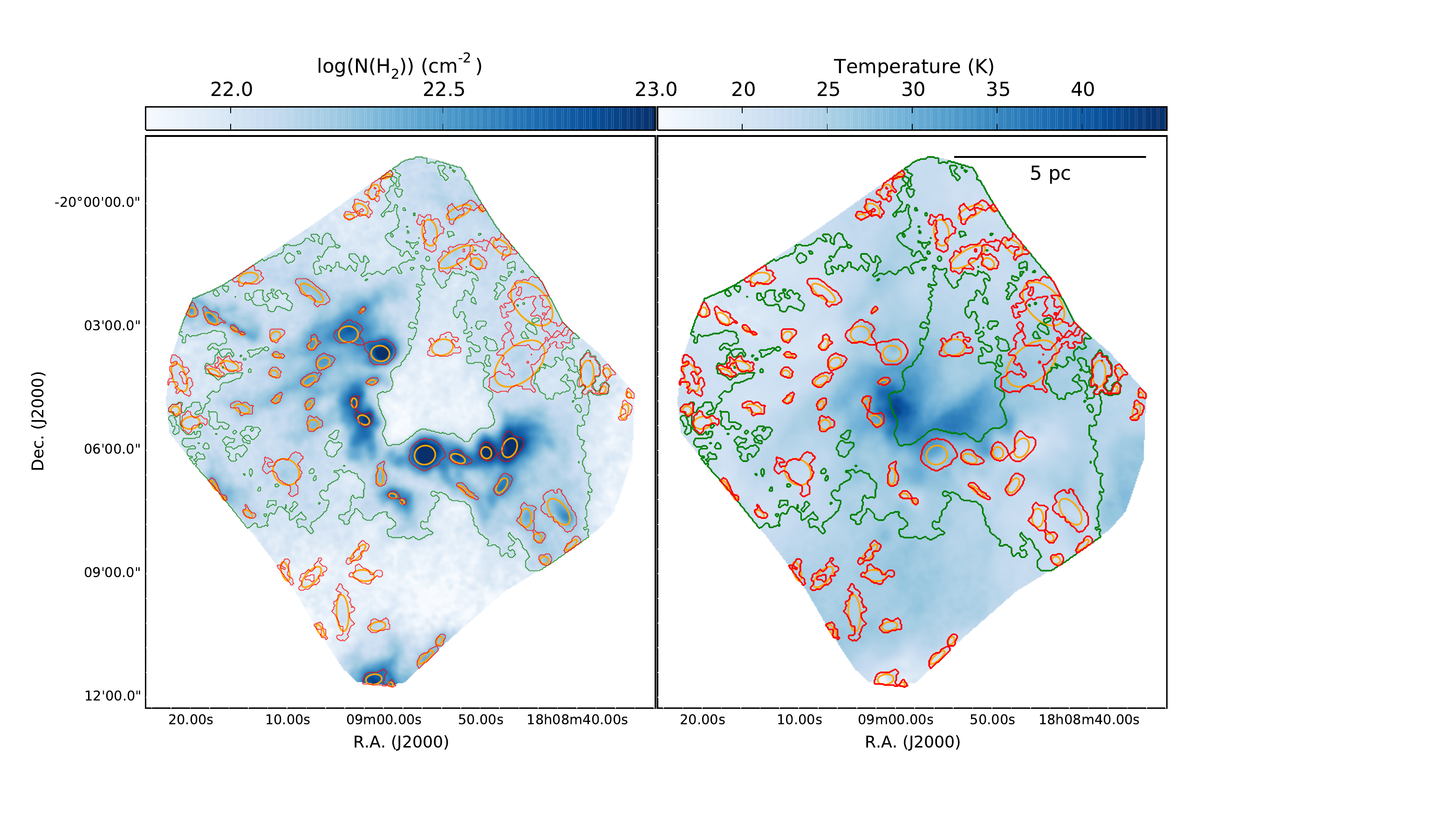}
\caption{Column density and dust temperature maps of G10.3-0.1 with red contours showing the leaves identified by {\tt dendrogram}. We fit ellipses to the leafs, which are shown in orange. Green contours show the parent structure of the leaves, i.e., the merge level that we subtracted from the leaves. }
\label{fig:w33_leaf}
\end{figure}

\begin{figure*}
\hspace{-0.3cm}
\begin{tabular}{ p{8.5cm} p{8.5cm} }
\includegraphics[width=9.5cm]{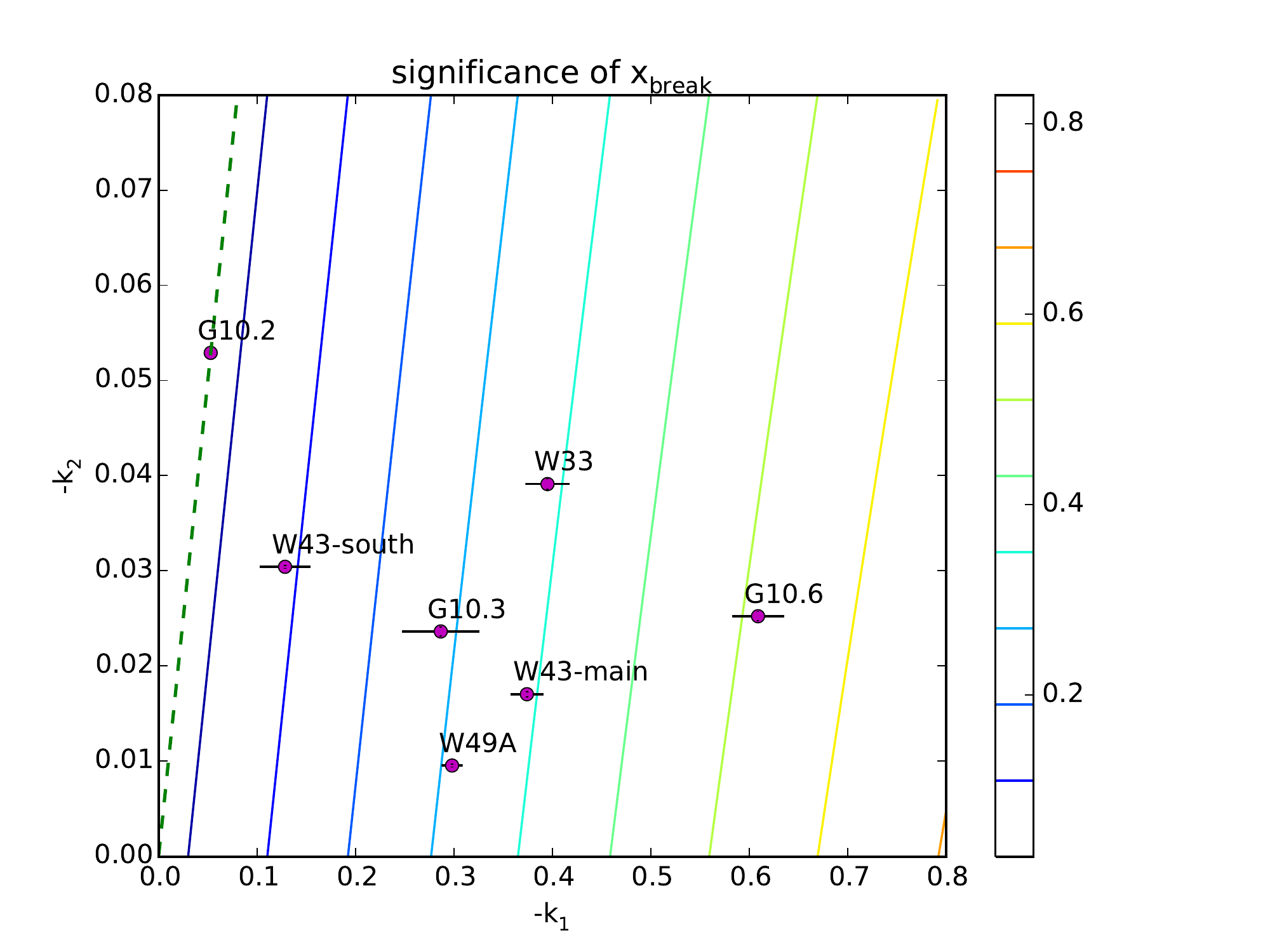}  & \includegraphics[width=9.5cm]{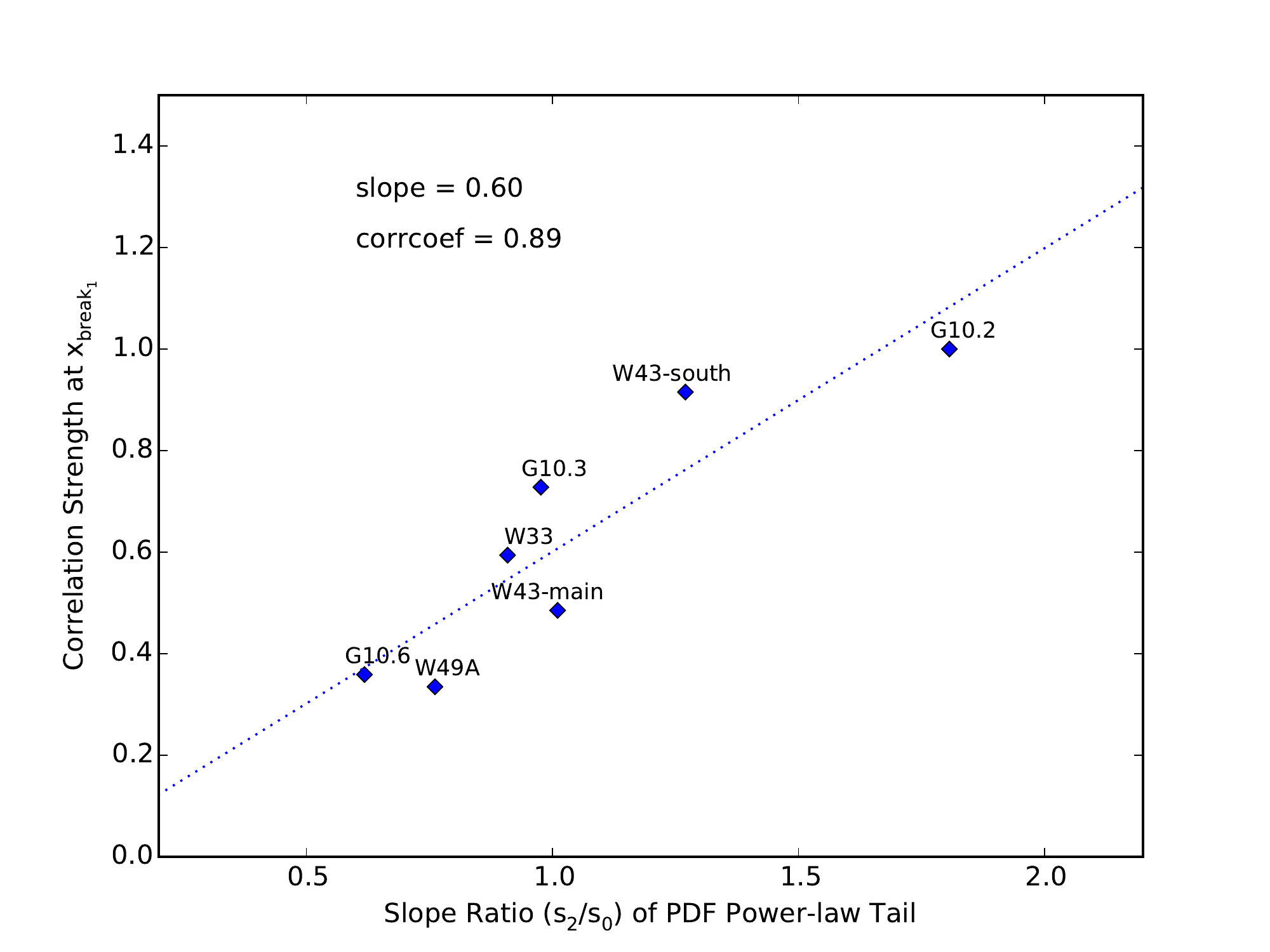} \\
\end{tabular}
\caption{Correlations of derived parameters of segmented linear fitted 2PT and N-PDFs for each source. {\it Left}: The slopes of first and second segmented components. Color contours show the levels of absolute difference between the two slopes with dashed line indicating equality. {\it Right}: The correlation between fitted N-PDF power-law slope ratios $s_{2}/s_{0}$ with correlation strength at x$_{break}$.
Fitted slope of the correlations and the Pearson correlation coefficient are indicated in each plot.}
\vspace{0.3cm}
\label{fig:2pt}
\end{figure*}

\section{Discussion}\label{section:discussion}

In this section, we provide tentative links of the derived properties and statistical quantities in Section \ref{section:results} to the physical environment of the OB cluster-forming regions, and to the evolutionary scenario. 
Our interpretation of the results of 2PT correlation functions, and the column density probability distribution functions, are given in Section \ref{discussion:2pt} and Section \ref{discussion:npdf}.
The comparison of the results of the 2PT correlation functions and the N-PDFs, are discussed in Section \ref{discussion:correlation}.
The properties and the spatial distribution of the dense molecular gas clumps/cores embedded in the observed OB cluster-forming regions, are presented in Section \ref{sub:dendro}.
We summarize the limitations of our study in Appendix C.
Finally, some brief hypotheses of physical conditions which lead to the observed structures are provided in Section \ref{discussion:physics}.

\subsection{Interpretation of 2PT correlation functions.}\label{discussion:2pt}
The power-law parametrization of the observed 2PT correlation function may be closely related to the fractal dimension (Mandelbrot 1983), which characterizes hierarchical or self-similar structures.
The hierarchical structure of a purely turbulence-dominated molecular cloud is expected to have a single power-law profile from its 2PT correlation function. 
On the other hand, gravity can also cause dense gas to collapse progressively into denser pockets with smaller spatial scales (e.g. Myers et al. 2013) and may introduce a self-similar pattern in the density distribution. 
The 2PT correlation function was also applied to studies of the clustering properties of star clusters (e.g., Elmegreen et al. (2006) on NGC 628, S{\'a}nchez et al. (2010) on M33, Gouliermis et al. (2014) on NGC 346). 
It has been proposed that the primordial spatial distribution of young stars or clusters may have an imprint of the structure of their natal molecular clouds (e.g. Hartman 2002; Kraus \& Hillenbrand 2008). 

For molecular clouds, the power-law index $\alpha$ of a parametrized 2PT correlation function (e.g. Table \ref{tab:2pt_fit_pl_seg3}), is related to the two-dimensional fractal dimension $D_{2}$ through $D_{2}$ = 2 + $\alpha$ (Mandelbrot 1983; Peebles 1980).
The relation between fractal dimension and 2PT correlation functions can be understood in how they are defined. In a fractal distribution of dimension $D_{2}$, the number of objects scales as $N \propto r^{D_{2}}$. Thus with a power-law scaling with correlation strength $\propto r^{\alpha}$, the numbers of objects within radius $r$ drops as $N\propto r^{2}\cdot r^{\alpha}$. 
The most naive de-projection of $D_{2}$ to the three-dimensional fractal dimension $D_{3}$, is simply adding one dimension (i.e. $D_{3} = D_{2} + 1$).
However, Gouliermis et al. (2014) argued that there is not a direct relation between $D_{2}$ and $D_{3}$, and have provided empirical conversion factors between $D_{2}$ and $D_{3}$ based on simulation results. 

The interstellar medium is observed to exhibit a three-dimensional fractal dimension of $D_{3} \sim$ 2.3 in various environments, which was proposed to be a consequence of it being dominated by scale-free turbulence (Elmegreen \& Falgarone 1996; Elmegreen et al. 2014). 
The fractal dimension values of the spatial distribution of young OB stars (of spectral type earlier than B4) in the Gould Belt is $D_{3} = 2.68 \pm 0.04$ (S{\'a}nchez et al. 2007a). 

The first power-law indices $\alpha_{1}$ of the 2PT correlation functions (see Table \ref{tab:2pt_fit_pl_seg3}) for the target sources W49A, G10.6-0.4, W43-Main, G10.3-0.1, and W33 correspond to $D_{2} \simeq 1.6 - 1.8$, which give $D_{3} \simeq1.6-2.3$ according to Gouliermis et al. (2014).

Break points of the 2PT correlation functions are seen from these sources, beyond which the correlation strengths remain roughly constant in separation scales for several parsecs (see Figure \ref{fig:2pt_all}, \ref{fig:2pt_G103}). This indicates that the density distribution is closer to being homogeneous on this spatial scale. 
The transition of the power-law indices of the 2PT correlation function may be related to different dominant  physical mechanism(s) at different spatial scales. 
We hypothesize that gravitational collapse dominates the smaller separations for the sources W49A, G10.6-0.4, W33 and G10.3-0.1. 
The flat transition with a power-law index shallower than $-0.1$ may indicate a characteristic scale beyond which turbulence begins to dominate, and induces a more dispersed density distribution that is characterized by an increase in the fractal dimension. 
It can also be because a more dispersed density distribution is not very evolved due to gravitational contraction.
The effect of gravitational collapse tends to convert gas structures into filaments or compact clumps/cores, which have $\lesssim$1 fractal dimensions in the most extreme cases.
This transition of physical mechanisms dominant at different spatial scales is consistent with numerical simulation results of column density power-spectra by Burkhart et al. (2015). They find that features of the power-spectra are closely related to the collapse stages of molecular clouds.
Our derived power-law indexes $\alpha_{2}$ for sources of W49A, G10.6-0.4, W43-Main, G10.3-0.1, and W33, correspond to $D_{3} \simeq 2.7-2.9$, which is close to the measurement found for the Gould Belt sources (S{\'a}nchez et al. 2007a).

\subsection{High column density tails of N-PDFs.}\label{discussion:npdf}
Our strategy of performing SED fittings to high angular resolution images that have little or no loss of extended structures is advantageous for measuring N-PDFs with high precision over a broad range of $N_{H_{2}}$ and $T_{d}$.
The resolved deviations of N-PDFs from power-laws at their high $N_{H_{2}}$ tails (Section \ref{subsection:pdf}) may provide key signatures of multiple physical mechanisms at work.
It has been suggested that when self-gravity becomes important, the resultant N-PDF shows a power-law tail at the high column density end (Klessen 2000; V{\'a}zquez-Semadeni et al. 2008; Kritsuk et al. 2011).
Tassis et al. (2010) has shown that for a singular isothermal-sphere, the N-PDF asymptotically approaches to a pure power-law form. 
They also showed that poor sampling can suppress the power-law tail and results in a distribution well-described by a log-normal distribution. 

Schneider et al. (2015a) report the detection of a second excess power-law tail for three high-mass star-forming regions based on dust emission observed by {\it Herschel}, and suggested that physical processes that inhibit the collapsing dense gas from flowing further inward could result in N-PDFs of this form.  
On the other hand, Girichidis et al. (2014) suggest, based on simulations that the power-law tail at high density extends to lower densities in the free-fall regime as time exceeds, and the steepening of slopes for power-law tails are due to physical processes that retard a free-fall collapse. The more shallow power-law tail slopes as clouds proceed to collapse is also found in many other studies (Kritsuk et al.2011; Federrath \& Klessen 2013; Burkhart et al.2015).

For W49A and G10.6-0.4, which show excesses at high column density end of their N-PDFs, they present highly centrally concentrated matter distributions.
Liu et al. (2011) has suggested that the radiative feedback and the pressure force of the ionized gas at the center of G10.6-0.4 are insufficient for halting the free-fall collapsing of this molecular cloud.
The central $\sim$1 pc scale massive molecular clump of G10.6-0.4, however, may be marginally rotationally supported (Liu et al. 2010a).
The likely marginally rotationally supported $\sim$1 pc scale massive molecular clump were also found at the center of G33.92+0.11 (Liu et al. 2012, 2015).

In contrast, the N-PDFs of W43-south and G10.2-0.3 have steeper high column density tails.
This may mean a relatively inefficient conversion of cloud material to high-density structures, or may be because of the dispersal of the dense structures from feedback.

\begin{figure*}
\hspace{-1cm}
\includegraphics[width=20cm]{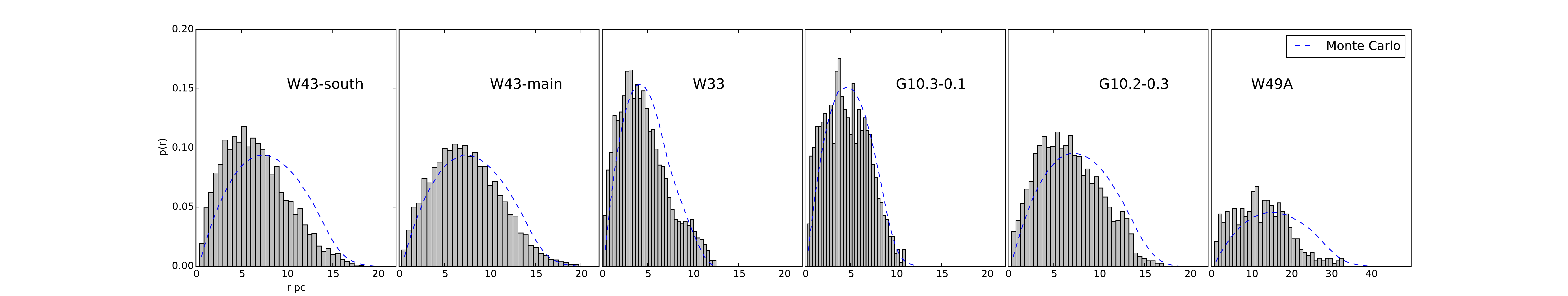}
\caption{Probability distribution functions for separations between clumps/cores in 5 regions.
Blue dashed lines indicate the Monte Carlo simulation results of random distribution of core/clumps within each source.}
\label{fig:sep_distri_func}
\end{figure*}

\begin{figure}
\hspace{-1cm}
\includegraphics[width=9.5cm]{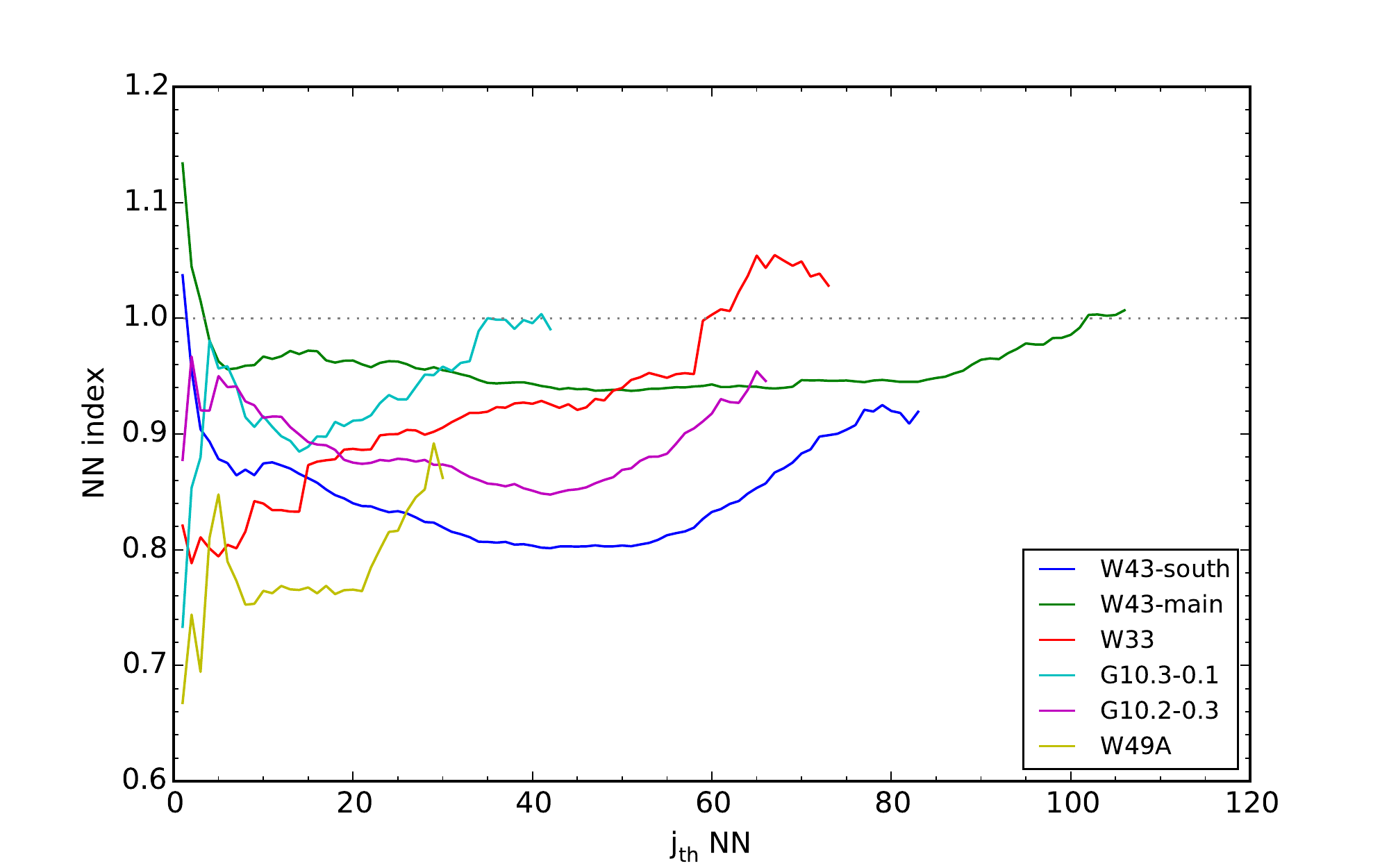}
\caption{$K_{th}$ nearest-neighbor index. Grey dashed line shows the index of value 1, which is complete randomness. The more identified clumps/cores residing in one target source, the larger order of NN index can be measured. }
\label{fig:knn_index_erosion_mc}
\end{figure}

\begin{figure}
\hspace{-1cm}
\includegraphics[width=9.5cm]{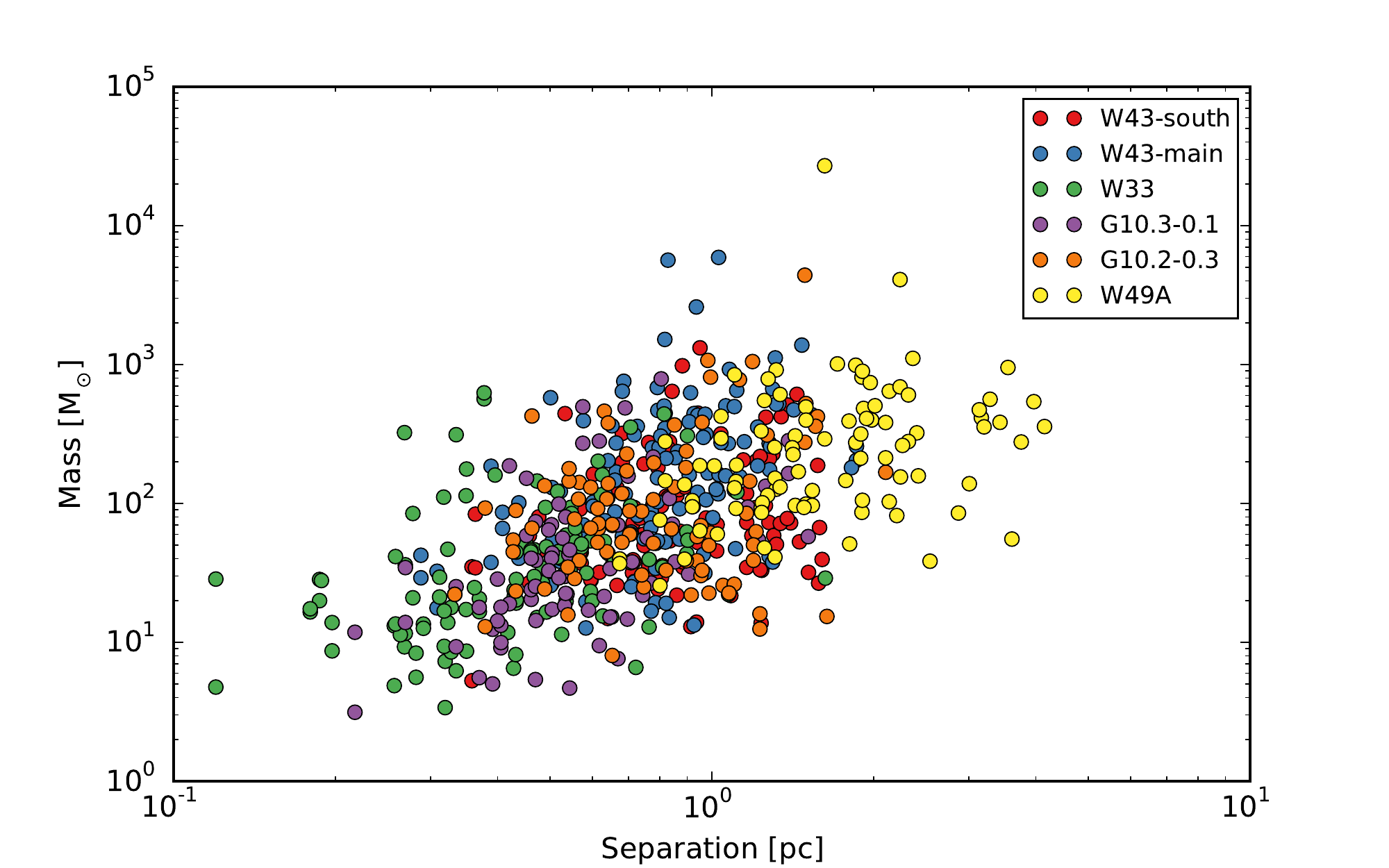}
\caption{Separation vs. clump/core mass. The separations are the nearest-neighbor distances for all the clumps/cores.}
\label{fig:mass_sep1}
\end{figure}

\subsection{Correlations of derived statistical quantities}\label{discussion:correlation}
Measurements of different statistical quantities may provide complementary information.
For example, the N-PDF characterizes the populations of low and high column density gas, while the 2PT functions and the enclosed mass profiles describe how they are spatially distributed.
The cross-comparison of these quantities may therefore help diagnose the physical mechanisms unambiguously.

The excess of high column density structures in W49A and G10.6-0.4 as seen from their N-PDFs (Figure \ref{fig:pdf}, appear to be dominantly from the centralized massive molecular clumps (Figures \ref{fig:TN_W49N}, \ref{fig:TN_G10p6}) in these clouds.
The characteristic, parsec scales of these massive molecular clumps, and the uniqueness of the massive molecular clump in these molecular clouds, result in the single rapidly decreasing component at short lags in each of their 2PT correlation functions.
Their relatively low correlation strengths at the first break points (i.e. $x_{break}$, see Figure \ref{fig:2pt_all}, \ref{fig:2pt} right, Table \ref{tab:2pt_fit_pl_seg3}, \ref{tab:2pt_fit_seg_2}) indicate that these centralized massive molecular clumps in fact contribute a significant fraction of the overall cloud mass. 
G10.6-0.4 additionally shows the largest difference between the fitted power-law/linear slopes before and after the first break point of its 2PT correlation function (Figure \ref{fig:2pt}, left), which indicates the sharpest transition of physical environments/mechanisms inside and exterior to its embedded centralized massive molecular clump, according to our discussion in Section \ref{discussion:2pt}. 

Molecular clouds W33, W43-Main and G10.3-0.1 also contain high $N_{H_{2}}$ molecular clumps (Figure \ref{fig:TN_W33}, \ref{fig:TN_W43M}, \ref{fig:TN_G10p3}): These clumps are not necessarily located close to the center of the parent molecular clouds, do not achieve exceptionally high column density (Figure \ref{fig:pdf}), and are less dominant as compared with the overall cloud masses.
The morphology of these molecular clouds looks relatively clumpy as compared to W49A and G10.6-0.4.

Finally, the most spatially diffused/fragmented clouds W43-South and G10.2-0.3 (Figure \ref{fig:TN_W43S}, \ref{fig:TN_G10p2}) show a deficit of high $N_{H_{2}}$ structures compared with the overall power-law fits to their N-PDFs.
Interestingly, the slopes of their 2PT correlation functions also do not vary significantly with spatial scale (Figure \ref{fig:2pt}, left).
This indicates that none of the localized dense structures contributes significantly to their overall cloud masses. 
In addition, there is no evidence of significant changes in the dominant physical mechanisms/environments over all resolved spatial scales. 

These observations are shown to have a strong correlation between the correlation strengths at the first turning points of their 2PT correlation functions, and the behavior of N-PDF at the high column density end, in the right panel of Figure \ref{fig:2pt}.

\subsection{Properties and spatial distribution of localized clumps/cores}\label{sub:dendro}
In this subsection, we investigate how the spatial distributions, or spatial separations of the overdensities identified by their dendrograms, are related to structures of their parent molecular clouds.
We exclude the source G10.6-0.4 from the quantitative analysis in this section, given the small number of identified dense clumps/cores, which leads to large statistical errors.

For the remaining six sources, we present the probability distribution functions of the clump/core separations in Figure \ref{fig:sep_distri_func}. We only select the cores/clumps that have a mean column density above 50-$\sigma$ for reliability.
The probability function, $p(r_{j})$ is calculated as the total number of clump/core-pair separations $N_{ij}$ that fall in the separation bin of $r_{j}-r_{j}+dr$, divided by the total number of pairs:

\begin{equation}\label{eq_dd}
 p(r_{j}) = \frac{2\sum_{i=1}^{N}N_{ij}}{N(N-1)dr}
\end{equation}

We ran Monte Carlo simulations by dropping the same number of clumps in each spatial random distribution within the perimeter for each source. This serves as a non-parametric way to compare between the true clump distribution and the simulated random distribution.
The simulated probability distribution functions were averaged from 1000 independent random realizations. 
The simulated results and the PDFs for the six sources are plotted in Figure \ref{fig:sep_distri_func}. 
In general, the distributions of identified core/clumps separations are skewed towards smaller separations, which points to an overall clustered situation.

While separation distribution functions provides a global view of clump/core separations, we relied on the nearest-neighbor (NN) method to estimate the local clump/core density by measuring the distance between each clump/core and its $j$th NN.  
For evaluating the clustering degree of the clumps, we compared the mean NN distance of the clump sample with a complete spatial randomness (CSR) pattern. 
The mean NN distance,

\begin{equation}
 \overline d =\frac{ \sum_{i=1}^{N}d_{i}}{N},
 \label{eq_knn}
\end{equation}
and the expected value of the NN distance in a random pattern,
\begin{equation}\label{eq_knn_r}
E(d_{i}) = 0.5\sqrt\frac{A_{f}}{N} + (0.0514 + \frac{0.041}{\sqrt N})\frac{P_{f}}{N}
 \end{equation}
where $A_{f}$ and $P_{f}$ are the area and perimeter of the field under consideration, respectively (Donnelly et al. 1978). The variance is,
  \begin{equation}\label{eq_knn_var}
Var(\overline d) = 0.070\frac{A_{f}}{N^{2}} + 0.037P_{f}\sqrt\frac{A_{f}}{N^{5}}
 \end{equation}
 and the z-score, which acts as a standardised value to enable the comparison between different sources, is
  \begin{equation}\label{eq_knn_z}
z_{score} = \frac{\overline d - E(d_{i})}{\sqrt{Var(\overline d)}}
 \end{equation}
We calculated z-score (standard normalised variate, Clark \& Evans 1954) for each region, which provides a measure of the degree of clustering.  
A positive z-score indicates dispersion or evenness while a negative z-score indicates clustering. The calculated z-scores are 0.26 for W43-main, -1.12 for W43-south, -2.71 for G10.2-0.3, -3.57 for W33, -3.67 for G10.3-0.1 and -4.08 for W49A, ordered from most dispersed to the most clustered.
In the calculation above, the area used is a rectangle with the correction factor of the boundary effect applied (Donnelly 1978). 

Nearest neighbor z-score estimated based on Monte-Carlo simulations (i.e. replace estimates of Equation \ref{eq_knn_r}, \ref{eq_knn_var} with measurements from Monte-Carlo simulation results) give results of 2.45 for W43-main, 0.64 for W43-south, -1.85 for G10.2-0.3, -2.77 for W33, -3.08 for G10.3-0.1, -5.75 for W49A. 

Since the NN (1st NN) is only an indicator of first-order spatial randomness, we also calculated the $K_{th}$ NN index, which is the ratio between actual mean $K_{th}$ NN distance and the simulated $K_{th}$ NN distance. 
This ratio serves as a measure of point distribution pattern, with values less than, equal to or greater than 1 indicating that the distribution pattern is more aggregated/clustering, standard as or of increasing dispersion (leading to a limiting case of regularity) as compared with a complete spatial randomness (CSR), respectively (Clark \& Evans 1954). 

For cases that show several clusters of clumps/cores distributed in the field, the $K_{th}$ NN index crosses 1.0, and together with the information of the mean clumps/cores separation, may provide a sense of the spatial scales where the dense gas structures lose coherence.
However, the interpretation of the high-order indexes needs to be taken with caution due to the dependence between different orders.
This requires more analytical studies, and comparison with numerical simulations.

We plotted these results in Figure \ref{fig:knn_index_erosion_mc}. 
The analysis of the 1st NN suggests that the distribution of the dense clumps/cores in G10.3-0.1, W33, and W49A are ordered or clustered to some extent.
The dense clumps/cores in W43-Main, W43-South and G10.2-0.3 are closer to being randomly distributed over the cloud areas.
These results are consistent with a visual impression of the column density maps (Figure \ref{fig:TN_W43M}, \ref{fig:TN_W43S}, \ref{fig:TN_W33}, \ref{fig:TN_G10p3}, \ref{fig:TN_G10p2}).
The analysis of Figure \ref{fig:knn_index_erosion_mc} may indicate that the distribution of dense clumps/cores in W43-Main may be spatially non-coherent on all scales. 
Physically, we hypothesize that the loss of spatial coherence may be due to the effect of stellar feedback, large-scale shocks due to the Galactic dynamics, or a combination of both effects, that dominate over the self-gravity.
We note that our present way of assessing spatial distribution functions and utilizing NN method takes no consideration of the area of clumps/cores, as we approximated them as point sources. 
Incorporating core properties will further improve our analyses and will be addressed in future work.

Finally, we plotted the separation versus clump/core mass in Figure \ref{fig:mass_sep1}, where the separation is defined as the distance between the centre of a clump/core to its nearest neighbour. 
Padoan \& Nordlund (2002) suggest that clump separation scales with clump mass as a consequence of turbulent fragmentation. 
However, gravitational fragmentation also induces such phenomenon, as pointed out by Bonnell et al. (2007). 
Our angular resolution is not ideal for investigating fragmentation models as it is also only reliably examined by high angular resolution molecular line analyses (e.g., Pillai et al. 2011). 
Despite this, we report the trend of clump/core separation scale with mass within our target sources, and this trend does not differ much between different sources.

\subsection{Physical hypothesis}\label{discussion:physics}
The sources W43-Main and W43-South are located around the intersection between the Galactic near 3 kpc arm, and the Galactic bar.
Previous works have suggested that these clouds show broader linewidths when compared with the rest of the observed samples.\footnote{This argument is based on linewidths measurements of W43 for the $\rm^{13}CO$ 1-0, $\rm ^{12}CO$ 2-1 molecular lines in Nguyen-Luong et
al. (2011), for G10.2-0.3 and G10.3-0.1 from $\rm^{13}CO$ 2-1 and $\rm C^{18}O$ 2-1 observations in Beuther et al. (2011), for W33 in multiple molecular lines including $\rm ^{12}CO$ 2-1, $\rm C^{18}O$ 2-1 in Immer et al. (2014), for W49A from $\rm ^{12}CO$ 1-0, $\rm ^{13}CO$ 1-0 in Galv{\'a}n-Madrid et al. (2013), and for G10.6-0.4 from $\rm ^{12}CO$ 1-0 in Liu et al. (2010a).} In fact, the FWHM velocity dispersion of $\rm \sim 20\ km^{-1}$ in W43 is among the largest dispersions determined for Galactic and extragalactic GMCs, as pointed out by Nguyen-Luong et
al. (2011).
We hypothesize that their large linewidths, and the randomized distribution of dense clumps/cores over the wide ares, are related to the continuous (in time) injection of the kinetic energy and supply of gas from diffuse molecular gas structures and some denser clumps and cores.
The evolution of these two objects can not be understood following the traditional picture of gravitationally bound and virialized molecular clouds with certain masses.
The continuous mass accumulation and the extended shock compression may aid the continuous formation of dense molecular clumps/cores and the formation of OB stars, despite the fact that some objects in the area may already be more evolved, and exerting extended radiative feedback.
Similar physical conditions may also occur in the Galactic central molecular zone (CMZ), where complicated gas dynamics can lead to interactions of molecular gas structures (i.e., Longmore et al. 2012; Liu et al. 2013b; Kauffmann et al. 2013; Rathborne et al. 2015).

Starbursts in external galaxies may also provide feedback to the adjacent giant molecular clouds and lead to similar physical conditions, although presently we lack sufficient spatial resolution in these extragalactic cases. 
It is possible that some dense molecular gas structures forming in situ or accumulated into these environment, are further coagulated due to self-gravity or shock compression into more condensed massive star-forming molecular gas complexes that permit the formation of gravitationally bound stellar clusters.
The `Z' shaped dense gas structures in W43-Main, or the Sgr B2 cloud in the Galactic CMZ, may be examples of this. 

From the low spatial resolution observations, the geometry and the gas kinematics of the W49A and G10.6-0.4 clouds, may be closest to the virialized molecular clouds.
However, in higher angular resolution observations, the dense molecular gas in these clouds present a hub-filament geometry, where several approximately radially orientated molecular gas filaments are connected to the centralized massive molecular clumps (Liu et al. 2012a; Galv\'an-Madrid et al. 2013).
Such a well organized cloud geometry is not easily explained with a model that is dominated by 
supersonic turbulence.
On the other hand, some of the simulations in Dale et al. (2012), (2013) that are turbulent clouds with initial virial ratios close to unity (but without artificially maintained the turbulence), the clouds can eventually form roughly centralized clusters with several radial accretion flows feeding them. 
We think this geometric configuration comes from (i) the clouds being approximately spherical, (ii) the clouds being approximately virial, so gravity is important and causes them to collapse as the turbulent field decays, and (iii) the energy of the turbulent field drains more quickly in the centre of the cloud because this is where most collisions between turbulent flows occur.

Observationally, on $\gtrsim$1 pc scales, the gas kinematics in these two molecular clouds appear to be dominated by the global collapse towards the centralized massive molecular clumps (Galv\'an-Madrid et al. 2009; Schneider et al. 2010; Liu et al. 2010a; Liu et al. 2013a; Galv\'an-Madrid et al. 2013; Kirk et al. 2013; Peretto et
al. 2013).
Their excess at the high column density end of the N-PDF (Figure \ref{fig:pdf}, see Section \ref{subsection:pdf}), and their significant rapidly decreasing components at the short-lags of their 2PT functions (Figure \ref{fig:2pt_all}, see Section \ref{subsection:2pt}, \ref{discussion:2pt}) are likely direct consequences of the global gravitational collapse.
The high-luminosity, high-mass concentration at the centers, and the ongoing global gravitational collapse make these two sources the most promising candidates to form gravitationally bound OB stellar clusters, which may represent the lowest mass ends of young massive clusters observed in external galaxies. 

It is clear that stellar feedback has had little effect on the large-scale structures of these clouds, since they are centrally-condensed and show no evidence that feedback is able to clear the gas from their potential gravitational wells or terminate star-formation. This is consistent with results form recent numerical simulations by Dale et al. (2012), (2013), (2014), that find that high-density clouds whose escape velocities are comparable to the sound speed of photo-ionized gas are largely immune to photo-ionization and/or wind feedback. It is likely therefore that these clouds will achieve high star formation efficiencies and produce dense and strongly-bound clusters (Ginsburg et al. 2012; Longmore 2014).

Both G10.2-0.3 and G10.3-0.1 are resolved with cool and dense massive molecular clumps/cores embedded with internal heating sources, indicating ongoing massive star-formation. 
However, stellar feedback, including the radiation and the expansion of the ionized gas, may already be sufficient to significantly change the morphology of the extended molecular gas structures. 
Part of the diffuse molecular gas may be dispersed/ionized, or is being dispersed/ionized by the stellar feedback.
The streaking dense molecular clumps/cores however can be better self-shielded, and so can survive the feedback for a longer time period.
Some of the detected cores in G10.2-0.3 may look like pillars if they were resolved in high angular resolution optical observations. Pillar structures are also observed in the simulations of Dale et al. (2012), (2013), (2014), where they result from the partial photo-evaporation of accretion flows carrying gas towards clusters.

The residual extended molecular gas structures in G10.2-0.3 are marginally resolved with curvatures closely following the shell of the H\textsc{ii} regions, although this remains uncertain due to the projection effects (Figure \ref{fig:TN_G10p2}).
The spatially well-ordered distribution of the molecular cores in G10.2-0.3 as compared with those in the W43-South (Figure \ref{fig:mass_sep1}; Section \ref{sub:dendro}) may be explained as dense cores formed in ordered parent molecular gas structures, which are later stripped away due to the stellar feedback.

\section{Conclusion}
\label{section:conclusion}
We acquired the (sub)millimeter continuum images of the {\it Herschel} and {\it Planck} space telescopes, and those taken by the ground based CSO, JCMT, and IRAM-30m telescopes bolometric observations, for seven OB cluster-forming regions in the Milky Way: W49A, W43-Main, W43-South, W33, G10.6-0.4, G10.2-0.3 and G10.3-0.1.
These target sources were analyzed because they are very luminous ($L$$\sim$10$^{6-7}$ $L_{\odot}$), their distance are relatively well determined, and we have access to the CSO SHARC2 350 \micron\ maps for most of them.
We have successfully linearly combined the space telescope images with those taken by the ground-based telescopes, which yielded images that have comparable angular resolutions with the ground-based observations and show little to no loss of extended structures. 
In addition, we have developed the procedure to iteratively fit a single-component modified black-body spectrum to the combined (sub)millimeter (350, 450, 850, 1200 \micron) images and the {\it Herschel} mid- and far- (70, 160, 250 \micron) infrared images, which yield precise dust temperature and column density maps with $\sim$10$''$ angular resolution.

In spite of the comparable bolometric luminosity of these sources, their derived column density maps on the $\gtrsim$10 pc scales show dramatically different morphologies that can be visually separated into classes of (1) amorphous cloud structures with widely scattered distribution of dense clumps and cores (W43-Main, W43-South, G10.2-0.3), (2) hub-filament systems with significant matter concentration at the centralized massive molecular clumps (W49A, G10.6-0.4), and (3) other morphological (e.g. the pearl ring-like structures of G10.3-0.1), or a morphology in between the first two classes.
They may represent different initial and boundary physical conditions, or the different evolutionary stages of the cluster-forming molecular clouds.
The dust temperature maps reveal different distributions of the heated dust and gas in these sources as well, although the temperature of the extended and diffused heated dust is not yet well measured due to the confusion with foreground/background emission, and because the assumption of a single temperature component no longer applies.

We have performed statistical analyses to quantify the observed morphology of the dense and massive components, of which the column density and temperature were well-determined by the SED fits.
The results of the statistical analyses are summarized as follows:

\begin{enumerate}
\item The enclosed mass profiles of the observed sources show a higher concentration than virialized molecular clouds. On 1-10 pc scales, the sources W49A, G10.6-0.4, and W33 present more concentrated distributions of mass than a gravitationally-bound gas sphere, with a 10 km\,s$^{-1}$ thermal sound speed. This may be a consequence of turbulence in the clouds decaying more rapidly in their centers, leading to the centers collapsing before the cloud envelopes.
\item The column density probability distribution functions (N-PDFs) of the observed sources can be approximated by power-laws, rather than log-normal distribution functions. However, the high column density ends of the derived N-PDFs deviate from the overall power laws. The centrally concentrated sources W49A and G10.6-0.4 show significant excesses at high column density. The extremely high column density clumps are likely not randomly located in these molecular clouds. However, this may be related to the global collapse of molecular clouds towards their centers which may not occur efficiently in all molecular clouds. On the other hand, the sources W43-South and G10.2-0.3 have close to randomized distributions of dense molecular gas, and show a deficit of high column density structures. The N-PDFs of these star-forming regions appear to be linked with the projected morphology of the molecular clouds. A power-law form of the N-PDF may also be evidence that the clouds are globally collapsing.
\item It is possible to quantify the visual impression of two dimensional column density distribution of molecular clouds using two-point correlation function. The two-point correlation functions (i.e., auto-correlation) of the column density distribution of the observed sources show a common feature: a rapidly decreasing correlation strength at short lags ($\sim$1 pc) connected with a shallower decreasing component at larger lags. The rapidly decreasing components are the most significant from the sources with high central concentrations, such as W49A and G10.6-0.4. The rapidly decreasing component cannot be clearly identified from W43-South and G10.2-0.3.
\item We identified a large number of dense molecular gas clumps/cores from all observed sources. The dense clumps/cores appear in clusters in W43-South, W33, and G10.3-0.1. Dense clumps/cores are only weakly clustered, or are nearly randomly distributed over the cloud area of W43-Main, W43-South, and G10.2-0.3. 
\end{enumerate}

This work demonstrates that a high spatial resolution ($\lesssim$0.3 pc) is the key to discriminate the morphology classes of the OB cluster-forming molecular clouds, and the spatial distribution of the embedded dense molecular gas clumps/cores using systematic statistical approaches.
The difference in the derived morphological classes and clump/core distributions, may be linked with the formation mechanism of the molecular clouds, or the imprint of stellar feedback.
We argue that detailed studies of these derived quantities, and the comparisons with the numerical hydrodynamics simulations, are crucial in advancing our understanding of the star-formation process,  
and the physics of high-mass star-formation in general.
Efforts have been made to archive our quantitative measurements of statistical properties, which can serve as templates for OB cluster-forming molecular clouds of different types or evolutionary statuses.

Finally, we remark that combining the {\it Herschel} and ground-based single-dish bolometric imaging at 350 and 450 \micron\, is presently the best way of providing short-spacing data for further combining with ALMA continuum observations at band 8 and 9.

\begin{acknowledgements}
We acknowledge Karl Menten, Friedrich Wyrowski, Jens Kauffmann, Thushara Pillai, Ke Wang, Attila Kov{\'a}cs, Bo Zhang (SHAO), and Bo Zhang (NAOC) for useful suggestions and discussion. Y.X.L. acknowledges Zhichen Pan for his remote observations on sources G10.2-0.3, G10.3-0.1 and G10.6-0.4 with SHARC2 instrument.
This research was supported by Technology under State Key Development Program for Basic Research (973 program) No. 2012CB821802 and the Guizhou Scientific Collaboration Program (No. 20130421). 
This research was also supported by the DFG cluster of excellence `Origin and Structure of the Universe' (JED). This research made use of astrodendro (http://www.dendrograms.org), Astropy, and APLpy (http://aplpy.github.com).
The ATLASGAL project is a collaboration between the Max-Planck-Gesellschaft, the European Southern Observatory (ESO) and the Universidad de Chile. It includes projects E-181.C-0885, E-078.F-9040(A), M-079.C-9501(A), M-081.C-9501(A) plus Chilean data. 
This material is based upon work at the Caltech Submillimeter Observatory, which is operated by the East Asian Observatory on behalf of The National Astronomical Observatory of Japan, Academia Sinica Institute of Astronomy and Astrophysics, the Korea Astronomy and Space Science Institute, the National Astronomical Observatories of China and the Chinese Academy of Sciences, and by the California Institute of Technology.

\end{acknowledgements}


\clearpage

\clearpage
\footnotesize  

\LongTables 

\clearpage
\begin{deluxetable*}{ l l l l l l l l lllll} 
\tablecaption{W43-main dendrogram leaves parameters.}
  \startdata
  \hline
  \hline
Leaf & RA & DEC & $r_{eff}$ & mass&corrected mass & luminosity & $\langle N_{\rm H_2} \rangle$ & merge level &  $\langle T_{d} \rangle$ & ${\beta}$ & $T_{bol}$&$ \langle N_{\rm H_2} \rangle \geq 50\ \sigma$$^{a}$ \\
 &(J2000)  &(J2000)   & $(\mathrm{pc})$ & $(\mathrm{M_{\odot}})$& $(\mathrm{M_{\odot}})$ & $(\mathrm{L_{\odot}})$ & $(\mathrm{cm^{-2}})$ &$(\mathrm{cm^{-2}})$ & $(\mathrm{K})$ &  & $(\mathrm{K})$& \\
 \hline
1 & 18:47:52.9 & -2:03:13 & 0.11 & 12.7 & 1.6 & $  4.5 \times 10^{ 2 }$ & $  1.6 \times 10^{ 22 }$ & $  1.4 \times 10^{ 22 }$ & 24.3 & 1.6 & 15.3 & -- \\
2 & 18:47:46.6 & -2:02:53 & 0.48 & 270.0 & 15.7 & $  1.3 \times 10^{ 4 }$ & $  1.6 \times 10^{ 22 }$ & $  1.5 \times 10^{ 22 }$ & 25.0 & 1.7 & 15.3 & -- \\
3 & 18:47:54.3 & -2:03:06 & 0.13 & 19.6 & 4.0 & $    6 \times 10^{ 2 }$ & $  1.8 \times 10^{ 22 }$ & $  1.4 \times 10^{ 22 }$ & 23.9 & 1.6 & 15.3 &  \\
4 & 18:47:52.4 & -2:02:42 & 0.49 & 236.0 & 11.7 & $  9.7 \times 10^{ 3 }$ & $  1.4 \times 10^{ 22 }$ & $  1.3 \times 10^{ 22 }$ & 25.4 & 1.6 & 15.5 & -- \\
5 & 18:47:39.6 & -2:02:29 & 0.46 & 270.1 & 28.5 & $  1.8 \times 10^{ 4 }$ & $  1.8 \times 10^{ 22 }$ & $  1.6 \times 10^{ 22 }$ & 27.2 & 1.6 & 16.1 &  \\
6 & 18:47:19.5 & -2:02:42 & 0.33 & 165.6 & 10.3 & $  4.3 \times 10^{ 3 }$ & $  2.1 \times 10^{ 22 }$ & $    2 \times 10^{ 22 }$ & 23.6 & 1.6 & 15.4 &  \\
\enddata
\tablenotetext{a}{Note: A blank means the mean column density of a leaf meets the $50 \sigma$ threshold while a `--' means not.}
\label{tab:tab_W43_main_leaves}

\end{deluxetable*}




\footnotesize  
\LongTables 

 \begin{deluxetable*}{ l l l l l l l l lllll} 
\tablecaption{W43-south dendrogram leaves parameters.}
  \startdata
  \hline
  \hline
Leaf & RA & DEC & $r_{eff}$ & mass&corrected mass & luminosity & $\langle N_{\rm H_2} \rangle$ & merge level &  $\langle T_{d} \rangle$ & ${\beta}$ & $T_{bol}$&$ \langle N_{\rm H_2} \rangle \geq 50\ \sigma$ \\
 &(J2000)  &(J2000)   & $(\mathrm{pc})$ & $(\mathrm{M_{\odot}})$& $(\mathrm{M_{\odot}})$ & $(\mathrm{L_{\odot}})$ & $(\mathrm{cm^{-2}})$ &$(\mathrm{cm^{-2}})$ & $(\mathrm{K})$ &  & $(\mathrm{K})$& \\
 \\
 \hline
1 & 18:45:52.2 & -2:47:12 & 0.21 & 26.7 & 1.5 & $    1 \times 10^{ 3 }$ & $  8.8 \times 10^{ 21 }$ & $  8.3 \times 10^{ 21 }$ & 21.8 & 2.0 & 15.0 & -- \\
2 & 18:45:56.1 & -2:47:08 & 0.22 & 44.9 & 8.3 & $  1.7 \times 10^{ 3 }$ & $  1.3 \times 10^{ 22 }$ & $  1.1 \times 10^{ 22 }$ & 20.9 & 2.1 & 15.0 &  \\
3 & 18:45:57.4 & -2:47:04 & 0.22 & 46.2 & 8.9 & $  1.8 \times 10^{ 3 }$ & $  1.3 \times 10^{ 22 }$ & $  1.1 \times 10^{ 22 }$ & 21.1 & 2.1 & 15.0 &  \\
4 & 18:46:02.1 & -2:46:58 & 0.36 & 95.2 & 11.0 & $  6.5 \times 10^{ 3 }$ & $  1.1 \times 10^{ 22 }$ & $  9.3 \times 10^{ 21 }$ & 23.2 & 2.0 & 14.8 &  \\
5 & 18:46:10.7 & -2:46:24 & 0.57 & 193.9 & 15.2 & $  1.3 \times 10^{ 4 }$ & $  8.4 \times 10^{ 21 }$ & $  7.8 \times 10^{ 21 }$ & 22.9 & 2.0 & 14.8 & -- \\
\enddata
\label{tab:tab_W43_south_leaves}
\end{deluxetable*}
 
\footnotesize  
\LongTables 

 \begin{deluxetable*}{ l l l l l l l l lllll} 
\tablecaption{W33 dendrogram leaves parameters.}
  \startdata
  \hline
  \hline

Leaf & RA & DEC & $r_{eff}$ & mass&corrected mass & luminosity & $\langle N_{\rm H_2} \rangle$ & merge level &  $\langle T_{d} \rangle$ & ${\beta}$ & $T_{bol}$&$ \langle N_{\rm H_2} \rangle \geq 50\ \sigma$ \\
 &(J2000)  &(J2000)   & $(\mathrm{pc})$ & $(\mathrm{M_{\odot}})$& $(\mathrm{M_{\odot}})$ & $(\mathrm{L_{\odot}})$ & $(\mathrm{cm^{-2}})$ &$(\mathrm{cm^{-2}})$ & $(\mathrm{K})$ &  & $(\mathrm{K})$& \\
\\
 \hline
1 & 18:14:22.7 & -18:04:34 & 0.13 & 13.0 & 0.9 & $  3.7 \times 10^{ 2 }$ & $  1.1 \times 10^{ 22 }$ & $    1 \times 10^{ 22 }$ & 21.2 & 1.9 & 15.3 & -- \\
2 & 18:14:17.4 & -18:04:08 & 0.30 & 93.7 & 16.3 & $  2.9 \times 10^{ 3 }$ & $  1.5 \times 10^{ 22 }$ & $  1.2 \times 10^{ 22 }$ & 22.5 & 1.8 & 15.2 & -- \\
3 & 18:14:20.5 & -18:04:02 & 0.10 & 7.3 & 0.7 & $  2.5 \times 10^{ 2 }$ & $  1.1 \times 10^{ 22 }$ & $  9.6 \times 10^{ 21 }$ & 22.7 & 1.8 & 15.1 & -- \\
4 & 18:14:23.0 & -18:03:14 & 0.26 & 59.5 & 6.5 & $  1.9 \times 10^{ 3 }$ & $  1.2 \times 10^{ 22 }$ & $  1.1 \times 10^{ 22 }$ & 22.0 & 1.9 & 15.1 & -- \\
5 & 18:14:18.3 & -18:03:08 & 0.16 & 23.3 & 1.7 & $  9.9 \times 10^{ 2 }$ & $  1.3 \times 10^{ 22 }$ & $  1.2 \times 10^{ 22 }$ & 24.1 & 1.7 & 15.2 & -- \\
\enddata
\label{tab:tab_W33_leaves}
\end{deluxetable*}
\hspace{-1 cm}
\footnotesize  
 \clearpage
\LongTables 

 \begin{deluxetable*}{ l l l l l l l l lllll} 
\tablecaption{G10.2-0.3 dendrogram leaves parameters.}
  \startdata
  \hline
  \hline
Leaf & RA & DEC & $r_{eff}$ & mass&corrected mass & luminosity & $\langle N_{\rm H_2} \rangle$ & merge level &  $\langle T_{d} \rangle$ & ${\beta}$ & $T_{bol}$&$ \langle N_{\rm H_2} \rangle \geq 50\ \sigma$ \\
 &(J2000)  &(J2000)   & $(\mathrm{pc})$ & $(\mathrm{M_{\odot}})$& $(\mathrm{M_{\odot}})$ & $(\mathrm{L_{\odot}})$ & $(\mathrm{cm^{-2}})$ &$(\mathrm{cm^{-2}})$ & $(\mathrm{K})$ &  & $(\mathrm{K})$& \\
\\
 \hline
1 & 18:09:22.3 & -20:21:43 & 0.18 & 38.7 & 4.5 & $  1.6 \times 10^{ 3 }$ & $  1.8 \times 10^{ 22 }$ & $  1.6 \times 10^{ 22 }$ & 27.3 & 1.4 & 16.1 &  \\
2 & 18:09:26.6 & -20:21:29 & 0.42 & 461.6 & 145.0 & $  2.9 \times 10^{ 4 }$ & $  3.7 \times 10^{ 22 }$ & $  2.5 \times 10^{ 22 }$ & 27.6 & 1.6 & 16.2 &  \\
3 & 18:09:25.4 & -20:21:10 & 0.12 & 28.8 & 2.2 & $    3 \times 10^{ 3 }$ & $  2.8 \times 10^{ 22 }$ & $  2.5 \times 10^{ 22 }$ & 30.4 & 1.6 & 17.8 &  \\
4 & 18:09:21.7 & -20:21:05 & 0.19 & 57.7 & 11.3 & $  3.8 \times 10^{ 3 }$ & $  2.2 \times 10^{ 22 }$ & $  1.8 \times 10^{ 22 }$ & 28.2 & 1.5 & 16.5 &  \\
5 & 18:09:31.4 & -20:21:03 & 0.09 & 8.0 & 0.3 & $  6.7 \times 10^{ 2 }$ & $  1.4 \times 10^{ 22 }$ & $  1.3 \times 10^{ 22 }$ & 34.2 & 1.2 & 20.3 & -- \\
\enddata
\label{tab:tab_G102_leaves}
\end{deluxetable*}
 
 \footnotesize  
\LongTables 

 \begin{deluxetable*}{ l l l l l l l l lllll} 
\tablecaption{G10.3-0.1 dendrogram leaves parameters.}
  \startdata
  \hline
  \hline

Leaf & RA & DEC & $r_{eff}$ & mass&corrected mass & luminosity & $\langle N_{\rm H_2} \rangle$ & merge level &  $\langle T_{d} \rangle$ & ${\beta}$ & $T_{bol}$&$ \langle N_{\rm H_2} \rangle \geq 50\ \sigma$ \\
 &(J2000)  &(J2000)   & $(\mathrm{pc})$ & $(\mathrm{M_{\odot}})$& $(\mathrm{M_{\odot}})$ & $(\mathrm{L_{\odot}})$ & $(\mathrm{cm^{-2}})$ &$(\mathrm{cm^{-2}})$ & $(\mathrm{K})$ &  & $(\mathrm{K})$& \\ \hline
1 & 18:08:59.3 & -20:11:41 & 0.07 & 19.0 & 9.0 & $  5.8 \times 10^{ 1 }$ & $  4.8 \times 10^{ 22 }$ & $  2.5 \times 10^{ 22 }$ & 16.2 & 2.0 & 24.9 &  \\
2 & 18:09:01.0 & -20:11:34 & 0.21 & 186.6 & 109.5 & $  2.1 \times 10^{ 3 }$ & $  6.1 \times 10^{ 22 }$ & $  2.5 \times 10^{ 22 }$ & 19.0 & 1.9 & 20.0 &  \\
3 & 18:08:55.7 & -20:11:02 & 0.17 & 43.9 & 8.7 & $  9.1 \times 10^{ 2 }$ & $  2.1 \times 10^{ 22 }$ & $  1.7 \times 10^{ 22 }$ & 20.6 & 1.9 & 15.8 &  \\
4 & 18:08:54.2 & -20:10:38 & 0.11 & 17.3 & 1.8 & $  4.7 \times 10^{ 2 }$ & $  1.9 \times 10^{ 22 }$ & $  1.7 \times 10^{ 22 }$ & 21.8 & 1.8 & 15.3 &  \\
5 & 18:09:04.2 & -20:09:57 & 0.32 & 70.1 & 9.3 & $    3 \times 10^{ 3 }$ & $  9.5 \times 10^{ 21 }$ & $  8.2 \times 10^{ 21 }$ & 25.0 & 1.6 & 15.4 & -- \\
6 & 18:09:06.5 & -20:10:23 & 0.12 & 7.6 & 0.7 & $  3.3 \times 10^{ 2 }$ & $  7.9 \times 10^{ 21 }$ & $  7.1 \times 10^{ 21 }$ & 25.3 & 1.6 & 15.5 & -- \\
\enddata
\label{tab:tab_G103_leaves}
\end{deluxetable*}
\footnotesize  
\LongTables 

\begin{deluxetable*}{ l l l l l l l l lllll} 
\tablecaption{G10.6-0.4 dendrogram leaves parameters.}
  \startdata
  \hline
  \hline
Leaf & RA & DEC & $r_{eff}$ & mass&corrected mass & luminosity & $\langle N_{\rm H_2} \rangle$ & merge level &  $\langle T_{d} \rangle$ & ${\beta}$ & $T_{bol}$&$ \langle N_{\rm H_2} \rangle \geq 50\ \sigma$ \\
 &(J2000)  &(J2000)   & $(\mathrm{pc})$ & $(\mathrm{M_{\odot}})$& $(\mathrm{M_{\odot}})$ & $(\mathrm{L_{\odot}})$ & $(\mathrm{cm^{-2}})$ &$(\mathrm{cm^{-2}})$ & $(\mathrm{K})$ &  & $(\mathrm{K})$& \\ \hline
1 & 18:10:20.0 & -19:59:51 & 0.64 & 189.4 & 31.5 & $  7.8 \times 10^{ 3 }$ & $  6.6 \times 10^{ 21 }$ & $  5.5 \times 10^{ 21 }$ & 20.9 & 2.1 & 15.0 & -- \\
2 & 18:10:34.7 & -19:58:19 & 0.73 & 370.8 & 53.5 & $    4 \times 10^{ 4 }$ & $  9.8 \times 10^{ 21 }$ & $  8.3 \times 10^{ 21 }$ & 25.3 & 2.0 & 15.4 & -- \\
3 & 18:10:46.2 & -19:58:31 & 0.46 & 123.7 & 28.6 & $  7.2 \times 10^{ 3 }$ & $  8.2 \times 10^{ 21 }$ & $  6.3 \times 10^{ 21 }$ & 21.7 & 2.1 & 14.7 & -- \\
4 & 18:10:18.2 & -19:58:03 & 0.18 & 19.3 & 1.3 & $  9.1 \times 10^{ 2 }$ & $  8.8 \times 10^{ 21 }$ & $  8.2 \times 10^{ 21 }$ & 21.4 & 2.1 & 14.8 & -- \\
5 & 18:10:41.6 & -19:57:50 & 0.41 & 238.7 & 38.7 & $  9.5 \times 10^{ 3 }$ & $    2 \times 10^{ 22 }$ & $  1.7 \times 10^{ 22 }$ & 22.3 & 1.9 & 15.0 &  \\
6 & 18:10:15.9 & -19:57:25 & 0.54 & 301.9 & 69.4 & $  9.9 \times 10^{ 3 }$ & $  1.5 \times 10^{ 22 }$ & $  1.1 \times 10^{ 22 }$ & 19.9 & 2.2 & 15.4 &  \\
\enddata
\label{tab:tab_G106_leaves}
\end{deluxetable*}
\footnotesize  
\LongTables 

\begin{deluxetable*}{ l l l l l l l l lllll} 
\tablecaption{W49A dendrogram leaves parameters.}
  \startdata
  \hline
  \hline

Leaf & RA & DEC & $r_{eff}$ & mass&corrected mass & luminosity & $\langle N_{\rm H_2} \rangle$ & merge level &  $\langle T_{d} \rangle$ & ${\beta}$ & $T_{bol}$&$ \langle N_{\rm H_2} \rangle \geq 50\ \sigma$ \\
 &(J2000)  &(J2000)   & $(\mathrm{pc})$ & $(\mathrm{M_{\odot}})$& $(\mathrm{M_{\odot}})$ & $(\mathrm{L_{\odot}})$ & $(\mathrm{cm^{-2}})$ &$(\mathrm{cm^{-2}})$ & $(\mathrm{K})$ &  & $(\mathrm{K})$& \\\hline
1 & 19:10:10.8 & 9:00:04 & 0.59 & 182.5 & 182.5 & $  6.4 \times 10^{ 3 }$ & $  7.5 \times 10^{ 21 }$ & $    0.0$ & 27.1 & 1.4 & 16.0 & -- \\
2 & 19:10:12.4 & 9:00:24 & 0.84 & 398.8 & 29.7 & $  1.4 \times 10^{ 4 }$ & $    8 \times 10^{ 21 }$ & $  7.4 \times 10^{ 21 }$ & 27.2 & 1.4 & 16.1 & -- \\
3 & 19:10:04.7 & 9:00:18 & 0.37 & 145.6 & 27.5 & $  3.7 \times 10^{ 3 }$ & $  1.5 \times 10^{ 22 }$ & $  1.3 \times 10^{ 22 }$ & 24.5 & 1.5 & 15.6 &  \\
4 & 19:10:05.1 & 9:00:31 & 0.54 & 279.1 & 21.6 & $  5.1 \times 10^{ 3 }$ & $  1.4 \times 10^{ 22 }$ & $  1.3 \times 10^{ 22 }$ & 23.3 & 1.5 & 15.6 &  \\
5 & 19:10:14.7 & 9:00:35 & 0.39 & 96.8 & 11.8 & $  2.8 \times 10^{ 3 }$ & $  9.2 \times 10^{ 21 }$ & $  8.1 \times 10^{ 21 }$ & 26.7 & 1.3 & 16.0 & -- \\
6 & 19:10:16.1 & 9:00:50 & 0.65 & 306.6 & 50.3 & $  2.2 \times 10^{ 4 }$ & $    1 \times 10^{ 22 }$ & $  8.5 \times 10^{ 21 }$ & 32.2 & 1.3 & 18.9 & -- \\
\enddata
\label{tab:tab_W49A_leaves}
\end{deluxetable*}
\vspace{3 cm}



\appendix
\color{black}
We summarize the data to help assess the quality of our images, and the errors in our SED analysis.
In addition, the present manuscript is intended to also serve as a high angular resolution survey of dense molecular gas clumps and cores.
While the dense molecular gas clumps and cores we identified are already summarized in tables in the main text, we additionally provide information to assess the ongoing or future star-formation of the observed sources.
We note that the dense gas clumps and cores will likely evolve in a much shorter ($\sim$$t_{\mbox{\scriptsize{free-fall}}}$) timescales, than the time scales for the evolution of the global cloud structures.
Observations of molecular clouds only provide snapshots of this in the time domain.
It may still be fair to compare the present properties of the cores identified in the observed molecular clouds. 
However, we argue that the links to the physics need to be understood in the context of the overall cloud evolution, including the stellar feedback, which are beyond the scope of our present works. 
For example, a molecular cloud which is presently not showing an abundance of dense cores/clumps over the cloud area, and is not showing active star-formation, may still quickly form many dense cores and stars in the near future, but before the overall cloud morphology has significantly evolved.
Nevertheless, the information we summarize here will hopefully help design future experiments that may or may not be directly relevant to our present study. 

The images we used for the SED analysis are provided in Section \ref{appendix:images}.
The correlations between the fitted dust opacity index, column density, and dust temperature, are summarized in Section \ref{appendix:fittings}.
The derived dendrograms of all observed sources, and the mass versus radius relations of all dense clumps/cores identified by dendrogram, are presented in Section \ref{appendix:dendrogram}.

\vspace{0.5cm}
\section{A. Images used for SED analysis}\label{appendix:images}
In this section, we present the {\it Herschel} PACS 70 and 160 \micron, the {\it Planck} 850 \micron, the CSO-SHARC2 350 \micron, the JCMT-SCUBA2 450 and 850 \micron, and the IRAM-30m-MAMBO2 1200 \micron\ images used for our analysis.
In addition, we present the combined {\it Herschel}+CSO 350 \micron, {\it Herschel}+JCMT 450 \micron, the {\it Planck}+JCMT 850 \micron, and the {\it Planck}+IRAM-30m 1200 \micron\ images, which were used together with the shorter wavelength data in our SED fittings.

\section{Far infrared and submillimeter images}

\clearpage
\vspace{0.5cm}
\section{B. The distribution of dust opacity index, column density and dust temperature}\label{appendix:fittings}
We summarize the pixel values of fitted dust opacity index ($\beta$), column density ($N_{H_{2}}$), and dust temperature ($T_{d}$), in the 2-dimensional histograms here.
To avoid confusion, we only present the pixels which have column densities higher than our given thresholds. 
Some sources show degeneracy of the fitted dust temperature and opacity index where the dust temperature is below $\sim$25 K.
An immediate diagnosis of such degeneracy may be the anti-correlated $\beta$ and $T_{d}$ values in the left column of Figure \ref{fig:hist2d}.
This might be alleviated by including better measurements in the long wavelength bands (e.g. 1200, 2000, and 3000 \micron), which are closer to the Rayleigh-Jeans limit, and therefore more sensitive to $\beta$.
This problem may also be alleviated by introducing more realistic dust models.

We found that including $\beta$ as a free parameter in our SED fittings is necessary for good convergence. 
The small uncertainties in $T_{d}$ and $\beta$ do not seriously bias our statistical comparisons of the $N_{H_{2}}$ distributions, given the large dynamic range of $N_{H_{2}}$ we are probing.
From the middle column of Figure \ref{fig:hist2d}, we see that the degeneracy of fits of $T_{d}$ and $\beta$ is the most serious when $N_{H_{2}}$ is low.
Therefore, part of the degeneracy is because of the impossibility of precisely determining the fit values when the signal to noise ratio is low. 
The scattered noisy fits would follow a trend due to the degeneracy.
Our present works do not analyze the dust opacity index.
Most of our quantitative analyses focus on the high-density structures, and therefore are not very biased by the degeneracy of fits.

In general, most of the observed sources show a weak anti-correlation between $N_{H_{2}}$ and $T_{d}$.
This may be interpreted as high $N_{H_{2}}$ structures also having high volume density, and therefore can be self-shielded from the external illumination from OB stars.
However, in the right column of Figure \ref{fig:hist2d}, we can still identify some 'trajectories' with positively correlated $N_{H_{2}}$ and $T_{d}$.
The most prominent ones can be seen in the cases of W33, W43-South, W49A, and G10.6-0.4.
For these cases, the deeply embedded luminous OB clusters in the dense molecular gas clumps/cores, may dominate the localized heating.
We think quantifying the features we observed in the $N_{H_{2}}$ and $T_{d}$ may be important for systematically diagnosing the evolutionary stages or star-forming activities of the OB cluster-forming molecular clouds in the high angular resolution observations.
However, it is not yet clear to us how to take the effect of spatial resolution into consideration.
More clues from the numerical simulations may be required to design the most meaningful quantification.


\section{C. Limitations and caveats}\label{section:caveat}
This work intends to systematically characterize the similarities and differences of the morphology of several most luminous OB cluster-forming molecular clouds in the Milky Way, using statistical approaches.
While some clues can indeed be provided, and may be linked to the physical mechanisms, the cross-comparisons of these results should still be regarded as preliminary concept experiments.
Our major difficulties are a direct consequence of the fact that the luminous OB cluster-forming regions are rare. Our presently yet small number of samples is not ideal for comparison in a statistical sense. 

In addition, the difficulty in collecting a significant sample with uniform distances results in different spatial (linear) resolutions and mass sensitivities of the derived $T_{d}$ and $N_{H_{2}}$ images.
The comparison of the N-PDF may be biased in the higher column density ends, because the higher angular resolution observations can resolve the denser, localized clumps and cores. 
Nevertheless, excluding the sources W33, G10.3-0.1, and W49A, which have rather different distances from the rest of the samples (Table \ref{tab:source_info}), will not significantly impact to our tentative conclusions made in the discussion of the N-PDF (Section \ref{subsection:pdf}).
In addition, the most distant ($\sim$11.4 kpc) source W49A shows the most significant excess at the high column density end of its N-PDF, which is expected to become even more significant if it were observed with the same spatial resolution with the $d$$\sim$5 kpc targets. 
Instrumentally, what is keeping us from achieving the uniform and high spatial resolution of all samples, is the limited angular resolution observations of the {\it Herschel} PACS bands. 
The ongoing developments of the ground based Terahertz bolometric instruments may improve the angular resolution of these wavelength bands after combining with the space-telescope observations, which will be critical to make the break-through observationally.
The observations of the 450-1100 \micron\ bands can already be combined with Atacama Large Millimeter Array (ALMA) 12 m-Array and Atacama Compact Array (ACA), and the Submillimeter Array (SMA) observations to achieve high quality images with high angular resolutions. 

Our fits of a modified black-body spectrum to constrain dust temperature, opacity index, and column density (Section \ref{subsection:SED}) are based on independent measurements of the SED at 70, 160, 250, 350, $\sim$500, and 850 \micron\ (and additionally 1200 \micron\ for the source G10.6-0.4, which is missing the 350 \micron\ band).
The available six wavelength bands fundamentally limit the robustness of the fits to more than one modified black-body component along the line of sight.
However, the assumption of a single dominant black-body component breaks down in some diffuse regions, which leads to jumps in the derived column density (Section \ref{subsection:maps}).
The accuracy of the derived $N_{\rm H_{2}}$ is therefore is uncertain in the low column density regime, and is excluded from all of our quantitative analysis.
This excludes a possible log-normal component at low column densities. 
We expect the same concern would apply to most of the similar previous works, and might be resolved by including observations of more wavelength bands.
For the regions where the SEDs are dominated by a single dense component (e.g. dense clumps, cores, or filaments), our derived $N_{\rm H_{2}}$ based on SED fittings is expected to be more accurate than those derived based on a single wavelength band, with assumptions of a constant temperature and dust opacity index.
We note that for sources showing crowded molecular gas structures, the high angular resolution ($\sim$10$''$) we achieved is particularly helpful for avoiding blending of distinct components in the line of sight, and therefore can aid to the accuracy of SED fits.
Converting the derived dust column density to $N_{\rm H_{2}}$ may be subject to the uncertainty of the gas-to-dust mass ratio. 
We expect this uncertainty to become serious on the $<$0.1 pc scale or smaller, which is not yet probed by our present images. 
The non-uniform mass sensitivities of our samples is essentially because the archival observational data were not designed for our particular scientific purposes. 
Nevertheless, the achieved sensitivities are already adequate for recovering the typical observed dense cores/clumps that can be resolved by our achieved spatial resolution (e.g. Section \ref{subsection:dendrogram}).
The issue of non-uniform mass sensitivities mainly impacts the comparison of the low column density, extended and diffused structures, which were excluded from most of our quantitative analysis due to confusion of foreground/background, and the uncertainties in the SED fittings.

In spite of these known issues, we argue that the resolved $T_{\rm d}$ and $N_{\rm H_{s}}$ images in this work provide invaluable information of individual sources. 
Observing molecular gas tracers with comparable spatial resolutions will provide information on whether the identified dense gas structures are gravitationally bound, how the (proto)stellar feedback shapes the velocity fields, density, and temperature distributions, and may help clarify how the dense gas structures are connected in the three dimensional space (Li et al. 2015; Qian et al. 2012).

\clearpage
\vspace{0.5cm}
\section{D. Dendrograms}\label{appendix:dendrogram}
We plotted the dendrogram-identified leaves in the left panel of each row. 
Middle panels show the dendrogram tree structure. The right panels show the mass vs. radius relation for all the identified structures, with red squares showing the uncorrected mass vs. radius for leaves, while purple squares show the corrected mass (merging level corrected) vs. radius for leaves. 
The green transparent lines link each leaf up to all its parental structures,  with a darker color indicating more substructures reside in it. 
We note the significant high column density leaves in G10.6-0.4 and W49A make their tree structures look different from the tree structures of the other molecular clouds we observe.

\begin{figure*}[h]
\vspace{-0.1cm}
\begin{tabular}{ p{0.8\linewidth} }
\hspace{0.5cm}\includegraphics[scale=0.3]{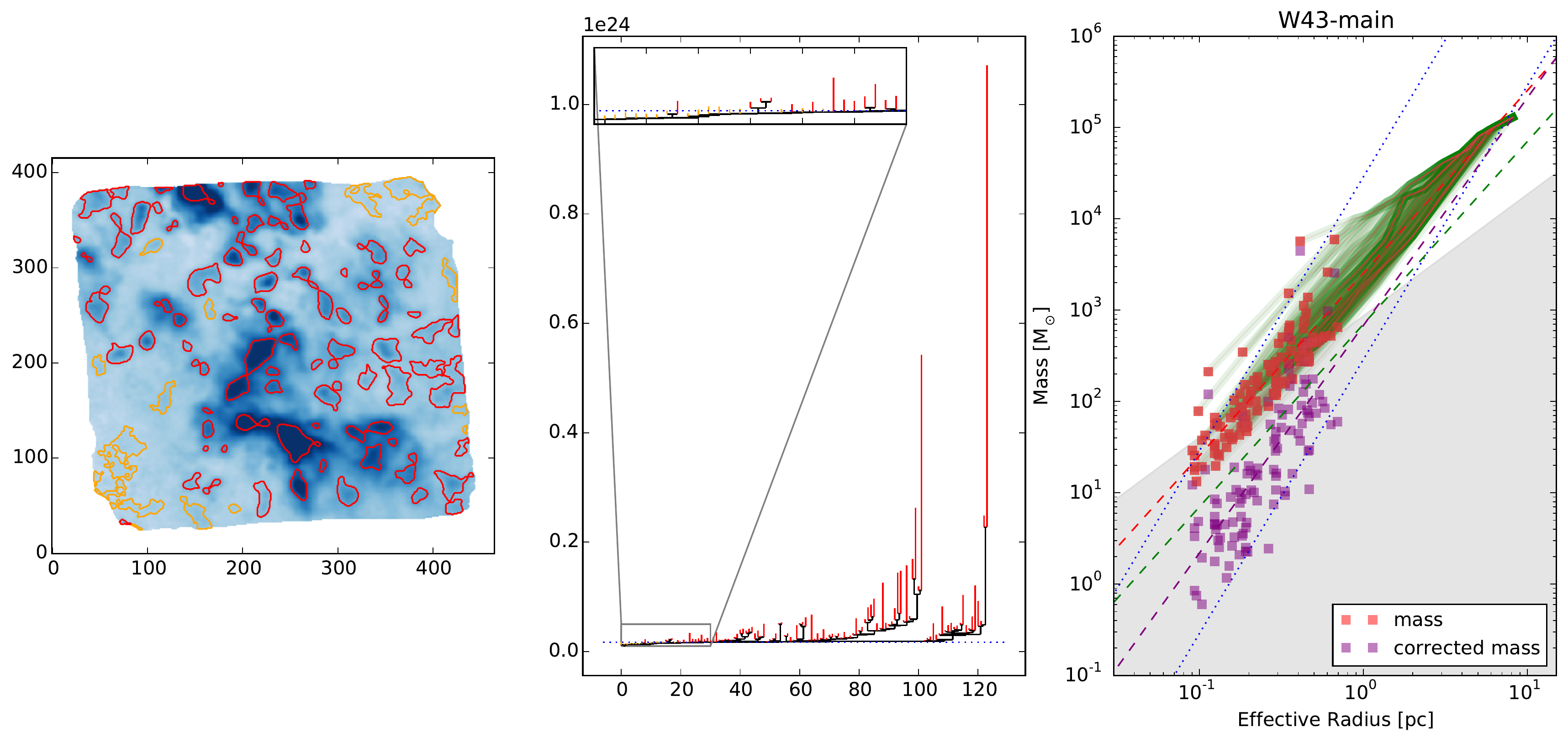} \\
\end{tabular}

\hspace{-0.3cm}
\vspace{-0.1cm}
\begin{tabular}{ p{0.8\linewidth} }
\hspace{0.5cm}\includegraphics[scale=0.3]{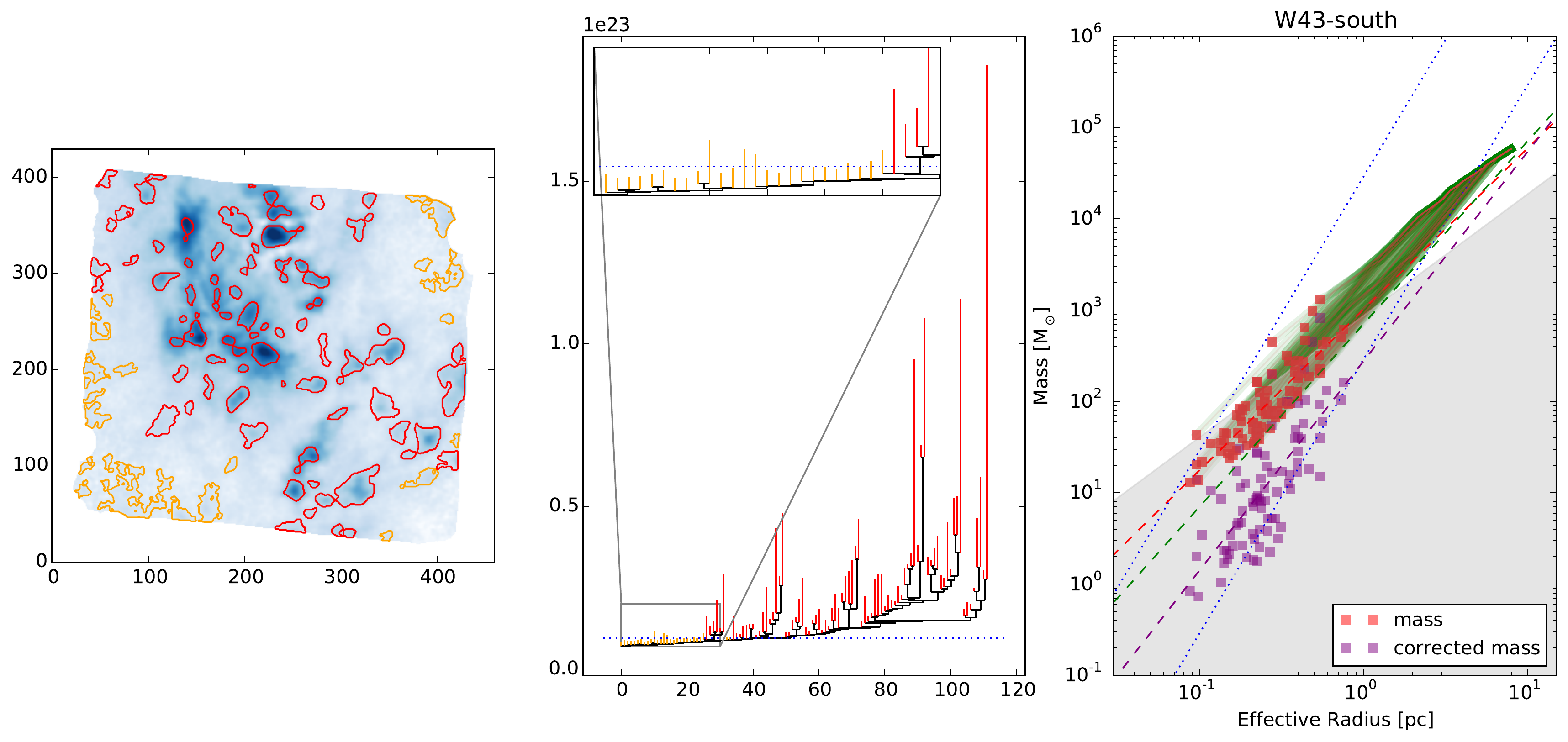} \\
\end{tabular}

\hspace{-0.3cm}
\vspace{-0.1cm}
\begin{tabular}{ p{0.8\linewidth} }
\hspace{0.5cm}\includegraphics[scale=0.3]{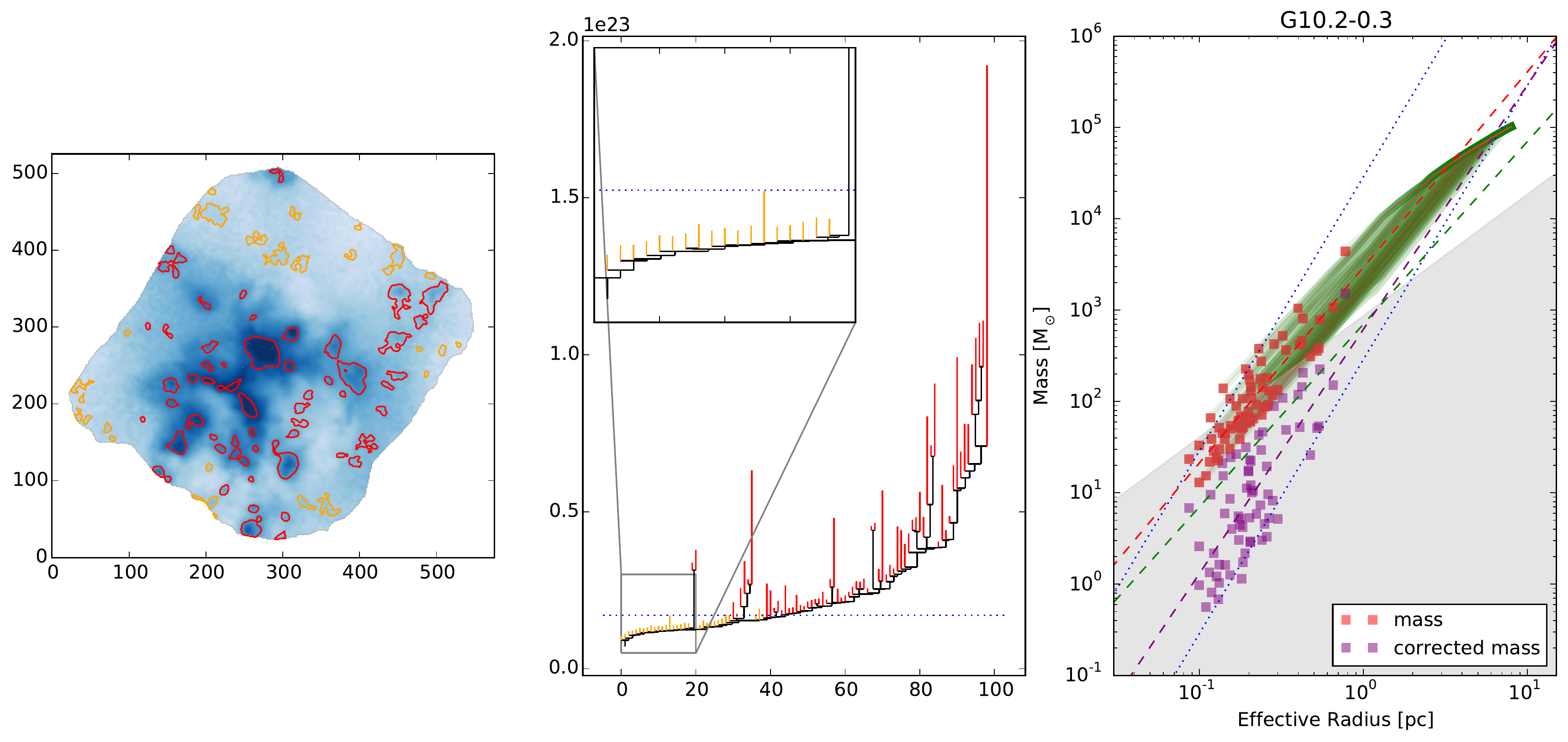} \\
\end{tabular}

\hspace{-0.3cm}
\vspace{-0.1cm}
\begin{tabular}{ p{0.8\linewidth} }
\hspace{0.5cm}\includegraphics[scale=0.3]{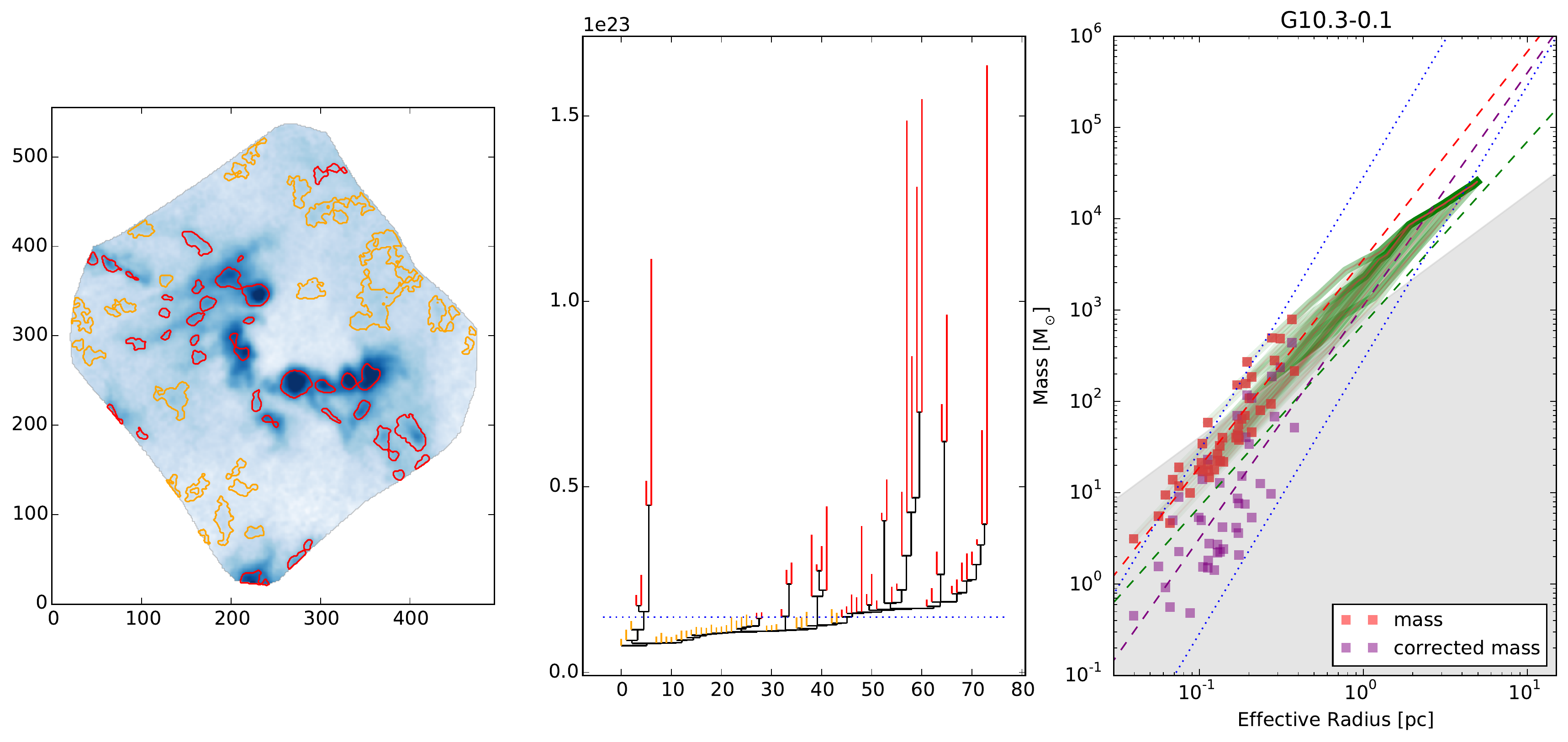} \\
\end{tabular}

\end{figure*}
\hspace{ -0.3cm}
\vspace{ -0.1cm}
\begin{figure*}
\hspace{-0.3cm}
\vspace{-0.1cm}
\begin{tabular}{ p{0.8\linewidth} }
\hspace{0.5cm}\includegraphics[scale=0.3]{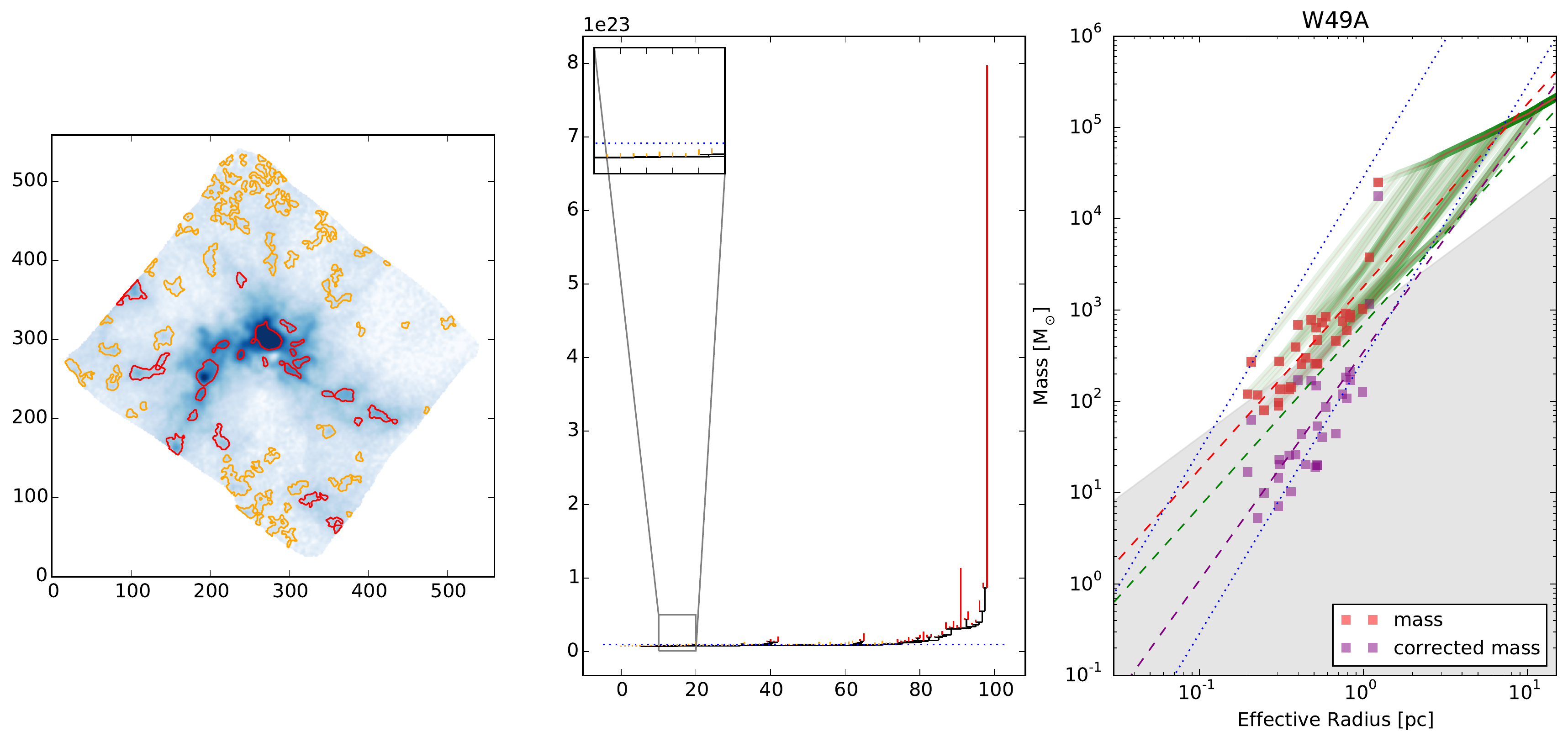} \\
\hspace{0.5cm}\includegraphics[scale=0.3]{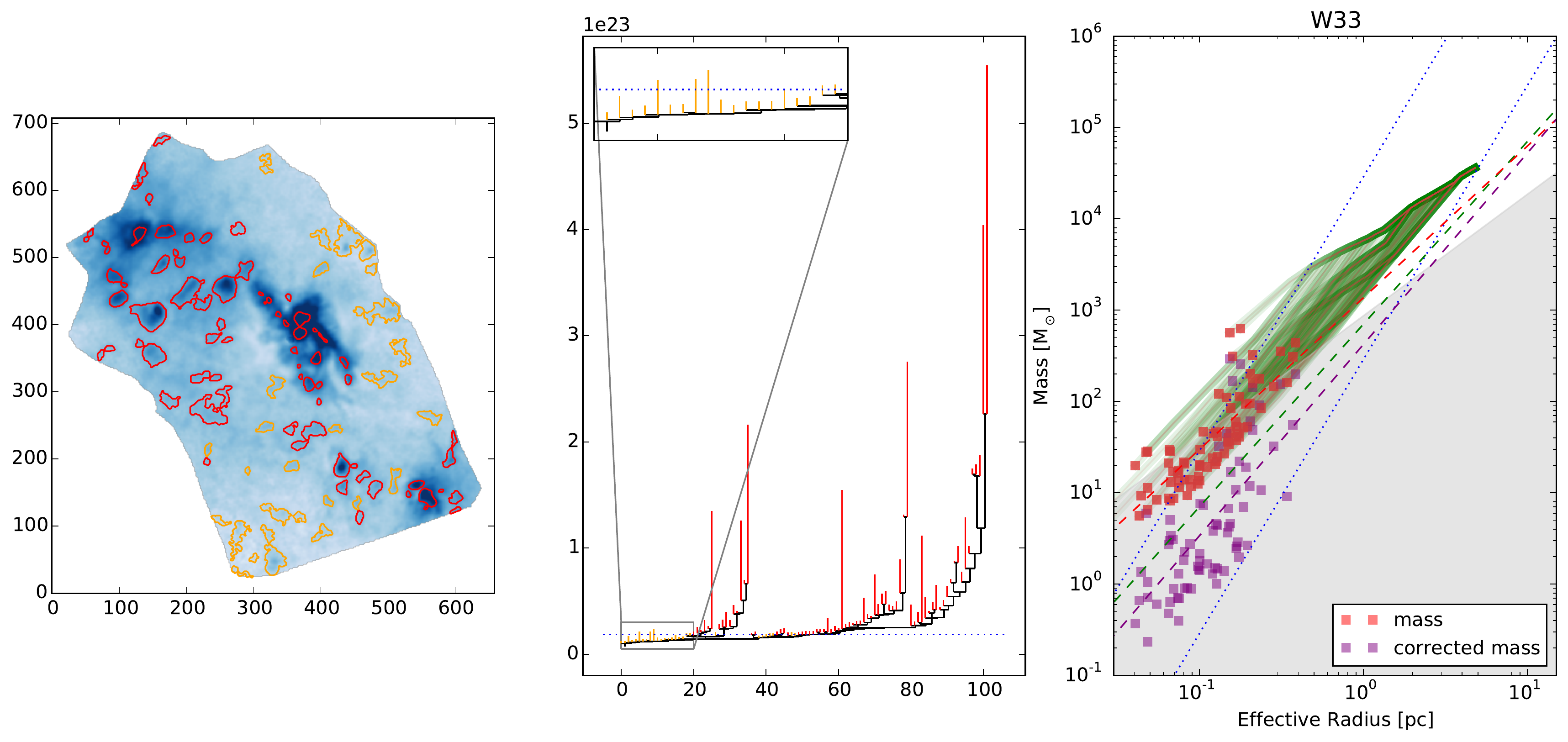} \\
\hspace{0.5cm}\includegraphics[scale=0.3]{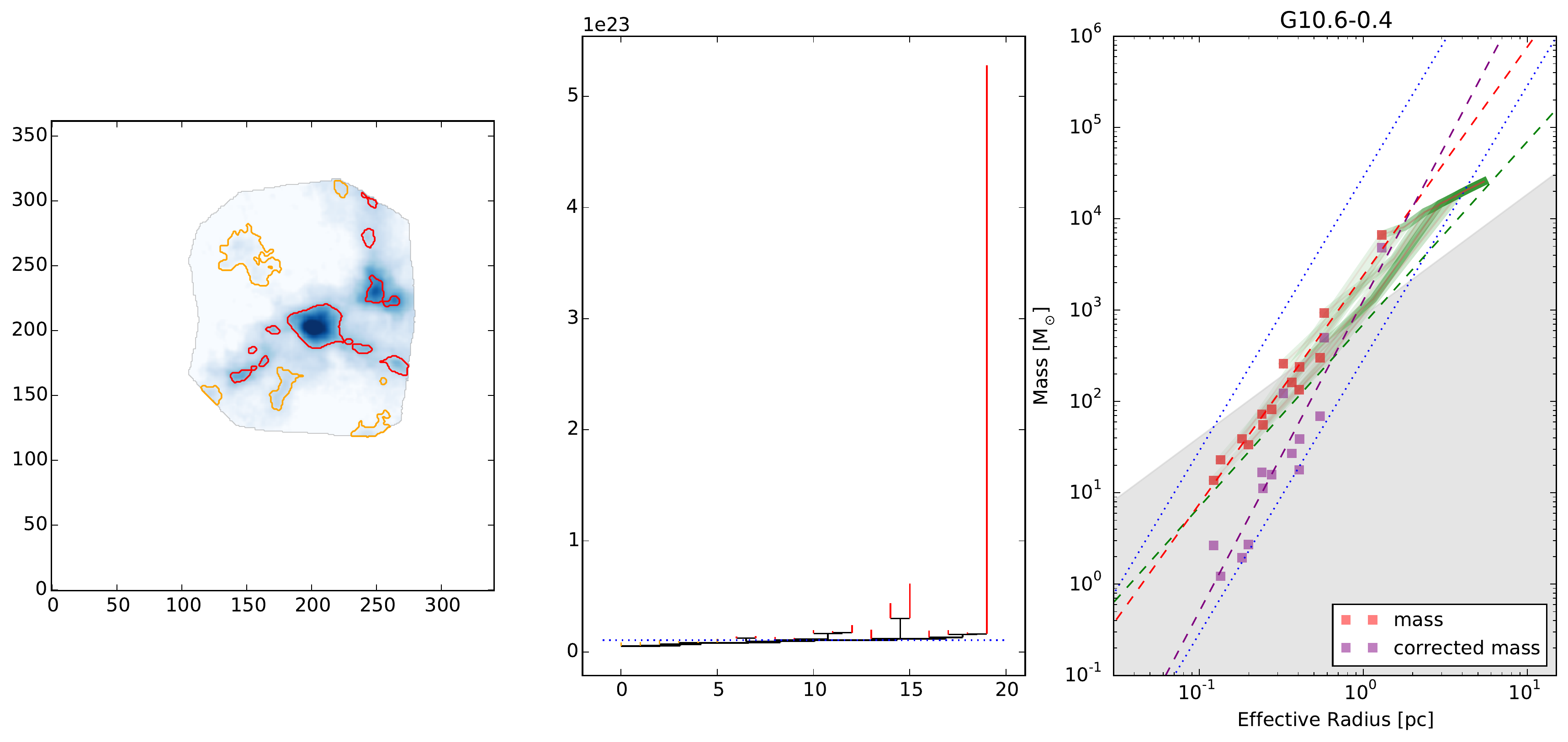}
\\
\end{tabular}

\caption{The dendrogram identified leaves are in the left panel of each row. }
\end{figure*}

\section{E. Comparing the results of SED fitting when including and excluding the 70 \textnormal{$\micron$} data}

In this section we compare the results of the SED fitting when including and exluding the 70 $\micron$ data. We note that there are previous papers discussing the effect of using shorter wavelength data for single component SED fits, and they suggest that this will lead to a certain bias on derived column densities and dust temperatures (e.g. Roy et al. 2013; Malinen et al. 2011). 
Empirically, we found that including the 70$\mu$m data on the short wavelength side of SED peak, helps with the convergence of fitting $\beta$ and the dust temperature simultaneously. Excluding the 70 $\micron$ data point leads to either an underestimation or unreasonably high dust temperatures ($\> 100$ K) in certain areas (as can be seen in the rightmost panel below, Figure 3).
Overall, the structures analyzed in the present manuscript have a much higher column density than those in Roy et al. (2013) and Malinen et al. (2011).
In addition, the regions we are studying are in general warmer than the case studied by e.g. Roy et al. (2013), so the SED of dust in the dense structures peaks towards shorter wavelengths.
These two factors significantly suppress the fractional contribution of very small grains (VSG) or other confusion at 70 $\mu$m.

From the plots below (Figure 22), it can be seen that with and without including the 70 $\mu$m data point in the fitting leads to a $\sim$20\% difference in the column density for most of the area we observed. As shown in the leftmost panel of Figure 3, the plots of the difference between the PACS 70 $\mu$m flux and the flux derived based on SED fits to all the other wavelengths are generally small and almost symmetric around zero. This shows that the inclusion or not of the 70 $\mu$m data point does not strongly bias our results. 
In the middle panels of Figure 3, we can see that the differences between the column densities derived with and without the 70 $\mu$m data are not strongly correlated with the column density, and therefore do not systematically bias our analysis of the N-PDF and 2PTcorrelation function.
Contrary to the hypothesis that including the $\sim$70 $\mu$m data would lead to overestimating the temperatures of the dense gas structures, we instead found that excluding the 70 $\mu$m data led to an overestimate of dust temperature around some relatively warm regions (e.g. close to the H\textsc{ii} regions), which then lead to an underestimation of the dust column density.

\end{document}